\documentclass[aps,prx,superscriptaddress,twocolumn]{revtex4} 
\usepackage{amsmath}
\usepackage{amssymb}
\usepackage{dsfont}
\usepackage{graphicx}
\usepackage{pstricks}
\usepackage{subfigure}
\usepackage{tikz}
\tikzset{>=latex}
\usetikzlibrary{arrows,matrix,decorations.markings}
\allowdisplaybreaks[1]

\newcommand{\arl}{\ar@{-}|@{>}}
\newcommand{\arr}{\ar@{-}|@{<}}
\newcommand{\aline}{\ar@{-}}
\newcommand{\rmd}{\mathrm{d}}
\newcommand{\rmu}{\mathrm{u}}
\newcommand{\Tmv}[2]{T_{#1 \rightarrow #2}}

\newcommand{\Tact}[1]{\mathcal{T}\Biggl(\begin{matrix}#1\end{matrix}\Biggr)}

\newcommand{\rmv}{\mathrm{v}}

\newcommand{\Z}{\mathbb{Z}}

\newcommand{\ds}[1]{\mathbb{#1}}

\newcommand{\Qb}{\overline{Q}}
\newcommand{\Bb}{\overline{B}}

\newcommand{\nbar}{\overline{n}}

\newcommand{\dual}[1]{{#1}^*}

\newcommand{\ket}[1]{\left|{#1}\right\rangle}

\newcommand{\bket}[1]{\Biggl|{\bmm #1 \emm}\Biggr\rangle}
\newcommand{\bbra}[1]{\Biggl\langle{\bmm #1 \emm}\Biggr|}

\newcommand{\bpm}{\begin{pmatrix}}
\newcommand{\epm}{\end{pmatrix}}
\newcommand{\bmm}{\begin{matrix}}
\newcommand{\emm}{\end{matrix}}

\usetikzlibrary{decorations.markings}
\usetikzlibrary{decorations.pathmorphing,positioning}
\tikzset{snake it/.style={decorate, decoration={snake,amplitude=0.15mm,segment length=1mm}}}
\tikzset{->-/.style={decoration={
			 markings,
			 mark=at position .55 with {\arrow{latex}}},postaction={decorate}}}

\newcommand{\ExpBimoduleDef}{
\Biggl|\;\begin{matrix}
\begin{tikzpicture}[scale=1]
\draw [ultra thick] (2.4,8.8) -- (2.4,9.6);
\draw [->-] (2.4,8.) -- (2.4,8.8);
\node [] at (2.18,8.4) {\scriptsize $j$};
\draw [->-] (3.2,9.4) -- (2.4,9.6);
\node [] at (2.74,9.27) {\scriptsize $a_{2}$};
\draw [->-] (3.2,8.6) -- (2.4,8.8);
\node [] at (2.74,8.47) {\scriptsize $a_{1}$};
\draw [->-] (1.6,9.) -- (2.4,8.8);
\node [] at (2.06,9.13) {\scriptsize $b_{1}$};
\draw [->-] (1.6,9.8) -- (2.4,9.6);
\node [] at (2.06,9.93) {\scriptsize $b_{2}$};
\draw [->-] (2.4,9.6) -- (2.4,10.4);
\node [] at (2.62,10.) {\scriptsize $k$};
\node [fill=white,draw,rounded corners=.06cm] at (2.4,9.6) {\scriptsize $P_M$};
\node [fill=white,draw,rounded corners=.06cm] at (2.4,8.8) {\scriptsize $P_M$};
\end{tikzpicture}\:
\end{matrix}\Biggr\rangle
}
\newcommand{\ExpBimoduleDefAA}{
\Biggl|\;\begin{matrix}
\begin{tikzpicture}[scale=1]
\draw [ultra thick] (3.4,9.1) -- (2.8,9.2);
\draw [ultra thick] (2.8,9.2) -- (2.2,9.3);
\draw [->-] (4.,9.) -- (3.4,9.1);
\node [] at (3.74,9.28) {\scriptsize $a_{2}$};
\draw [->-] (3.8,8.4) -- (3.4,9.1);
\node [] at (3.35,8.61) {\scriptsize $a_{1}$};
\draw [->-] (1.6,9.4) -- (2.2,9.3);
\node [] at (1.86,9.12) {\scriptsize $b_{1}$};
\draw [->-] (1.8,10.) -- (2.2,9.3);
\node [] at (2.25,9.79) {\scriptsize $b_{2}$};
\draw [->-] (2.8,9.2) -- (2.8,10.4);
\node [] at (3.02,9.8) {\scriptsize $k$};
\draw [->-] (2.8,8.) -- (2.8,9.2);
\node [] at (2.58,8.6) {\scriptsize $j$};
\draw [fill] (2.2,9.3) circle [x radius=0.08, y radius=0.08];
\draw [fill] (3.4,9.1) circle [x radius=0.08, y radius=0.08];
\node [fill=white,draw,rounded corners=.06cm] at (2.8,9.2) {\scriptsize $P_M$};
\end{tikzpicture}\:
\end{matrix}\Biggr\rangle
}
\newcommand{\ExpBimoduleDefAB}{
\Biggl|\;\begin{matrix}
\begin{tikzpicture}[scale=0.8]
\draw [->-] (2.4,9.2) -- (2.4,10.4);
\node [] at (2.62,9.8) {\scriptsize $k$};
\draw [->-] (3.2,8.6) -- (2.4,9.2);
\node [] at (2.93,9.08) {\scriptsize $a$};
\draw [->-] (1.6,9.8) -- (2.4,9.2);
\node [] at (1.87,9.32) {\scriptsize $b$};
\draw [->-] (2.4,8.) -- (2.4,9.2);
\node [] at (2.18,8.6) {\scriptsize $i$};
\node [fill=white,draw,rounded corners=.06cm] at (2.4,9.2) {\scriptsize $P_M$};
\end{tikzpicture}\:
\end{matrix}\Biggr\rangle
}
\newcommand{\ExpBimoduleDefAC}{
\Biggl|\;\begin{matrix}
\begin{tikzpicture}[scale=0.8]
\draw [->-] (2.8,8.4) -- (2.2,8.9);
\node [] at (2.64,8.82) {\scriptsize $a$};
\draw [->-] (1.6,10.) -- (2.2,9.5);
\node [] at (1.76,9.58) {\scriptsize $b$};
\draw [->-] (2.2,9.5) -- (2.2,10.4);
\node [] at (2.42,9.95) {\scriptsize $k$};
\draw [->-] (2.2,8.) -- (2.2,8.9);
\node [] at (1.98,8.45) {\scriptsize $i$};
\draw [->-] (2.2,8.9) -- (2.2,9.5);
\node [] at (2.42,9.2) {\scriptsize $j$};
\end{tikzpicture}\:
\end{matrix}\Biggr\rangle
}
\newcommand{\ExpBimoduleDefAD}{
\Biggl|\;\begin{matrix}
\begin{tikzpicture}[scale=1]
\draw [->-] (2.4,9.2) -- (2.4,10.4);
\node [] at (2.62,9.8) {\scriptsize $k$};
\draw [->-] (3.2,8.6) -- (2.4,9.2);
\node [] at (2.93,9.08) {\scriptsize $a$};
\draw [->-] (1.6,9.8) -- (2.4,9.2);
\node [] at (1.87,9.32) {\scriptsize $b$};
\draw [->-] (2.4,8.) -- (2.4,9.2);
\node [] at (2.18,8.6) {\scriptsize $i$};
\node [fill=white,draw,rounded corners=.06cm] at (2.4,9.2) {\scriptsize $P_M$};
\node [] at (2.2,8.9) {\scriptsize $\alpha$};
\node [] at (2.6,9.5) {\scriptsize $\beta$};
\draw [fill] (2.4,9.) circle [x radius=0.08, y radius=0.08];
\draw [fill] (2.4,9.4) circle [x radius=0.08, y radius=0.08];
\end{tikzpicture}\:
\end{matrix}\Biggr\rangle
}
\newcommand{\ExpBimoduleOrtho}{
\Biggl|\;\begin{matrix}
\begin{tikzpicture}[scale=1]
\draw [ultra thick] (3.,8.6) ..controls (2.983,8.817) and (2.967,9.033) .. (2.9,9.2)
			 ..controls (2.833,9.367) and (2.717,9.483) .. (2.6,9.6);
\draw [ultra thick] (3.,8.6) ..controls (2.9,8.633) and (2.8,8.667) .. (2.7,8.7)
			 ..controls (2.6,8.733) and (2.5,8.767) .. (2.4,8.8);
\draw [ultra thick] (2.4,8.8) ..controls (2.233,8.917) and (2.067,9.033) .. (2.,9.2)
			 ..controls (1.933,9.367) and (1.967,9.583) .. (2.,9.8);
\draw [ultra thick] (2.6,9.6) ..controls (2.5,9.683) and (2.4,9.767) .. (2.3,9.8)
			 ..controls (2.2,9.833) and (2.1,9.817) .. (2.,9.8);
\draw [->-] (2.2,8.) -- (2.4,8.8);
\node [] at (2.09,8.45) {\scriptsize $j$};
\draw [->-] (2.4,8.8) -- (2.6,9.6);
\node [] at (2.29,9.25) {\scriptsize $j$};
\draw [->-] (2.6,9.6) -- (2.8,10.4);
\node [] at (2.49,10.05) {\scriptsize $j$};
\draw [->-] (3.4,8.2) -- (3.,8.6);
\node [] at (3.36,8.56) {\scriptsize $a$};
\draw [->-] (2.,9.8) -- (1.6,10.3);
\node [] at (1.97,10.19) {\scriptsize $b$};
\draw [fill] (3.,8.6) circle [x radius=0.08, y radius=0.08];
\draw [fill] (2.,9.8) circle [x radius=0.08, y radius=0.08];
\node [fill=white,draw,rounded corners=.06cm] at (2.6,9.6) {\scriptsize $P_M$};
\node [fill=white,draw,rounded corners=.06cm] at (2.4,8.8) {\scriptsize $P_N$};
\end{tikzpicture}\:
\end{matrix}\Biggr\rangle
}
\newcommand{\ExpBimoduleOrthoAA}{
\Biggl|\;\begin{matrix}
\begin{tikzpicture}[scale=1]
\draw [->-] (2.6,8.6) -- (2.1,9.2);
\node [] at (2.52,9.04) {\scriptsize $a$};
\draw [->-] (2.1,9.2) -- (1.6,9.8);
\node [] at (1.68,9.36) {\scriptsize $b$};
\draw [->-] (2.1,9.2) -- (2.4,10.4);
\node [] at (2.46,9.75) {\scriptsize $j$};
\draw [->-] (1.8,8.) -- (2.1,9.2);
\node [] at (2.16,8.55) {\scriptsize $j$};
\node [fill=white,draw,rounded corners=.06cm] at (2.1,9.2) {\scriptsize $P_M$};
\end{tikzpicture}\:
\end{matrix}\Biggr\rangle
}
\newcommand{\ExpBoundaryBbpAA}{
\Biggl|\;\begin{matrix}
\begin{tikzpicture}[scale=0.7]
\path [lightgray, fill] (1.6,7.2) -- (2.5,7.2) -- (2.7,8.) -- (2.5,8.8) -- (2.7,9.6) -- (2.5,10.4) -- (1.6,10.4) -- cycle;
\draw [->-] (1.9,8.8) -- (2.5,8.8);
\node [] at (2.2,8.58) {\scriptsize $j_{5}$};
\draw [->-] (2.7,8.) -- (3.5,8.);
\node [] at (3.1,7.78) {\scriptsize $a_{1}$};
\draw [->-] (2.7,9.6) -- (2.5,10.4);
\node [] at (2.88,10.07) {\scriptsize $j_{4}$};
\draw [->-] (2.7,9.6) -- (3.5,9.6);
\node [] at (3.1,9.38) {\scriptsize $a_{2}$};
\draw [->-] (2.5,8.8) -- (2.7,9.6);
\node [] at (2.32,9.27) {\scriptsize $j_{3}$};
\draw [->-] (2.7,8.) -- (2.5,8.8);
\node [] at (2.88,8.47) {\scriptsize $j_{2}$};
\draw [->-] (2.5,7.2) -- (2.7,8.);
\node [] at (2.32,7.67) {\scriptsize $j_{1}$};
\end{tikzpicture}\:
\end{matrix}\Biggr\rangle
}
\newcommand{\BoundaryBbpAF}{
\begin{tikzpicture}[scale=0.7]
\path [lightgray, fill] (1.6,7.2) -- (2.5,7.2) -- (2.7,8.) -- (2.5,8.8) -- (2.7,9.6) -- (2.7,9.6) -- (2.5,10.4) -- (1.6,10.4) -- cycle;
\draw [->-] (1.9,8.8) -- (2.5,8.8);
\node [] at (2.2,8.58) {\scriptsize $j_{5}$};
\draw [->-] (2.5,8.8) -- (2.7,9.6);
\node [] at (2.25,9.29) {\scriptsize $j'_{3}$};
\draw [->-] (2.7,9.6) -- (2.5,10.4);
\node [] at (2.88,10.07) {\scriptsize $j_{4}$};
\draw [->-] (2.7,9.6) -- (3.3,9.6);
\node [] at (3.,9.38) {\scriptsize $a'_{2}$};
\draw [->-] (2.5,7.2) -- (2.7,8.);
\node [] at (2.32,7.67) {\scriptsize $j_{1}$};
\draw [->-] (2.7,8.) -- (2.5,8.8);
\node [] at (2.95,8.49) {\scriptsize $j'_{2}$};
\draw [->-] (2.7,8.) -- (3.3,8.);
\node [] at (3.,7.78) {\scriptsize $a'_{1}$};
\end{tikzpicture}
}

\newcommand{\ExpBoundaryBbpAG}{
\Biggl|\;\begin{matrix}
\begin{tikzpicture}[scale=0.7]
\path [lightgray, fill] (1.6,7.2) -- (2.5,7.2) -- (2.7,8.) -- (2.5,8.8) -- (2.7,9.6) -- (2.5,10.4) -- (1.6,10.4) -- cycle;
\draw [->-] (1.9,8.8) -- (2.5,8.8);
\node [] at (2.2,8.58) {\scriptsize $j_{5}$};
\draw [->-] (2.5,8.8) -- (2.7,9.6);
\node [] at (2.32,9.27) {\scriptsize $j_{3}$};
\draw [->-] (2.7,9.6) -- (2.5,10.4);
\node [] at (2.32,9.93) {\scriptsize $j_{4}$};
\draw [->-] (2.7,9.6) -- (3.5,9.6);
\node [] at (3.1,9.82) {\scriptsize $a_{2}$};
\draw [->-] (3.5,8.) -- (4.3,8.);
\node [] at (3.9,8.22) {\scriptsize $a'_{1}$};
\draw [->-] (3.5,9.6) -- (4.3,9.6);
\node [] at (3.9,9.82) {\scriptsize $a'_{2}$};
\draw [->-] (2.5,7.2) -- (2.7,8.);
\node [] at (2.32,7.67) {\scriptsize $j_{1}$};
\draw [->-] (2.7,8.) -- (2.5,8.8);
\node [] at (2.88,8.47) {\scriptsize $j_{2}$};
\draw [->-] (2.7,8.) -- (3.5,8.);
\node [] at (3.1,7.78) {\scriptsize $a_{1}$};
\draw [ultra thick] (3.5,9.6) -- (3.5,8.);
\draw [fill] (3.5,9.6) circle [x radius=0.08, y radius=0.08];
\draw [fill] (3.5,8.) circle [x radius=0.08, y radius=0.08];
\node [] at (2.9,8.8) {\scriptsize $\times$};
\end{tikzpicture}\:
\end{matrix}\Biggr\rangle
}
\newcommand{\ExpBoundaryBbpAH}{
\displaystyle\sum_{t,a'_{2}}\frac{f_{t^* {a'_{2}}^* a_{2}}\mathrm{u}_{t}\mathrm{u}_{a'_{2}}}{\mathrm{u}_{a_{2}}\sqrt{\mathrm{d}_A}}\displaystyle\sum_{a'_{1}}\frac{f_{a_{1} t {a'_{1}}^*}\mathrm{u}_{a'_{1}}\sqrt{D}}{\mathrm{u}_{a_{1}}\mathrm{u}_{t}\sqrt{\mathrm{d}_A}}
\Biggl|\;\begin{matrix}
\begin{tikzpicture}[scale=0.7]
\path [lightgray, fill] (1.6,7.2) -- (2.5,7.2) -- (2.7,8.) -- (3.5,8.) -- (3.3,9.6) -- (2.7,9.6) -- (2.5,10.4) -- (1.6,10.4) -- cycle;
\draw [->-] (1.9,8.8) -- (2.5,8.8);
\node [] at (2.2,8.58) {\scriptsize $j_{5}$};
\draw [->-] (2.5,8.8) -- (2.7,9.6);
\node [] at (2.32,9.27) {\scriptsize $j_{3}$};
\draw [->-] (2.7,9.6) -- (2.5,10.4);
\node [] at (2.32,9.93) {\scriptsize $j_{4}$};
\draw [->-] (3.3,9.6) -- (4.1,9.6);
\node [] at (3.7,9.38) {\scriptsize $a'_{2}$};
\draw [->-] (3.5,8.) -- (4.2,8.);
\node [] at (3.85,7.78) {\scriptsize $a'_{1}$};
\draw [->-] (2.7,9.6) -- (3.3,9.6);
\node [] at (3.,9.82) {\scriptsize $a_{2}$};
\draw [->-] (2.5,7.2) -- (2.7,8.);
\node [] at (2.32,7.67) {\scriptsize $j_{1}$};
\draw [->-] (2.7,8.) -- (2.5,8.8);
\node [] at (2.88,8.47) {\scriptsize $j_{2}$};
\draw [->-] (2.7,8.) -- (3.5,8.);
\node [] at (3.1,7.78) {\scriptsize $a_{1}$};
\draw [->-] (3.3,9.6) -- (3.5,8.);
\node [] at (3.62,8.83) {\scriptsize $t$};
\end{tikzpicture}\:
\end{matrix}\Biggr\rangle
}
\newcommand{\ExpBoundaryBbpAI}{
\displaystyle\sum_{t,a'_{2}}\frac{f_{t^* {a'_{2}}^* a_{2}}\mathrm{u}_{t}\mathrm{u}_{a'_{2}}}{\mathrm{u}_{a_{2}}\sqrt{\mathrm{d}_A}}\displaystyle\sum_{a'_{1}}\frac{f_{a_{1} t {a'_{1}}^*}\mathrm{u}_{a'_{1}}\sqrt{D}}{\mathrm{u}_{a_{1}}\mathrm{u}_{t}\sqrt{\mathrm{d}_A}}\displaystyle\sum_{j'_{3}}G^{j_{4}^* j_{3} a_{2}^*}_{t^* {a'_{2}}^* j'_{3}}\mathrm{v}_{a_{2}}\mathrm{v}_{j'_{3}}
\Biggl|\;\begin{matrix}
\begin{tikzpicture}[scale=0.7]
\path [lightgray, fill] (1.6,7.2) -- (2.5,7.2) -- (2.9,8.) -- (3.7,8.) -- (3.,9.3) -- (3.,10.) -- (2.5,10.4) -- (1.6,10.4) -- cycle;
\draw [] (3.,9.3) -- (3.,10.);
\node [] at (3.37,9.65) {\scriptsize $j'_{3}$};
\draw [->-] (1.9,8.8) -- (2.5,8.8);
\node [] at (2.2,8.58) {\scriptsize $j_{5}$};
\draw [->-] (2.5,7.2) -- (2.9,8.);
\node [] at (2.96,7.47) {\scriptsize $j_{1}$};
\draw [->-] (2.5,8.8) -- (3.,9.3);
\node [] at (2.56,9.24) {\scriptsize $j_{3}$};
\draw [->-] (2.9,8.) -- (2.5,8.8);
\node [] at (2.44,8.27) {\scriptsize $j_{2}$};
\draw [->-] (2.9,8.) -- (3.7,8.);
\node [] at (3.3,7.78) {\scriptsize $a_{1}$};
\draw [->-] (3.,10.) -- (2.5,10.4);
\node [] at (2.58,9.99) {\scriptsize $j_{4}$};
\draw [->-] (3.,10.) -- (4.3,10.);
\node [] at (3.65,10.22) {\scriptsize $a'_{2}$};
\draw [->-] (3.7,8.) -- (4.4,8.);
\node [] at (4.05,7.78) {\scriptsize $a'_{1}$};
\draw [->-] (3.,9.3) -- (3.7,8.);
\node [] at (3.54,8.75) {\scriptsize $t$};
\end{tikzpicture}\:
\end{matrix}\Biggr\rangle
}
\newcommand{\ExpBoundaryBbpAJ}{
\displaystyle\sum_{t,a'_{2}}\frac{f_{t^* {a'_{2}}^* a_{2}}\mathrm{u}_{t}\mathrm{u}_{a'_{2}}}{\mathrm{u}_{a_{2}}\sqrt{\mathrm{d}_A}}\displaystyle\sum_{a'_{1}}\frac{f_{a_{1} t {a'_{1}}^*}\mathrm{u}_{a'_{1}}\sqrt{D}}{\mathrm{u}_{a_{1}}\mathrm{u}_{t}\sqrt{\mathrm{d}_A}}\displaystyle\sum_{j'_{3}}G^{j_{4}^* j_{3} a_{2}^*}_{t^* {a'_{2}}^* j'_{3}}\mathrm{v}_{a_{2}}\mathrm{v}_{j'_{3}}\displaystyle\sum_{j'_{2}}G^{j_{5} j_{2} j_{3}^*}_{t^* {j'_{3}}^* j'_{2}}\mathrm{v}_{j_{3}}\mathrm{v}_{j'_{2}}
\Biggl|\;\begin{matrix}
\begin{tikzpicture}[scale=0.7]
\path [lightgray, fill] (1.6,7.2) -- (2.5,7.2) -- (2.9,7.8) -- (3.7,8.) -- (2.9,8.5) -- (2.5,8.9) -- (2.9,9.8) -- (2.5,10.4) -- (1.6,10.4) -- cycle;
\draw [] (2.5,8.9) -- (2.9,9.8);
\node [] at (2.38,9.49) {\scriptsize $j'_{3}$};
\draw [->-] (1.9,8.8) -- (2.5,8.9);
\node [] at (2.24,8.62) {\scriptsize $j_{5}$};
\draw [->-] (2.5,7.2) -- (2.9,7.8);
\node [] at (2.93,7.34) {\scriptsize $j_{1}$};
\draw [->-] (2.9,7.8) -- (2.9,8.5);
\node [] at (2.61,8.15) {\scriptsize $j_{2}$};
\draw [->-] (2.9,7.8) -- (3.7,8.);
\node [] at (3.36,7.67) {\scriptsize $a_{1}$};
\draw [->-] (2.9,8.5) -- (2.5,8.9);
\node [] at (2.93,8.93) {\scriptsize $j'_{2}$};
\draw [->-] (2.9,8.5) -- (3.7,8.);
\node [] at (3.42,8.44) {\scriptsize $t$};
\draw [->-] (2.9,9.8) -- (2.5,10.4);
\node [] at (2.93,10.26) {\scriptsize $j_{4}$};
\draw [->-] (2.9,9.8) -- (4.3,9.8);
\node [] at (3.6,10.02) {\scriptsize $a'_{2}$};
\draw [->-] (3.7,8.) -- (4.4,8.);
\node [] at (4.05,7.78) {\scriptsize $a'_{1}$};
\end{tikzpicture}\:
\end{matrix}\Biggr\rangle
}
\newcommand{\ExpBoundaryBbpAM}{
\displaystyle\sum_{t,a'_{2}}\frac{f_{t^* {a'_{2}}^* a_{2}}\mathrm{u}_{t}\mathrm{u}_{a'_{2}}}{\mathrm{u}_{a_{2}}\sqrt{\mathrm{d}_A}}\displaystyle\sum_{a'_{1}}\frac{f_{a_{1} t {a'_{1}}^*}\mathrm{u}_{a'_{1}}\sqrt{D}}{\mathrm{u}_{a_{1}}\mathrm{u}_{t}\sqrt{\mathrm{d}_A}}\displaystyle\sum_{j'_{3}}G^{j_{4}^* j_{3} a_{2}^*}_{t^* {a'_{2}}^* j'_{3}}\mathrm{v}_{a_{2}}\mathrm{v}_{j'_{3}}\displaystyle\sum_{j'_{2}}G^{j_{5} j_{2} j_{3}^*}_{t^* {j'_{3}}^* j'_{2}}\mathrm{v}_{j_{3}}\mathrm{v}_{j'_{2}}G^{t^* {j'_{2}}^* j_{2}}_{j_{1} a_{1}^* a'_{1}}\mathrm{v}_{j_{2}}\mathrm{v}_{a'_{1}}\frac{\mathrm{v}_{t}\mathrm{v}_{a_{1}}}{\mathrm{v}_{a'_{1}}\sqrt{D}}
\Biggl|\;\begin{matrix}
\begin{tikzpicture}[scale=0.7]
\path [lightgray, fill] (1.6,7.2) -- (2.6,7.2) -- (2.8,8.) -- (2.6,8.8) -- (2.8,9.6) -- (2.6,10.4) -- (1.6,10.4) -- cycle;
\draw [] (2.6,8.8) -- (2.8,9.6);
\node [] at (2.35,9.29) {\scriptsize $j'_{3}$};
\draw [->-] (2.,8.8) -- (2.6,8.8);
\node [] at (2.3,8.58) {\scriptsize $j_{5}$};
\draw [->-] (2.8,9.6) -- (3.6,9.6);
\node [] at (3.2,9.82) {\scriptsize $a'_{2}$};
\draw [->-] (2.8,8.) -- (3.6,8.);
\node [] at (3.2,7.78) {\scriptsize $a'_{1}$};
\draw [->-] (2.6,7.2) -- (2.8,8.);
\node [] at (2.42,7.67) {\scriptsize $j_{1}$};
\draw [->-] (2.8,8.) -- (2.6,8.8);
\node [] at (3.05,8.49) {\scriptsize $j'_{2}$};
\draw [->-] (2.8,9.6) -- (2.6,10.4);
\node [] at (2.42,9.93) {\scriptsize $j_{4}$};
\end{tikzpicture}\:
\end{matrix}\Biggr\rangle
}
\newcommand{\ExpBoundaryBbpSimple}{
\Biggl|\;\begin{matrix}
\begin{tikzpicture}[scale=0.7]
\path [lightgray, fill] (1.6,8.) -- (2.3,8.) -- (2.3,10.4) -- (1.6,10.4) -- cycle;
\draw [->-] (2.3,8.) -- (2.3,8.8);
\node [] at (2.01,8.4) {\scriptsize $j_{1}$};
\draw [->-] (2.3,8.8) -- (2.3,9.6);
\node [] at (2.01,9.2) {\scriptsize $j_{2}$};
\draw [->-] (2.3,9.6) -- (2.3,10.4);
\node [] at (2.01,10.) {\scriptsize $j_{3}$};
\draw [->-] (2.3,8.8) -- (2.8,8.8);
\node [] at (2.55,8.58) {\scriptsize $a_{1}$};
\draw [->-] (2.3,9.6) -- (2.8,9.6);
\node [] at (2.55,9.38) {\scriptsize $a_{2}$};
\end{tikzpicture}\:
\end{matrix}\Biggr\rangle
}
\newcommand{\BoundaryDiskEff}[3]{
\begin{tikzpicture}[scale=0.7]
\draw [lightgray, fill] (2.2,9.) ..controls (2.2,8.771) and (2.257,8.543) .. (2.4,8.4)
			 ..controls (2.543,8.257) and (2.771,8.2) .. (3.,8.2)
			 ..controls (3.229,8.2) and (3.457,8.257) .. (3.6,8.4)
			 ..controls (3.743,8.543) and (3.8,8.771) .. (3.8,9.)
			 ..controls (3.8,9.229) and (3.743,9.457) .. (3.6,9.6)
			 ..controls (3.457,9.743) and (3.229,9.8) .. (3.,9.8)
			 ..controls (2.771,9.8) and (2.543,9.743) .. (2.4,9.6)
			 ..controls (2.257,9.457) and (2.2,9.229) .. (2.2,9.);
\draw [->-] (2.2,9.) ..controls (2.233,8.767) and (2.267,8.533) .. (2.4,8.4)
			 ..controls (2.533,8.267) and (2.767,8.233) .. (3.,8.2);
\draw [->-] (3.,8.2) ..controls (3.233,8.233) and (3.467,8.267) .. (3.6,8.4)
			 ..controls (3.733,8.533) and (3.767,8.767) .. (3.8,9.);
\draw [->-] (3.8,9.) ..controls (3.767,9.233) and (3.733,9.467) .. (3.6,9.6)
			 ..controls (3.467,9.733) and (3.233,9.767) .. (3.,9.8);
\draw [dashed] (3.,9.8) ..controls (2.767,9.767) and (2.533,9.733) .. (2.4,9.6)
			 ..controls (2.267,9.467) and (2.233,9.233) .. (2.2,9.);
\draw [->-] (3.,9.8) -- (3.,10.4);
\draw [->-] (3.8,9.) -- (4.4,9.);
\draw [->-] (3.,8.2) -- (3.,7.6);
\draw [->-] (2.2,9.) -- (1.6,9.);
\node [] at (3.4,7.6) {\scriptsize $a_{1}$};
\node [] at (4.4,8.6) {\scriptsize $a_{2}$};
\node [] at (3.4,10.4) {\scriptsize $a_{3}$};
\node [] at (1.6,8.6) {\scriptsize $a_N$};
\node [] at (3.8,8.2) {\scriptsize $#1$};
\node [] at (3.8,9.8) {\scriptsize $#2$};
\node [] at (2.2,8.2) {\scriptsize $#3$};
\end{tikzpicture}
}

\newcommand{\BoundaryTransform}{
\begin{tikzpicture}[scale=1]
\path [lightgray, fill] (1.6,6.8) -- (2.2,6.8) -- (2.2,10.4) -- (1.6,10.4) -- cycle;
\draw [->-] (2.2,6.8) -- (2.2,7.4);
\draw [->-] (2.2,7.4) -- (2.2,8.);
\draw [->-] (2.2,8.) -- (2.2,8.6);
\draw [->-] (2.2,8.6) -- (2.2,9.2);
\draw [->-] (2.2,9.2) -- (2.2,9.8);
\draw [->-] (2.2,9.8) -- (2.2,10.4);
\draw [->-] (5.2,7.4) -- (5.5,7.4);
\draw [->-] (5.7,7.4) -- (6.,7.4);
\draw [->-] (5.,8.) -- (5.502,8.1);
\draw [->-] (5.698,8.14) -- (6.,8.2);
\draw [->-] (5.2,8.8) -- (5.5,8.8);
\draw [->-] (5.7,8.8) -- (6.,8.8);
\draw [->-] (5.4,9.4) -- (5.5,9.4);
\draw [->-] (5.7,9.4) -- (6.,9.4);
\draw [gray, dashed, very thick] (3.,6.8) -- (5.6,6.8);
\draw [gray, dashed, very thick] (5.6,6.8) -- (5.6,10.4);
\draw [gray, dashed, very thick] (5.6,10.4) -- (3.,10.4);
\draw [ultra thick] (2.8,9.8) -- (2.911,9.745);
\draw [ultra thick] (3.089,9.655) -- (3.2,9.6);
\draw [ultra thick] (3.9,8.5) -- (4.4,8.6);
\draw [ultra thick] (2.8,9.2) -- (2.929,9.329);
\draw [ultra thick] (3.071,9.471) -- (3.2,9.6);
\draw [ultra thick] (2.8,7.4) -- (2.903,7.426);
\draw [ultra thick] (3.097,7.474) -- (3.6,7.6);
\draw [ultra thick] (3.6,7.6) -- (4.2,7.4);
\draw [ultra thick] (4.6,7.8) -- (5.2,7.4);
\draw [ultra thick] (4.4,8.6) -- (4.8,8.8);
\draw [ultra thick] (4.8,8.8) -- (5.2,8.8);
\draw [ultra thick] (2.8,8.6) -- (2.9,8.6);
\draw [ultra thick] (3.1,8.6) -- (3.4,8.6);
\draw [ultra thick] (2.8,8.) -- (2.929,8.129);
\draw [ultra thick] (3.071,8.271) -- (3.4,8.6);
\draw [ultra thick] (3.7,8.1) -- (3.9,8.5);
\draw [ultra thick] (3.6,9.2) -- (4.,9.);
\draw [ultra thick] (4.,9.) -- (4.3,9.5);
\draw [ultra thick] (4.6,7.8) -- (5.,8.);
\draw [ultra thick] (4.2,7.4) -- (5.2,7.4);
\draw [ultra thick] (5.2,8.8) -- (5.4,9.4);
\draw [ultra thick] (5.2,9.8) -- (5.4,9.4);
\draw [ultra thick] (2.2,9.8) -- (2.8,9.8);
\draw [ultra thick] (2.2,9.2) -- (2.8,9.2);
\draw [ultra thick] (2.2,8.6) -- (2.8,8.6);
\draw [ultra thick] (2.2,8.) -- (2.8,8.);
\draw [ultra thick] (2.2,7.4) -- (2.8,7.4);
\draw [gray, dashed, very thick] (3.,10.4) -- (3.,6.8);
\draw [very thick, ultra thick] (3.6,9.2) -- (3.4,8.6);
\draw [very thick, ultra thick] (3.4,8.6) -- (3.9,8.5);
\draw [very thick, ultra thick] (3.7,8.1) -- (3.6,7.6);
\draw [very thick, ultra thick] (3.7,8.1) -- (4.2,8.);
\draw [very thick, ultra thick] (4.2,8.) -- (4.6,7.8);
\draw [very thick, ultra thick] (4.2,7.4) -- (4.2,8.);
\draw [very thick, ultra thick] (4.3,9.5) -- (5.2,9.8);
\draw [very thick, ultra thick] (4.8,8.8) -- (5.2,9.8);
\draw [very thick, ultra thick] (4.,9.) -- (4.4,8.6);
\draw [very thick, ultra thick] (3.2,9.6) -- (3.7,9.6);
\draw [very thick, ultra thick] (3.7,9.6) -- (4.3,9.5);
\draw [very thick, ultra thick] (3.6,9.2) -- (3.7,9.6);
\draw [fill] (3.2,9.6) circle [x radius=0.08, y radius=0.08];
\draw [fill] (3.6,9.2) circle [x radius=0.08, y radius=0.08];
\draw [fill] (4.,9.) circle [x radius=0.08, y radius=0.08];
\draw [fill] (3.4,8.6) circle [x radius=0.08, y radius=0.08];
\draw [fill] (3.6,7.6) circle [x radius=0.08, y radius=0.08];
\draw [fill] (3.7,8.1) circle [x radius=0.08, y radius=0.08];
\draw [fill] (3.9,8.5) circle [x radius=0.08, y radius=0.08];
\draw [fill] (4.4,8.6) circle [x radius=0.08, y radius=0.08];
\draw [fill] (4.8,8.8) circle [x radius=0.08, y radius=0.08];
\draw [fill] (5.2,8.8) circle [x radius=0.08, y radius=0.08];
\draw [fill] (5.,8.) circle [x radius=0.08, y radius=0.08];
\draw [fill] (4.2,7.4) circle [x radius=0.08, y radius=0.08];
\draw [fill] (5.2,7.4) circle [x radius=0.08, y radius=0.08];
\draw [fill] (4.2,8.) circle [x radius=0.08, y radius=0.08];
\draw [fill] (4.6,7.8) circle [x radius=0.08, y radius=0.08];
\draw [fill] (3.7,9.6) circle [x radius=0.08, y radius=0.08];
\draw [fill] (4.3,9.5) circle [x radius=0.08, y radius=0.08];
\draw [fill] (5.2,9.8) circle [x radius=0.08, y radius=0.08];
\end{tikzpicture}
}

\newcommand{\BpBasisEff}[5]{
\begin{tikzpicture}[scale=0.8]
\path [lightgray, fill] (1.6,8.) -- (2.4,8.) -- (2.4,10.4) -- (1.6,10.4) -- cycle;
\draw [->-] (2.4,8.) -- (2.4,8.8);
\draw [->-] (2.4,8.8) -- (2.4,9.6);
\draw [->-] (2.4,9.6) -- (2.4,10.4);
\draw [->-] (2.4,8.8) -- (3.,8.8);
\draw [->-] (2.4,9.6) -- (3.,9.6);
\node [] at (3.,9.) {\scriptsize $#1$};
\node [] at (3.,9.8) {\scriptsize $#2$};
\node [] at (2.,8.4) {\scriptsize $#3$};
\node [] at (2.,9.2) {\scriptsize $#4$};
\node [] at (2.,10.) {\scriptsize $#5$};
\end{tikzpicture}
}

\newcommand{\ExpBpBpeqBp}{
\Biggl|\;\begin{matrix}
\begin{tikzpicture}[scale=0.7]
\path [lightgray, fill] (1.6,7.2) -- (2.2,7.2) -- (2.4,8.) -- (2.2,8.8) -- (2.4,9.6) -- (2.2,10.4) -- (1.6,10.4) -- cycle;
\draw [->-] (1.8,8.8) -- (2.2,8.8);
\draw [->-] (2.2,7.2) -- (2.4,8.);
\node [] at (2.02,7.67) {\scriptsize $j_{1}$};
\draw [->-] (2.4,8.) -- (2.2,8.8);
\node [] at (2.02,8.33) {\scriptsize $j_{2}$};
\draw [->-] (2.2,8.8) -- (2.4,9.6);
\node [] at (2.02,9.27) {\scriptsize $j_{3}$};
\draw [->-] (2.4,9.6) -- (2.2,10.4);
\node [] at (2.02,9.93) {\scriptsize $j_{4}$};
\draw [->-] (2.4,8.) -- (3.,8.);
\node [] at (2.7,7.78) {\scriptsize $a_{1}$};
\draw [->-] (2.4,9.6) -- (3.,9.6);
\node [] at (2.7,9.38) {\scriptsize $a_{2}$};
\end{tikzpicture}\:
\end{matrix}\Biggr\rangle
}
\newcommand{\ExpBpBpeqBpAA}{
\Biggl|\;\begin{matrix}
\begin{tikzpicture}[scale=0.7]
\path [lightgray, fill] (1.6,7.2) -- (2.2,7.2) -- (2.3,8.) -- (2.2,8.8) -- (2.3,9.6) -- (2.2,10.4) -- (1.6,10.4) -- cycle;
\draw [->-] (1.8,8.8) -- (2.2,8.8);
\draw [->-] (2.2,7.2) -- (2.3,8.);
\node [] at (1.96,7.64) {\scriptsize $j_{1}$};
\draw [->-] (2.3,8.) -- (2.2,8.8);
\node [] at (1.96,8.36) {\scriptsize $j_{2}$};
\draw [->-] (2.2,8.8) -- (2.3,9.6);
\node [] at (1.96,9.24) {\scriptsize $j_{3}$};
\draw [->-] (2.3,9.6) -- (2.2,10.4);
\node [] at (1.96,9.96) {\scriptsize $j_{4}$};
\draw [->-] (2.3,8.) -- (2.9,8.);
\node [] at (2.6,7.78) {\scriptsize $a_{1}$};
\draw [->-] (3.3,8.) -- (3.9,8.);
\node [] at (3.6,7.78) {\scriptsize $a'_{1}$};
\draw [->-] (2.3,9.6) -- (2.9,9.6);
\node [] at (2.6,9.82) {\scriptsize $a_{2}$};
\draw [->-] (3.3,9.6) -- (3.9,9.6);
\node [] at (3.6,9.82) {\scriptsize $a'_{2}$};
\draw [ultra thick] (2.9,8.) -- (2.9,9.6);
\draw [ultra thick] (2.9,8.) -- (3.3,8.);
\draw [ultra thick] (3.3,8.) -- (3.3,9.6);
\draw [ultra thick] (2.9,9.6) -- (3.3,9.6);
\draw [fill] (2.9,9.6) circle [x radius=0.08, y radius=0.08];
\draw [fill] (2.9,8.) circle [x radius=0.08, y radius=0.08];
\draw [fill] (3.3,9.6) circle [x radius=0.08, y radius=0.08];
\draw [fill] (3.3,8.) circle [x radius=0.08, y radius=0.08];
\node [] at (2.5,8.7) {\scriptsize $\times$};
\node [] at (3.1,8.8) {\scriptsize $\times$};
\end{tikzpicture}\:
\end{matrix}\Biggr\rangle
}
\newcommand{\ExpBpBpeqBpAB}{
\Biggl|\;\begin{matrix}
\begin{tikzpicture}[scale=0.7]
\path [lightgray, fill] (1.6,7.2) -- (2.2,7.2) -- (2.3,8.) -- (2.2,8.8) -- (2.3,9.6) -- (2.2,10.4) -- (1.6,10.4) -- cycle;
\draw [ultra thick] (2.9,8.4) ..controls (3.02,8.444) and (3.14,8.489) .. (3.2,8.6)
			 ..controls (3.26,8.711) and (3.26,8.889) .. (3.2,9.)
			 ..controls (3.14,9.111) and (3.02,9.156) .. (2.9,9.2);
\draw [ultra thick] (2.9,8.4) ..controls (2.78,8.444) and (2.66,8.489) .. (2.6,8.6)
			 ..controls (2.54,8.711) and (2.54,8.889) .. (2.6,9.)
			 ..controls (2.66,9.111) and (2.78,9.156) .. (2.9,9.2);
\draw [->-] (1.8,8.8) -- (2.2,8.8);
\draw [->-] (2.2,7.2) -- (2.3,8.);
\node [] at (1.96,7.64) {\scriptsize $j_{1}$};
\draw [->-] (2.3,8.) -- (2.2,8.8);
\node [] at (1.96,8.36) {\scriptsize $j_{2}$};
\draw [->-] (2.2,8.8) -- (2.3,9.6);
\node [] at (1.96,9.24) {\scriptsize $j_{3}$};
\draw [->-] (2.3,9.6) -- (2.2,10.4);
\node [] at (1.96,9.96) {\scriptsize $j_{4}$};
\draw [->-] (2.3,8.) -- (2.9,8.);
\node [] at (2.6,7.78) {\scriptsize $a_{1}$};
\draw [->-] (2.9,8.) -- (3.5,8.);
\node [] at (3.2,7.78) {\scriptsize $a'_{1}$};
\draw [->-] (2.3,9.6) -- (2.9,9.6);
\node [] at (2.6,9.82) {\scriptsize $a_{2}$};
\draw [->-] (2.9,9.6) -- (3.5,9.6);
\node [] at (3.2,9.82) {\scriptsize $a'_{2}$};
\draw [ultra thick] (2.9,8.) -- (2.9,8.4);
\draw [ultra thick] (2.9,9.2) -- (2.9,9.6);
\draw [fill] (2.9,9.6) circle [x radius=0.08, y radius=0.08];
\draw [fill] (2.9,8.) circle [x radius=0.08, y radius=0.08];
\draw [fill] (2.9,8.4) circle [x radius=0.08, y radius=0.08];
\draw [fill] (2.9,9.2) circle [x radius=0.08, y radius=0.08];
\node [] at (2.9,8.8) {\scriptsize $\times$};
\node [] at (2.4,8.5) {\scriptsize $\times$};
\end{tikzpicture}\:
\end{matrix}\Biggr\rangle
}
\newcommand{\ExpBpBpeqBpAC}{
\Biggl|\;\begin{matrix}
\begin{tikzpicture}[scale=0.7]
\path [lightgray, fill] (1.6,7.2) -- (2.2,7.2) -- (2.3,8.) -- (2.2,8.8) -- (2.3,9.6) -- (2.2,10.4) -- (1.6,10.4) -- cycle;
\draw [->-] (1.8,8.8) -- (2.2,8.8);
\draw [->-] (2.2,7.2) -- (2.3,8.);
\node [] at (1.96,7.64) {\scriptsize $j_{1}$};
\draw [->-] (2.3,8.) -- (2.2,8.8);
\node [] at (1.96,8.36) {\scriptsize $j_{2}$};
\draw [->-] (2.2,8.8) -- (2.3,9.6);
\node [] at (1.96,9.24) {\scriptsize $j_{3}$};
\draw [->-] (2.3,9.6) -- (2.2,10.4);
\node [] at (1.96,9.96) {\scriptsize $j_{4}$};
\draw [->-] (2.3,8.) -- (2.9,8.);
\node [] at (2.6,7.78) {\scriptsize $a_{1}$};
\draw [->-] (2.3,9.6) -- (2.9,9.6);
\node [] at (2.6,9.38) {\scriptsize $a_{2}$};
\draw [->-] (2.9,9.6) -- (3.5,9.6);
\node [] at (3.2,9.38) {\scriptsize $a'_{2}$};
\draw [->-] (2.9,8.) -- (3.5,8.);
\node [] at (3.2,7.78) {\scriptsize $a'_{1}$};
\draw [ultra thick] (2.9,8.) -- (2.9,9.6);
\draw [fill] (2.9,9.6) circle [x radius=0.08, y radius=0.08];
\draw [fill] (2.9,8.) circle [x radius=0.08, y radius=0.08];
\node [] at (2.5,8.8) {\scriptsize $\times$};
\end{tikzpicture}\:
\end{matrix}\Biggr\rangle
}

\newcommand{\ExpBpGraphicalAA}{
\Biggl|\;\begin{matrix}
\begin{tikzpicture}[scale=0.5]
\draw [] (1.6,8.8) -- (2.,9.);
\draw [] (2.,9.) -- (2.,9.6);
\draw [] (2.,9.6) -- (1.6,9.8);
\draw [] (2.,9.6) -- (2.4,9.8);
\draw [] (2.4,8.2) -- (2.4,8.8);
\draw [] (2.4,8.8) -- (2.,9.);
\draw [] (2.4,9.8) -- (2.4,10.4);
\draw [] (2.4,9.8) -- (2.8,9.6);
\draw [] (2.8,9.) -- (2.4,8.8);
\draw [] (2.8,9.6) -- (2.8,9.);
\draw [] (2.8,9.6) -- (3.2,9.8);
\draw [] (3.2,8.8) -- (2.8,9.);
\draw [] (5.7,8.8) -- (6.1,9.);
\draw [] (6.1,9.) -- (6.1,9.6);
\draw [] (6.1,9.6) -- (5.7,9.8);
\draw [] (6.1,9.6) -- (6.5,9.8);
\draw [] (6.5,8.2) -- (6.5,8.8);
\draw [] (6.5,8.8) -- (6.1,9.);
\draw [] (6.5,9.8) -- (6.5,10.4);
\draw [] (6.5,9.8) -- (6.9,9.6);
\draw [] (6.9,9.) -- (6.5,8.8);
\draw [] (6.9,9.6) -- (6.9,9.);
\draw [] (6.9,9.6) -- (7.3,9.8);
\draw [] (7.3,8.8) -- (6.9,9.);
\draw [->, very thick] (3.9,9.3) -- (4.7,9.3);
\node [] at (2.4,9.3) {\scriptsize $\times$};
\node [] at (6.5,9.3) {\scriptsize $\cdot$};
\end{tikzpicture}\:
\end{matrix}\Biggr\rangle
}
\newcommand{\BpsActionAA}{
\begin{tikzpicture}[scale=0.7]
\draw [->-] (1.6,10.4) -- (2.,10.);
\node [] at (1.61,10.01) {\scriptsize $j_{5}$};
\draw [->-] (2.,10.) -- (2.6,9.1);
\node [] at (2.07,9.39) {\scriptsize $j_{1}$};
\draw [->-] (2.6,8.6) -- (2.6,9.1);
\node [] at (2.89,8.85) {\scriptsize $j_{4}$};
\draw [->-] (2.6,9.1) -- (3.2,10.);
\node [] at (3.13,9.39) {\scriptsize $j_{2}$};
\draw [->-] (3.2,10.) -- (2.,10.);
\node [] at (2.6,10.22) {\scriptsize $j_{3}$};
\draw [->-] (3.6,10.4) -- (3.2,10.);
\node [] at (3.59,10.01) {\scriptsize $j_{6}$};
\end{tikzpicture}
}

\newcommand{\BpsActionAB}{
\begin{tikzpicture}[scale=0.7]
\draw [->-] (1.6,10.4) -- (2.,10.);
\node [] at (1.61,10.01) {\scriptsize $j_{5}$};
\draw [->-] (2.6,8.6) -- (2.6,9.);
\node [] at (2.89,8.8) {\scriptsize $j_{4}$};
\draw [->-] (2.6,9.) -- (2.,10.);
\node [] at (2.,9.32) {\scriptsize $j'_{1}$};
\draw [->-] (3.2,10.) -- (2.,10.);
\node [] at (2.6,10.22) {\scriptsize $j'_{3}$};
\draw [->-] (3.2,10.) -- (2.6,9.);
\node [] at (3.2,9.32) {\scriptsize $j'_{2}$};
\draw [->-] (3.6,10.4) -- (3.2,10.);
\node [] at (3.59,10.01) {\scriptsize $j_{6}$};
\end{tikzpicture}
}

\newcommand{\ExpEquivalentStrongFrobeniusAA}{
\frac{\sqrt{D}}{\mathrm{d}_A}\displaystyle\sum_{a_{2},a_{4},a_{5}}f_{a_{4}^* a_{5} a_{6}^*} f_{a_{1} a_{2}^* a_{4}} f_{a_{2} a_{3} a_{5}^*}
\Biggl|\;\begin{matrix}
\begin{tikzpicture}[scale=0.7]
\draw [->-] (2.1,8.7) -- (3.1,8.7);
\node [] at (2.6,8.48) {\scriptsize $a_{2}$};
\draw [->-] (3.6,8.2) -- (3.1,8.7);
\node [] at (3.54,8.64) {\scriptsize $a_{3}$};
\draw [->-] (2.6,9.7) -- (2.1,8.7);
\node [] at (2.09,9.33) {\scriptsize $a_{4}$};
\draw [->-] (3.1,8.7) -- (2.6,9.7);
\node [] at (3.11,9.33) {\scriptsize $a_{5}$};
\draw [->-] (2.6,9.7) -- (2.6,10.4);
\node [] at (2.89,10.05) {\scriptsize $a_{6}$};
\draw [->-] (1.6,8.2) -- (2.1,8.7);
\node [] at (1.66,8.64) {\scriptsize $a_{1}$};
\end{tikzpicture}\:
\end{matrix}\Biggr\rangle
}
\newcommand{\ExpEquivalentStrongFrobeniusAB}{
\frac{\sqrt{D}}{\mathrm{d}_A}\displaystyle\sum_{a_{2},a_{4},a_{5}}f_{a_{4}^* a_{5} a_{6}^*} f_{a_{1} a_{2}^* a_{4}}f_{a_{2} a_{3} a_{5}^*}\nonumber\\&\qquad\times\frac{\mathrm{v}_{a_{2}}\mathrm{v}_{a_{4}}\mathrm{v}_{a_{5}}}{D^{1/2}}G^{a_{6} a_{3}^* a_{1}^*}_{a_{2}^* a_{4} a_{5}^*}
\Biggl|\;\begin{matrix}
\begin{tikzpicture}[scale=0.5]
\draw [->-] (1.6,8.2) -- (2.6,9.);
\node [] at (1.93,8.81) {\scriptsize $a_{1}$};
\draw [->-] (3.6,8.2) -- (2.6,9.);
\node [] at (3.27,8.81) {\scriptsize $a_{3}$};
\draw [->-] (2.6,9.) -- (2.6,10.4);
\node [] at (2.89,9.7) {\scriptsize $a_{6}$};
\end{tikzpicture}\:
\end{matrix}\Biggr\rangle
}
\newcommand{\ExpEquivalentStrongFrobeniusAC}{
f_{a_{6}^* a_{1} a_{3}}
\Biggl|\;\begin{matrix}
\begin{tikzpicture}[scale=0.5]
\draw [->-] (3.6,8.2) -- (2.6,9.);
\node [] at (3.27,8.81) {\scriptsize $a_{3}$};
\draw [->-] (2.6,9.) -- (2.6,10.4);
\node [] at (2.89,9.7) {\scriptsize $a_{6}$};
\draw [->-] (1.6,8.2) -- (2.6,9.);
\node [] at (1.93,8.81) {\scriptsize $a_{1}$};
\end{tikzpicture}\:
\end{matrix}\Biggr\rangle
}
\newcommand{\ExpEquivalentStrongFrobeniusCompactAA}{
\Biggl|\;\begin{matrix}
\begin{tikzpicture}[scale=0.8]
\draw [->-] (1.6,8.2) -- (2.2,8.8);
\node [] at (2.09,8.31) {\scriptsize $j_{1}$};
\draw [->-] (2.6,9.6) -- (2.6,10.4);
\node [] at (2.89,10.) {\scriptsize $j_{6}$};
\draw [->-] (3.6,8.2) -- (3.,8.8);
\node [] at (3.49,8.69) {\scriptsize $j_{3}$};
\draw [ultra thick] (2.2,8.8) -- (3.,8.8);
\draw [ultra thick] (3.,8.8) -- (2.6,9.6);
\draw [ultra thick] (2.6,9.6) -- (2.2,8.8);
\draw [fill] (2.6,9.6) circle [x radius=0.08, y radius=0.08];
\draw [fill] (2.2,8.8) circle [x radius=0.08, y radius=0.08];
\draw [fill] (3.,8.8) circle [x radius=0.08, y radius=0.08];
\end{tikzpicture}\:
\end{matrix}\Biggr\rangle
}
\newcommand{\ExpEquivalentStrongFrobeniusCompactAB}{
\Biggl|\;\begin{matrix}
\begin{tikzpicture}[scale=0.8]
\draw [->-] (2.6,9.) -- (2.6,10.4);
\node [] at (2.89,9.7) {\scriptsize $j_{6}$};
\draw [->-] (1.6,8.2) -- (2.6,9.);
\node [] at (2.27,8.39) {\scriptsize $j_{1}$};
\draw [->-] (3.6,8.2) -- (2.6,9.);
\node [] at (3.27,8.81) {\scriptsize $j_{3}$};
\draw [fill] (2.6,9.) circle [x radius=0.08, y radius=0.08];
\end{tikzpicture}\:
\end{matrix}\Biggr\rangle
}
\newcommand{\FigBBcommute}{
\begin{tikzpicture}[scale=0.8]
\path [lightgray, fill] (1.6,6.4) -- (4.4,6.4) -- (4.8,7.2) -- (4.4,8.) -- (4.8,8.8) -- (4.4,9.6) -- (4.8,10.4) -- (1.6,10.4) -- cycle;
\draw [->-] (4.4,6.4) -- (4.8,7.2);
\draw [->-] (4.8,7.2) -- (4.4,8.);
\draw [->-] (4.4,8.) -- (4.8,8.8);
\draw [->-] (4.8,8.8) -- (4.4,9.6);
\draw [->-] (4.4,9.6) -- (4.8,10.4);
\draw [->-] (3.6,8.) -- (4.4,8.);
\draw [->-] (3.6,9.6) -- (4.4,9.6);
\draw [->-] (4.8,8.8) -- (5.6,8.8);
\draw [->-] (4.8,7.2) -- (5.6,7.2);
\draw [->-] (3.6,8.) -- (3.2,8.8);
\draw [->-] (3.2,8.8) -- (3.6,9.6);
\draw [->-] (3.2,7.2) -- (3.6,8.);
\draw [->-] (2.8,8.8) -- (3.2,8.8);
\draw [->-] (3.6,9.6) -- (3.2,10.4);
\draw [->-] (3.6,6.4) -- (3.2,7.2);
\draw [->-] (2.8,7.2) -- (3.2,7.2);
\node [] at (4.,8.8) {\scriptsize $p$};
\node [] at (5.2,8.) {\scriptsize $p'$};
\end{tikzpicture}
}

\newcommand{\FigBBcommuteAA}{
\begin{tikzpicture}[scale=0.4]
\path [lightgray, fill] (-6.4,6.4) -- (-4.3,6.4) -- (-3.9,7.2) -- (-4.3,8.) -- (-3.9,8.8) -- (-4.3,9.6) -- (-3.9,10.4) -- (-6.4,10.4) -- cycle;
\path [lightgray, fill] (-0.7,6.4) -- (1.4,6.4) -- (1.8,7.2) -- (1.4,8.) -- (1.8,8.8) -- (1.4,9.6) -- (1.8,10.4) -- (-0.7,10.4) -- cycle;
\path [lightgray, fill] (5.9,6.4) -- (8.,6.4) -- (8.4,7.2) -- (8.,8.) -- (8.4,8.8) -- (8.,9.6) -- (8.4,10.4) -- (5.9,10.4) -- cycle;
\draw [] (-5.9,7.2) -- (-5.5,7.2);
\draw [] (-5.9,8.8) -- (-5.5,8.8);
\draw [] (-5.5,7.2) -- (-5.1,8.);
\draw [] (-5.5,8.8) -- (-5.1,9.6);
\draw [] (-5.1,6.4) -- (-5.5,7.2);
\draw [] (-5.1,8.) -- (-5.5,8.8);
\draw [] (-5.1,8.) -- (-4.3,8.);
\draw [] (-5.1,9.6) -- (-5.5,10.4);
\draw [] (-5.1,9.6) -- (-4.3,9.6);
\draw [] (-4.3,6.4) -- (-3.9,7.2);
\draw [] (-4.3,8.) -- (-3.9,8.8);
\draw [] (-4.3,9.6) -- (-3.9,10.4);
\draw [] (-3.9,7.2) -- (-4.3,8.);
\draw [] (-3.9,7.2) -- (-3.1,7.2);
\draw [] (-3.9,8.8) -- (-4.3,9.6);
\draw [] (-3.9,8.8) -- (-3.1,8.8);
\draw [] (-0.2,7.2) -- (0.2,7.2);
\draw [] (-0.2,8.8) -- (0.2,8.8);
\draw [] (0.2,7.2) -- (0.6,8.);
\draw [] (0.2,8.8) -- (0.6,9.6);
\draw [] (0.6,6.4) -- (0.2,7.2);
\draw [] (0.6,8.) -- (0.2,8.8);
\draw [] (0.6,8.) -- (1.4,8.);
\draw [] (0.6,9.6) -- (0.2,10.4);
\draw [] (0.6,9.6) -- (1.4,9.6);
\draw [] (1.4,6.4) -- (1.8,7.2);
\draw [] (1.4,8.) -- (1.8,8.8);
\draw [] (1.4,9.6) -- (1.8,10.4);
\draw [] (1.8,7.2) -- (1.4,8.);
\draw [] (1.8,7.2) -- (2.4,7.2);
\draw [] (1.8,8.8) -- (1.4,9.6);
\draw [] (1.8,8.8) -- (2.4,8.8);
\draw [] (6.4,7.2) -- (6.8,7.2);
\draw [] (6.4,8.8) -- (6.8,8.8);
\draw [] (6.8,7.2) -- (7.2,8.);
\draw [] (6.8,8.8) -- (7.2,9.6);
\draw [] (7.2,6.4) -- (6.8,7.2);
\draw [] (7.2,8.) -- (6.8,8.8);
\draw [] (7.2,8.) -- (8.,8.);
\draw [] (7.2,9.6) -- (6.8,10.4);
\draw [] (7.2,9.6) -- (8.,9.6);
\draw [] (8.,6.4) -- (8.4,7.2);
\draw [] (8.,8.) -- (8.4,8.8);
\draw [] (8.,9.6) -- (8.4,10.4);
\draw [] (8.4,7.2) -- (8.,8.);
\draw [] (8.4,7.2) -- (9.2,7.2);
\draw [] (8.4,8.8) -- (8.,9.6);
\draw [] (8.4,8.8) -- (9.2,8.8);
\draw [very thick, ->] (-2.7,8.4) -- (-1.6,8.4);
\draw [very thick, ->] (3.7,8.4) -- (4.9,8.4);
\draw [ultra thick] (2.4,8.8) -- (2.4,7.2);
\draw [ultra thick] (2.4,8.8) -- (3.2,8.8);
\draw [ultra thick] (2.4,7.2) -- (3.2,7.2);
\draw [fill] (2.4,8.8) circle [x radius=0.1, y radius=0.1];
\draw [fill] (2.4,7.2) circle [x radius=0.1, y radius=0.1];
\node [] at (1.,8.8) {\scriptsize $\times$};
\node [] at (2.,7.9) {\scriptsize $\times$};
\node [] at (7.6,8.8) {\scriptsize $\cdot$};
\end{tikzpicture}
}

\newcommand{\FigBBcommuteAB}{
\begin{tikzpicture}[scale=0.4]
\path [lightgray, fill] (-7.5,6.4) -- (-5.5,6.4) -- (-5.1,7.2) -- (-5.5,8.) -- (-5.1,8.8) -- (-5.5,9.6) -- (-5.1,10.4) -- (-5.5,11.2) -- (-7.5,11.2) -- cycle;
\path [lightgray, fill] (-2.5,6.4) -- (-0.5,6.4) -- (-0.1,7.2) -- (-0.5,8.) -- (-0.1,8.8) -- (-0.5,9.6) -- (-0.1,10.4) -- (-0.5,11.2) -- (-2.5,11.2) -- cycle;
\path [lightgray, fill] (4.6,6.4) -- (6.6,6.4) -- (7.,7.2) -- (6.6,8.) -- (7.,8.8) -- (6.6,9.6) -- (7.,10.4) -- (6.6,11.2) -- (4.6,11.2) -- cycle;
\draw [] (-7.1,7.2) -- (-6.7,7.2);
\draw [] (-7.1,8.8) -- (-6.7,8.8);
\draw [] (-6.7,7.2) -- (-6.3,8.);
\draw [] (-6.7,8.8) -- (-6.3,9.6);
\draw [] (-6.3,6.4) -- (-6.7,7.2);
\draw [] (-6.3,8.) -- (-6.7,8.8);
\draw [] (-6.3,8.) -- (-5.5,8.);
\draw [] (-6.3,9.6) -- (-6.7,10.4);
\draw [] (-6.3,9.6) -- (-5.5,9.6);
\draw [] (-5.5,6.4) -- (-5.1,7.2);
\draw [] (-5.5,8.) -- (-5.1,8.8);
\draw [] (-5.5,9.6) -- (-5.1,10.4);
\draw [] (-5.1,7.2) -- (-5.5,8.);
\draw [] (-5.1,7.2) -- (-4.3,7.2);
\draw [] (-5.1,8.8) -- (-5.5,9.6);
\draw [] (-5.1,8.8) -- (-4.3,8.8);
\draw [very thick, ->] (-3.9,8.4) -- (-2.8,8.4);
\draw [very thick, ->] (2.5,8.4) -- (3.7,8.4);
\draw [] (-7.1,10.4) -- (-6.7,10.4);
\draw [] (-6.7,10.4) -- (-6.3,11.2);
\draw [] (-5.1,10.4) -- (-5.5,11.2);
\draw [] (-5.1,10.4) -- (-4.3,10.4);
\draw [] (-2.1,7.2) -- (-1.7,7.2);
\draw [] (-2.1,8.8) -- (-1.7,8.8);
\draw [] (-1.7,7.2) -- (-1.3,8.);
\draw [] (-1.7,8.8) -- (-1.3,9.6);
\draw [] (-1.3,6.4) -- (-1.7,7.2);
\draw [] (-1.3,8.) -- (-1.7,8.8);
\draw [] (-1.3,8.) -- (-0.5,8.);
\draw [] (-1.3,9.6) -- (-1.7,10.4);
\draw [] (-1.3,9.6) -- (-0.5,9.6);
\draw [] (-0.5,6.4) -- (-0.1,7.2);
\draw [] (-0.5,8.) -- (-0.1,8.8);
\draw [] (-0.5,9.6) -- (-0.1,10.4);
\draw [] (-0.1,7.2) -- (-0.5,8.);
\draw [] (-0.1,7.2) -- (0.7,7.2);
\draw [] (-0.1,8.8) -- (-0.5,9.6);
\draw [] (-0.1,8.8) -- (0.3,8.8);
\draw [] (-2.1,10.4) -- (-1.7,10.4);
\draw [] (-1.7,10.4) -- (-1.3,11.2);
\draw [] (-0.1,10.4) -- (-0.5,11.2);
\draw [] (-0.1,10.4) -- (0.7,10.4);
\draw [] (5.,7.2) -- (5.4,7.2);
\draw [] (5.,8.8) -- (5.4,8.8);
\draw [] (5.4,7.2) -- (5.8,8.);
\draw [] (5.4,8.8) -- (5.8,9.6);
\draw [] (5.8,6.4) -- (5.4,7.2);
\draw [] (5.8,8.) -- (5.4,8.8);
\draw [] (5.8,8.) -- (6.6,8.);
\draw [] (5.8,9.6) -- (5.4,10.4);
\draw [] (5.8,9.6) -- (6.6,9.6);
\draw [] (6.6,6.4) -- (7.,7.2);
\draw [] (6.6,8.) -- (7.,8.8);
\draw [] (6.6,9.6) -- (7.,10.4);
\draw [] (7.,7.2) -- (6.6,8.);
\draw [] (7.,7.2) -- (7.8,7.2);
\draw [] (7.,8.8) -- (6.6,9.6);
\draw [] (7.,8.8) -- (7.8,8.8);
\draw [] (5.,10.4) -- (5.4,10.4);
\draw [] (5.4,10.4) -- (5.8,11.2);
\draw [] (7.,10.4) -- (6.6,11.2);
\draw [] (7.,10.4) -- (7.8,10.4);
\draw [ultra thick] (0.7,7.2) -- (0.3,8.8);
\draw [ultra thick] (0.3,8.8) -- (1.1,8.8);
\draw [ultra thick] (1.1,8.8) -- (0.7,10.4);
\draw [ultra thick] (0.7,10.4) -- (1.5,10.4);
\draw [ultra thick] (1.1,8.8) -- (1.5,8.8);
\draw [ultra thick] (0.7,7.2) -- (1.5,7.2);
\draw [fill] (0.7,10.4) circle [x radius=0.08, y radius=0.08];
\draw [fill] (1.1,8.8) circle [x radius=0.08, y radius=0.08];
\draw [fill] (0.3,8.8) circle [x radius=0.08, y radius=0.08];
\draw [fill] (0.7,7.2) circle [x radius=0.08, y radius=0.08];
\node [] at (-4.8,9.6) {\scriptsize $p$};
\node [] at (-4.8,7.9) {\scriptsize $p'$};
\node [] at (0.2,9.6) {\scriptsize $\times$};
\node [] at (0.1,7.9) {\scriptsize $\times$};
\end{tikzpicture}
}

\newcommand{\FigBBcommuteAC}{
\begin{tikzpicture}[scale=0.4]
\path [lightgray, fill] (-2.4,6.3) -- (-0.4,6.3) -- (0.,7.1) -- (-0.4,7.9) -- (0.,8.7) -- (-0.4,9.5) -- (0.,10.3) -- (-0.4,11.1) -- (-2.4,11.1) -- cycle;
\path [lightgray, fill] (2.6,6.3) -- (4.6,6.3) -- (5.,7.1) -- (4.6,7.9) -- (5.,8.7) -- (4.6,9.5) -- (5.,10.3) -- (4.6,11.1) -- (2.6,11.1) -- cycle;
\path [lightgray, fill] (9.7,6.3) -- (11.7,6.3) -- (12.1,7.1) -- (11.7,7.9) -- (12.1,8.7) -- (11.7,9.5) -- (12.1,10.3) -- (11.7,11.1) -- (9.7,11.1) -- cycle;
\draw [] (-2.,7.1) -- (-1.6,7.1);
\draw [] (-2.,8.7) -- (-1.6,8.7);
\draw [] (-1.6,7.1) -- (-1.2,7.9);
\draw [] (-1.6,8.7) -- (-1.2,9.5);
\draw [] (-1.2,6.3) -- (-1.6,7.1);
\draw [] (-1.2,7.9) -- (-1.6,8.7);
\draw [] (-1.2,7.9) -- (-0.4,7.9);
\draw [] (-1.2,9.5) -- (-1.6,10.3);
\draw [] (-1.2,9.5) -- (-0.4,9.5);
\draw [] (-0.4,6.3) -- (0.,7.1);
\draw [] (-0.4,7.9) -- (0.,8.7);
\draw [] (-0.4,9.5) -- (0.,10.3);
\draw [] (0.,7.1) -- (-0.4,7.9);
\draw [] (0.,7.1) -- (0.8,7.1);
\draw [] (0.,8.7) -- (-0.4,9.5);
\draw [] (0.,8.7) -- (0.8,8.7);
\draw [very thick, ->] (1.2,8.3) -- (2.3,8.3);
\draw [very thick, ->] (7.6,8.3) -- (8.8,8.3);
\draw [] (-2.,10.3) -- (-1.6,10.3);
\draw [] (-1.6,10.3) -- (-1.2,11.1);
\draw [] (0.,10.3) -- (-0.4,11.1);
\draw [] (0.,10.3) -- (0.8,10.3);
\draw [] (3.,7.1) -- (3.4,7.1);
\draw [] (3.,8.7) -- (3.4,8.7);
\draw [] (3.4,7.1) -- (3.8,7.9);
\draw [] (3.4,8.7) -- (3.8,9.5);
\draw [] (3.8,6.3) -- (3.4,7.1);
\draw [] (3.8,7.9) -- (3.4,8.7);
\draw [] (3.8,7.9) -- (4.6,7.9);
\draw [] (3.8,9.5) -- (3.4,10.3);
\draw [] (3.8,9.5) -- (4.6,9.5);
\draw [] (4.6,6.3) -- (5.,7.1);
\draw [] (4.6,7.9) -- (5.,8.7);
\draw [] (4.6,9.5) -- (5.,10.3);
\draw [] (5.,7.1) -- (4.6,7.9);
\draw [] (5.,7.1) -- (5.8,7.1);
\draw [] (5.,8.7) -- (4.6,9.5);
\draw [] (5.,8.7) -- (5.4,8.7);
\draw [] (3.,10.3) -- (3.4,10.3);
\draw [] (3.4,10.3) -- (3.8,11.1);
\draw [] (5.,10.3) -- (4.6,11.1);
\draw [] (5.,10.3) -- (5.8,10.3);
\draw [] (10.1,7.1) -- (10.5,7.1);
\draw [] (10.1,8.7) -- (10.5,8.7);
\draw [] (10.5,7.1) -- (10.9,7.9);
\draw [] (10.5,8.7) -- (10.9,9.5);
\draw [] (10.9,6.3) -- (10.5,7.1);
\draw [] (10.9,7.9) -- (10.5,8.7);
\draw [] (10.9,7.9) -- (11.7,7.9);
\draw [] (10.9,9.5) -- (10.5,10.3);
\draw [] (10.9,9.5) -- (11.7,9.5);
\draw [] (11.7,6.3) -- (12.1,7.1);
\draw [] (11.7,7.9) -- (12.1,8.7);
\draw [] (11.7,9.5) -- (12.1,10.3);
\draw [] (12.1,7.1) -- (11.7,7.9);
\draw [] (12.1,7.1) -- (12.9,7.1);
\draw [] (12.1,8.7) -- (11.7,9.5);
\draw [] (12.1,8.7) -- (12.9,8.7);
\draw [] (10.1,10.3) -- (10.5,10.3);
\draw [] (10.5,10.3) -- (10.9,11.1);
\draw [] (12.1,10.3) -- (11.7,11.1);
\draw [] (12.1,10.3) -- (12.9,10.3);
\draw [ultra thick] (5.4,8.7) -- (6.2,8.7);
\draw [ultra thick] (5.8,10.3) -- (6.6,10.3);
\draw [ultra thick] (6.2,8.7) -- (6.6,8.7);
\draw [ultra thick] (5.8,7.1) -- (6.6,7.1);
\draw [ultra thick] (5.8,7.1) -- (6.2,8.7);
\draw [ultra thick] (5.4,8.7) -- (5.8,10.3);
\draw [fill] (5.4,8.7) circle [x radius=0.08, y radius=0.08];
\draw [fill] (6.2,8.7) circle [x radius=0.08, y radius=0.08];
\draw [fill] (5.8,10.3) circle [x radius=0.08, y radius=0.08];
\draw [fill] (5.8,7.1) circle [x radius=0.08, y radius=0.08];
\node [] at (0.3,9.5) {\scriptsize $p$};
\node [] at (0.3,7.8) {\scriptsize $p'$};
\node [] at (5.3,9.5) {\scriptsize $\times$};
\node [] at (5.2,7.8) {\scriptsize $\times$};
\end{tikzpicture}
}

\newcommand{\FigBoundary}{
\begin{tikzpicture}[scale=0.7]
\path [lightgray, fill] (2.1,5.) -- (4.1,5.) -- (3.9,5.8) -- (4.1,6.6) -- (3.9,7.4) -- (4.1,8.2) -- (3.9,9.) -- (4.1,9.8) -- (3.9,10.6) -- (4.1,11.4) -- (2.1,11.4) -- cycle;
\draw [->-] (4.1,5.) -- (3.9,5.8);
\draw [->-] (3.9,5.8) -- (4.1,6.6);
\draw [->-] (4.1,6.6) -- (3.9,7.4);
\draw [->-] (3.9,7.4) -- (4.1,8.2);
\draw [->-] (4.1,8.2) -- (3.9,9.);
\draw [->-] (3.9,9.) -- (4.1,9.8);
\draw [->-] (4.1,9.8) -- (3.9,10.6);
\draw [->-] (3.9,10.6) -- (4.1,11.4);
\draw [->-] (3.1,5.8) -- (3.9,5.8);
\draw [->-] (3.1,7.4) -- (3.9,7.4);
\draw [->-] (3.1,9.) -- (3.9,9.);
\draw [->-] (3.1,10.6) -- (3.9,10.6);
\draw [->-] (4.1,9.8) -- (4.9,9.8);
\draw [->-] (4.1,8.2) -- (4.9,8.2);
\draw [->-] (4.1,6.6) -- (4.9,6.6);
\node [] at (4.2,5.3) {\scriptsize $j_1$};
\node [] at (3.3,5.5) {\scriptsize $j_2$};
\node [] at (3.7,6.4) {\scriptsize $j_3$};
\node [] at (4.2,7.1) {\scriptsize $j_4$};
\node [] at (3.3,7.1) {\scriptsize $j_5$};
\node [] at (3.7,8.) {\scriptsize $j_6$};
\node [] at (4.2,8.7) {\scriptsize $j_7$};
\node [] at (3.3,8.7) {\scriptsize $j_8$};
\node [] at (3.7,9.6) {\scriptsize $j_9$};
\node [] at (4.7,6.2) {\scriptsize $a_1$};
\node [] at (4.7,7.8) {\scriptsize $a_2$};
\node [] at (4.7,9.4) {\scriptsize $a_3$};
\node [] at (4.2,10.2) {\scriptsize $j_{10}$};
\node [] at (3.3,10.3) {\scriptsize $j_{11}$};
\node [] at (3.7,11.2) {\scriptsize $j_{12}$};
\node [] at (2.5,8.4) {\scriptsize $\text{bulk}$};
\end{tikzpicture}
}

\newcommand{\FigDiskGraph}{
\begin{tikzpicture}[scale=1]
\draw [lightgray, fill] (3.9,6.1) ..controls (4.294,6.131) and (4.572,6.239) .. (4.9,6.3)
			 ..controls (5.228,6.361) and (5.604,6.374) .. (5.8,6.6)
			 ..controls (5.996,6.826) and (6.01,7.265) .. (6.,7.6)
			 ..controls (5.99,7.935) and (5.955,8.167) .. (5.9,8.5)
			 ..controls (5.845,8.833) and (5.769,9.266) .. (5.6,9.5)
			 ..controls (5.431,9.734) and (5.168,9.77) .. (4.9,9.8)
			 ..controls (4.632,9.83) and (4.357,9.856) .. (4.1,9.9)
			 ..controls (3.843,9.944) and (3.603,10.007) .. (3.3,10.)
			 ..controls (2.997,9.993) and (2.63,9.916) .. (2.4,9.7)
			 ..controls (2.17,9.484) and (2.078,9.128) .. (2.,8.8)
			 ..controls (1.922,8.472) and (1.859,8.172) .. (1.9,7.9)
			 ..controls (1.941,7.628) and (2.085,7.385) .. (2.2,7.1)
			 ..controls (2.315,6.815) and (2.402,6.486) .. (2.7,6.3)
			 ..controls (2.998,6.114) and (3.506,6.069) .. (3.9,6.1);
\draw [] (3.9,6.1) ..controls (4.294,6.131) and (4.572,6.239) .. (4.9,6.3)
			 ..controls (5.228,6.361) and (5.604,6.374) .. (5.8,6.6)
			 ..controls (5.996,6.826) and (6.01,7.265) .. (6.,7.6)
			 ..controls (5.99,7.935) and (5.955,8.167) .. (5.9,8.5)
			 ..controls (5.845,8.833) and (5.769,9.266) .. (5.6,9.5)
			 ..controls (5.431,9.734) and (5.168,9.77) .. (4.9,9.8)
			 ..controls (4.632,9.83) and (4.357,9.856) .. (4.1,9.9)
			 ..controls (3.843,9.944) and (3.603,10.007) .. (3.3,10.)
			 ..controls (2.997,9.993) and (2.63,9.916) .. (2.4,9.7)
			 ..controls (2.17,9.484) and (2.078,9.128) .. (2.,8.8)
			 ..controls (1.922,8.472) and (1.859,8.172) .. (1.9,7.9)
			 ..controls (1.941,7.628) and (2.085,7.385) .. (2.2,7.1)
			 ..controls (2.315,6.815) and (2.402,6.486) .. (2.7,6.3)
			 ..controls (2.998,6.114) and (3.506,6.069) .. (3.9,6.1);
\draw [] (3.9,6.1) -- (3.8,5.7);
\draw [] (5.8,6.6) -- (6.1,6.2);
\draw [] (5.9,8.5) -- (6.5,8.7);
\draw [] (4.9,9.8) -- (5.1,10.3);
\draw [] (3.3,10.) -- (3.1,10.4);
\draw [] (2.,8.8) -- (1.6,8.9);
\draw [] (2.2,7.1) -- (1.8,6.9);
\draw [] (2.4,9.7) -- (2.7,9.2);
\draw [] (4.1,9.9) -- (4.,9.4);
\draw [] (5.6,9.5) -- (5.,9.);
\draw [] (6.,7.6) -- (5.3,7.8);
\draw [] (4.9,6.3) -- (4.5,6.8);
\draw [] (2.7,6.3) -- (3.2,6.9);
\draw [] (1.9,7.9) -- (2.4,7.9);
\draw [] (3.2,6.9) -- (3.8,6.9);
\draw [] (3.8,6.9) -- (4.5,6.8);
\draw [] (4.5,6.8) -- (4.7,7.5);
\draw [] (4.7,7.5) -- (5.3,7.8);
\draw [] (5.3,7.8) -- (4.8,8.2);
\draw [] (4.8,8.2) -- (5.,9.);
\draw [] (5.,9.) -- (4.3,8.9);
\draw [] (4.3,8.9) -- (4.,9.4);
\draw [] (4.,9.4) -- (3.5,9.);
\draw [] (3.5,9.) -- (2.7,9.2);
\draw [] (2.7,9.2) -- (2.7,8.6);
\draw [] (2.7,8.6) -- (2.4,7.9);
\draw [] (2.4,7.9) -- (2.9,7.6);
\draw [] (2.9,7.6) -- (3.2,6.9);
\draw [] (2.7,8.6) -- (3.2,8.3);
\draw [] (3.2,8.3) -- (3.7,8.4);
\draw [] (3.7,8.4) -- (3.8,7.7);
\draw [] (2.9,7.6) -- (3.2,8.3);
\draw [] (3.5,9.) -- (3.7,8.4);
\draw [] (3.8,6.9) -- (3.8,7.7);
\draw [] (4.3,8.9) -- (4.3,8.3);
\draw [] (4.3,8.3) -- (4.8,8.2);
\draw [] (3.8,7.7) -- (4.1,7.8);
\draw [] (4.1,7.8) -- (4.3,8.3);
\draw [] (4.7,7.5) -- (4.1,7.8);
\end{tikzpicture}
}

\newcommand{\FigTmoveInOutAA}{
\begin{tikzpicture}[scale=0.5]
\draw [] (1.6,8.) -- (2.,8.5);
\draw [] (1.6,9.2) -- (2.,9.9);
\draw [] (1.6,9.2) -- (2.4,9.2);
\draw [] (2.,8.5) -- (1.6,9.2);
\draw [] (2.,8.5) -- (2.8,8.5);
\draw [] (2.,9.9) -- (1.6,10.4);
\draw [] (2.,9.9) -- (2.8,9.9);
\draw [] (2.4,9.2) -- (2.8,8.5);
\draw [] (2.4,9.2) -- (2.8,9.9);
\draw [] (2.8,8.5) -- (3.4,9.2);
\draw [] (2.8,9.9) -- (3.4,9.2);
\draw [] (3.4,9.2) -- (4.,9.2);
\draw [] (6.8,10.3) -- (7.2,10.);
\draw [] (7.2,8.8) -- (6.8,8.5);
\draw [] (7.2,10.) -- (7.2,8.8);
\draw [] (7.2,10.) -- (8.,9.9);
\draw [] (8.,8.9) -- (7.2,8.8);
\draw [] (8.,8.9) -- (8.8,9.4);
\draw [] (8.,9.9) -- (8.,8.9);
\draw [] (8.,9.9) -- (8.8,9.4);
\draw [] (8.8,9.4) -- (9.4,9.4);
\draw [->, very thick] (4.8,9.2) -- (5.8,9.2);
\node [] at (8.3,9.4) {\scriptsize $\cdot$};
\node [] at (2.2,8.9) {\scriptsize $\times$};
\node [] at (2.2,9.5) {\scriptsize $\times$};
\end{tikzpicture}
}

\newcommand{\FigTmoveInOutAB}{
\begin{tikzpicture}[scale=0.5]
\draw [] (1.6,7.9) -- (2.,8.5);
\draw [] (1.6,7.9) -- (2.4,7.9);
\draw [] (1.7,6.6) -- (2.,7.2);
\draw [] (1.7,9.2) -- (2.5,8.6);
\draw [] (2.,7.2) -- (1.6,7.9);
\draw [] (2.,7.2) -- (2.8,7.2);
\draw [] (2.,8.5) -- (2.5,8.6);
\draw [] (2.4,7.9) -- (2.,8.5);
\draw [] (2.5,8.6) -- (3.4,8.);
\draw [] (2.8,7.2) -- (2.4,7.9);
\draw [] (2.8,7.2) -- (3.4,8.);
\draw [] (3.1,3.) -- (3.5,3.5);
\draw [] (3.1,4.2) -- (3.5,4.9);
\draw [] (3.1,4.2) -- (3.9,4.2);
\draw [] (3.4,8.) -- (4.,8.);
\draw [] (3.5,3.5) -- (3.1,4.2);
\draw [] (3.5,3.5) -- (4.3,3.5);
\draw [] (3.5,4.9) -- (3.1,5.4);
\draw [] (3.5,4.9) -- (4.3,4.9);
\draw [] (3.9,4.2) -- (4.3,3.5);
\draw [] (3.9,4.2) -- (4.3,4.9);
\draw [] (4.3,3.5) -- (4.9,4.2);
\draw [] (4.3,4.9) -- (4.9,4.2);
\draw [] (4.9,4.2) -- (5.5,4.2);
\draw [] (5.8,7.8) -- (6.1,8.4);
\draw [] (5.8,10.4) -- (6.6,9.8);
\draw [] (6.1,8.4) -- (6.4,9.3);
\draw [] (6.1,8.4) -- (6.9,8.4);
\draw [] (6.4,9.3) -- (6.6,9.8);
\draw [] (6.6,9.8) -- (7.5,9.2);
\draw [] (6.9,8.4) -- (6.4,9.3);
\draw [] (6.9,8.4) -- (7.5,9.2);
\draw [] (7.5,9.2) -- (8.1,9.2);
\draw [] (8.3,5.3) -- (8.7,5.);
\draw [] (8.7,3.8) -- (8.3,3.5);
\draw [] (8.7,5.) -- (8.7,3.8);
\draw [] (8.7,5.) -- (9.5,4.9);
\draw [] (9.5,3.9) -- (8.7,3.8);
\draw [] (9.5,3.9) -- (10.3,4.4);
\draw [] (9.5,4.9) -- (9.5,3.9);
\draw [] (9.5,4.9) -- (10.3,4.4);
\draw [] (10.2,9.) -- (10.6,8.7);
\draw [] (10.3,4.4) -- (10.9,4.4);
\draw [] (10.6,7.5) -- (10.2,7.2);
\draw [] (10.6,8.7) -- (10.6,7.5);
\draw [] (10.6,8.7) -- (11.5,8.1);
\draw [] (11.5,8.1) -- (10.6,7.5);
\draw [] (11.5,8.1) -- (12.1,8.1);
\draw [->, very thick] (3.2,5.8) -- (2.8,6.6);
\draw [->, very thick] (4.,8.8) -- (5.,9.2);
\draw [->, very thick] (8.8,8.8) -- (9.8,8.4);
\draw [->, very thick] (10.4,6.6) -- (10.,5.8);
\node [] at (9.8,4.4) {\scriptsize $\cdot$};
\node [] at (3.7,3.9) {\scriptsize $\times$};
\node [] at (3.7,4.5) {\scriptsize $\times$};
\node [] at (2.1,7.6) {\scriptsize $\times$};
\node [] at (2.1,8.1) {\scriptsize $\times$};
\node [] at (6.4,8.7) {\scriptsize $\times$};
\end{tikzpicture}
}

\newcommand{\FigTrivalentGraphAA}{
\begin{tikzpicture}[scale=0.75]
\draw [->-] (4.,6.) -- (4.,6.4);
\draw [->-] (4.,6.4) -- (5.,6.6);
\draw [->-] (5.,6.6) -- (5.,7.4);
\draw [->-] (5.,7.4) -- (4.2,7.6);
\draw [->-] (4.2,7.6) -- (4.4,8.4);
\draw [->-] (4.4,8.4) -- (5.,9.);
\draw [->-] (5.,9.) -- (4.6,9.6);
\draw [->-] (4.6,9.6) -- (4.,9.2);
\draw [->-] (4.,9.2) -- (3.2,9.6);
\draw [->-] (3.2,9.6) -- (2.2,9.4);
\draw [->-] (2.2,9.4) -- (1.6,9.6);
\draw [->-] (4.,6.4) -- (3.6,7.);
\draw [->-] (3.6,7.) -- (2.6,6.8);
\draw [->-] (3.6,7.) -- (4.2,7.6);
\draw [->-] (2.4,6.) -- (2.6,6.8);
\draw [->-] (2.6,6.8) -- (2.2,7.4);
\draw [->-] (2.2,7.4) -- (2.6,8.4);
\draw [->-] (2.6,8.4) -- (3.4,8.4);
\draw [->-] (3.4,8.4) -- (4.4,8.4);
\draw [->-] (3.4,8.4) -- (4.,9.2);
\draw [->-] (2.6,8.4) -- (2.2,9.4);
\draw [->-] (2.2,7.4) -- (1.6,7.6);
\draw [->-] (4.6,9.6) -- (4.8,10.2);
\draw [->-] (3.2,9.6) -- (3.4,10.4);
\draw [->-] (5.4,6.) -- (5.,6.6);
\draw [->-] (5.,7.4) -- (5.6,7.8);
\draw [->-] (5.6,7.8) -- (5.6,8.8);
\draw [->-] (5.6,8.8) -- (6.,9.6);
\draw [->-] (5.,9.) -- (5.6,8.8);
\draw [->-] (5.6,7.8) -- (6.2,7.8);
\node [] at (3.8,6.) {\scriptsize $j_{1}$};
\node [] at (4.6,6.3) {\scriptsize $j_{2}$};
\node [] at (5.4,6.4) {\scriptsize $j_{3}$};
\node [] at (4.,6.8) {\scriptsize $j_{4}$};
\node [] at (5.3,7.) {\scriptsize $j_{5}$};
\node [] at (6.,7.6) {\scriptsize $j_{6}$};
\node [] at (2.3,6.5) {\scriptsize $j_{7}$};
\node [] at (3.1,6.7) {\scriptsize $j_{8}$};
\node [] at (2.2,7.1) {\scriptsize $j_{9}$};
\node [] at (2.,7.7) {\scriptsize $j_{10}$};
\node [] at (2.8,7.8) {\scriptsize $j_{11}$};
\node [] at (3.7,7.5) {\scriptsize $j_{12}$};
\node [] at (4.,8.) {\scriptsize $j_{13}$};
\node [] at (4.7,7.7) {\scriptsize $j_{14}$};
\node [] at (5.9,8.3) {\scriptsize $j_{15}$};
\node [] at (2.2,8.7) {\scriptsize $j_{16}$};
\node [] at (3.,8.7) {\scriptsize $j_{17}$};
\node [] at (3.5,8.9) {\scriptsize $j_{18}$};
\node [] at (4.,8.6) {\scriptsize $j_{19}$};
\node [] at (4.9,8.5) {\scriptsize $j_{20}$};
\node [] at (5.4,9.1) {\scriptsize $j_{21}$};
\node [] at (6.,9.1) {\scriptsize $j_{22}$};
\node [] at (4.6,9.2) {\scriptsize $j_{23}$};
\node [] at (1.8,9.3) {\scriptsize $j_{24}$};
\node [] at (2.5,9.8) {\scriptsize $j_{25}$};
\node [] at (3.5,10.1) {\scriptsize $j_{26}$};
\node [] at (3.3,9.4) {\scriptsize $j_{27}$};
\node [] at (4.1,9.7) {\scriptsize $j_{28}$};
\node [] at (5.,9.9) {\scriptsize $j_{29}$};
\end{tikzpicture}
}

\newcommand{\FigTrivalentGraphAB}{
\begin{tikzpicture}[scale=0.75]
\draw [->-] (2.2,9.4) -- (1.6,9.6);
\draw [->-] (2.2,9.4) -- (2.6,8.4);
\draw [->-] (2.2,7.4) -- (1.6,7.6);
\draw [->-] (2.2,7.4) -- (2.6,6.8);
\draw [->-] (2.6,8.4) -- (2.2,7.4);
\draw [->-] (3.2,9.6) -- (2.2,9.4);
\draw [->-] (3.2,9.6) -- (3.4,10.4);
\draw [->-] (4.,6.) -- (4.,6.4);
\draw [->-] (3.4,8.4) -- (2.6,8.4);
\draw [->-] (3.4,8.4) -- (4.4,8.4);
\draw [->-] (3.6,7.) -- (2.6,6.8);
\draw [->-] (4.,9.2) -- (3.2,9.6);
\draw [->-] (4.,9.2) -- (3.4,8.4);
\draw [->-] (4.,6.4) -- (3.6,7.);
\draw [->-] (4.,6.4) -- (5.,6.6);
\draw [->-] (4.2,7.6) -- (3.6,7.);
\draw [->-] (4.2,7.6) -- (5.,7.4);
\draw [->-] (4.4,8.4) -- (4.2,7.6);
\draw [->-] (4.6,9.6) -- (4.,9.2);
\draw [->-] (5.,7.4) -- (5.6,7.8);
\draw [->-] (5.,9.) -- (4.4,8.4);
\draw [->-] (5.,9.) -- (4.6,9.6);
\draw [->-] (5.4,6.) -- (5.,6.6);
\draw [->-] (5.6,7.8) -- (5.6,8.8);
\draw [->-] (5.6,7.8) -- (6.2,7.8);
\draw [->-] (5.6,8.8) -- (6.,9.6);
\draw [->-] (2.6,6.8) -- (2.4,6.);
\draw [->-] (5.,7.4) -- (5.,6.6);
\draw [->-] (4.8,10.2) -- (4.6,9.6);
\draw [->-] (5.6,8.8) -- (5.,9.);
\node [] at (3.8,6.) {\scriptsize $j_{1}$};
\node [] at (2.,7.7) {\scriptsize $j_{10}$};
\node [] at (2.8,7.8) {\scriptsize $j_{11}^*$};
\node [] at (3.7,7.5) {\scriptsize $j_{12}^*$};
\node [] at (4.,8.) {\scriptsize $j_{13}^*$};
\node [] at (4.7,7.7) {\scriptsize $j_{14}^*$};
\node [] at (5.9,8.3) {\scriptsize $j_{15}$};
\node [] at (2.2,8.7) {\scriptsize $j_{16}^*$};
\node [] at (3.,8.7) {\scriptsize $j_{17}^*$};
\node [] at (3.5,8.9) {\scriptsize $j_{18}^*$};
\node [] at (4.,8.6) {\scriptsize $j_{19}$};
\node [] at (4.6,6.3) {\scriptsize $j_{2}$};
\node [] at (4.9,8.5) {\scriptsize $j_{20}^*$};
\node [] at (6.,9.1) {\scriptsize $j_{22}$};
\node [] at (4.6,9.2) {\scriptsize $j_{23}$};
\node [] at (1.8,9.3) {\scriptsize $j_{24}$};
\node [] at (2.5,9.8) {\scriptsize $j_{25}$};
\node [] at (3.5,10.1) {\scriptsize $j_{26}$};
\node [] at (3.3,9.4) {\scriptsize $j_{27}$};
\node [] at (4.1,9.7) {\scriptsize $j_{28}$};
\node [] at (5.4,6.4) {\scriptsize $j_{3}$};
\node [] at (4.,6.8) {\scriptsize $j_{4}$};
\node [] at (6.,7.6) {\scriptsize $j_{6}$};
\node [] at (3.1,6.7) {\scriptsize $j_{8}$};
\node [] at (2.3,6.5) {\scriptsize $j_{7}^*$};
\node [] at (2.2,7.1) {\scriptsize $j_{9}^*$};
\node [] at (5.3,7.) {\scriptsize $j_{5}^*$};
\node [] at (5.,9.9) {\scriptsize $j_{29}^*$};
\node [] at (5.4,9.1) {\scriptsize $j_{21}^*$};
\end{tikzpicture}
}

\newcommand{\ExpFrobeniusAlgebraAA}{
\displaystyle\sum_{c}f_{b a c^*}f_{c d e^*}
\Biggl|\;\begin{matrix}
\begin{tikzpicture}[scale=0.7]
\draw [->-] (1.6,10.4) -- (2.,9.8);
\node [] at (1.62,9.98) {\scriptsize $b$};
\draw [->-] (3.2,9.2) -- (2.8,9.8);
\node [] at (3.18,9.62) {\scriptsize $d$};
\draw [->-] (2.8,9.8) -- (3.2,10.4);
\node [] at (3.18,9.98) {\scriptsize $e$};
\draw [->-] (2.,9.8) -- (2.8,9.8);
\node [] at (2.4,9.58) {\scriptsize $c$};
\draw [->-] (1.6,9.2) -- (2.,9.8);
\node [] at (1.62,9.62) {\scriptsize $a$};
\end{tikzpicture}\:
\end{matrix}\Biggr\rangle
}
\newcommand{\ExpFrobeniusAlgebraAB}{
\displaystyle\sum_{c}f_{a b c^*}f_{c d e^*}\displaystyle\sum_{g}G^{a b c^*}_{d e^* g}\mathrm{v}_{c}\mathrm{v}_{g}
\Biggl|\;\begin{matrix}
\begin{tikzpicture}[scale=0.7]
\draw [->-] (1.6,10.4) -- (2.4,10.2);
\node [] at (1.95,10.09) {\scriptsize $a$};
\draw [->-] (2.4,10.2) -- (3.2,10.4);
\node [] at (2.85,10.09) {\scriptsize $e$};
\draw [->-] (3.2,9.2) -- (2.4,9.4);
\node [] at (2.85,9.51) {\scriptsize $d$};
\draw [->-] (2.4,9.4) -- (2.4,10.2);
\node [] at (2.62,9.8) {\scriptsize $g$};
\draw [->-] (1.6,9.2) -- (2.4,9.4);
\node [] at (1.95,9.51) {\scriptsize $b$};
\end{tikzpicture}\:
\end{matrix}\Biggr\rangle
}
\newcommand{\ExpFrobeniusAlgebraAC}{
\displaystyle\sum_{g}f_{b d g^*}f_{a g e^*}
\Biggl|\;\begin{matrix}
\begin{tikzpicture}[scale=0.7]
\draw [->-] (1.6,10.4) -- (2.4,10.2);
\node [] at (1.95,10.09) {\scriptsize $a$};
\draw [->-] (2.4,10.2) -- (3.2,10.4);
\node [] at (2.85,10.09) {\scriptsize $e$};
\draw [->-] (3.2,9.2) -- (2.4,9.4);
\node [] at (2.85,9.51) {\scriptsize $d$};
\draw [->-] (2.4,9.4) -- (2.4,10.2);
\node [] at (2.62,9.8) {\scriptsize $g$};
\draw [->-] (1.6,9.2) -- (2.4,9.4);
\node [] at (1.95,9.51) {\scriptsize $b$};
\end{tikzpicture}\:
\end{matrix}\Biggr\rangle
}
\newcommand{\ExpFrobeniusAlgebraCompactAA}{
\Biggl|\;\begin{matrix}
\begin{tikzpicture}[scale=0.8]
\draw [->-] (1.6,10.4) -- (2.,9.8);
\node [] at (1.62,9.98) {\scriptsize $i$};
\draw [->-] (2.8,9.8) -- (3.2,10.4);
\node [] at (3.18,9.98) {\scriptsize $l$};
\draw [->-] (3.2,9.2) -- (2.8,9.8);
\node [] at (3.18,9.62) {\scriptsize $k$};
\draw [->-] (1.6,9.2) -- (2.,9.8);
\node [] at (1.62,9.62) {\scriptsize $j$};
\draw [ultra thick] (2.,9.8) -- (2.8,9.8);
\draw [fill] (2.,9.8) circle [x radius=0.08, y radius=0.08];
\draw [fill] (2.8,9.8) circle [x radius=0.08, y radius=0.08];
\end{tikzpicture}\:
\end{matrix}\Biggr\rangle
}
\newcommand{\ExpFrobeniusAlgebraCompactAB}{
\Biggl|\;\begin{matrix}
\begin{tikzpicture}[scale=0.8]
\draw [->-] (1.6,10.4) -- (2.4,10.2);
\node [] at (1.95,10.09) {\scriptsize $i$};
\draw [->-] (2.4,10.2) -- (3.2,10.4);
\node [] at (2.85,10.09) {\scriptsize $l$};
\draw [->-] (3.2,9.2) -- (2.4,9.4);
\node [] at (2.85,9.51) {\scriptsize $k$};
\draw [->-] (1.6,9.2) -- (2.4,9.4);
\node [] at (1.95,9.51) {\scriptsize $j$};
\draw [ultra thick] (2.4,9.4) -- (2.4,10.2);
\draw [fill] (2.4,10.2) circle [x radius=0.08, y radius=0.08];
\draw [fill] (2.4,9.4) circle [x radius=0.08, y radius=0.08];
\end{tikzpicture}\:
\end{matrix}\Biggr\rangle
}
\newcommand{\GroundStateCylinderAA}{
\begin{tikzpicture}[scale=1]
\draw [dashed] (3.3,9.8) ..controls (2.992,9.783) and (2.683,9.767) .. (2.5,9.6)
			 ..controls (2.317,9.433) and (2.258,9.117) .. (2.2,8.8);
\draw [->-] (2.2,8.8) ..controls (2.225,8.658) and (2.25,8.517) .. (2.3,8.4)
			 ..controls (2.35,8.283) and (2.425,8.192) .. (2.5,8.1);
\draw [->-] (2.5,8.1) ..controls (2.583,8.025) and (2.667,7.95) .. (2.8,7.9)
			 ..controls (2.933,7.85) and (3.117,7.825) .. (3.3,7.8);
\draw [->-] (3.3,7.8) ..controls (3.483,7.825) and (3.667,7.85) .. (3.8,7.9)
			 ..controls (3.933,7.95) and (4.017,8.025) .. (4.1,8.1);
\draw [->-] (4.1,8.1) ..controls (4.175,8.192) and (4.25,8.283) .. (4.3,8.4)
			 ..controls (4.35,8.517) and (4.375,8.658) .. (4.4,8.8);
\draw [->-] (4.4,8.8) ..controls (4.375,8.942) and (4.35,9.083) .. (4.3,9.2)
			 ..controls (4.25,9.317) and (4.175,9.408) .. (4.1,9.5);
\draw [->-] (4.1,9.5) ..controls (4.017,9.575) and (3.933,9.65) .. (3.8,9.7)
			 ..controls (3.667,9.75) and (3.483,9.775) .. (3.3,9.8);
\draw [->-] (1.6,8.8) -- (2.2,8.8);
\node [] at (1.9,8.58) {\scriptsize $a_{1}$};
\draw [->-] (3.3,7.2) -- (3.3,7.8);
\node [] at (3.59,7.5) {\scriptsize $a_{2}$};
\draw [->-] (3.3,10.4) -- (3.3,9.8);
\node [] at (3.01,10.1) {\scriptsize $a_{4}$};
\draw [->-] (5.,8.8) -- (4.4,8.8);
\node [] at (4.7,9.02) {\scriptsize $a_{3}$};
\draw [->-] (3.,8.5) -- (2.5,8.1);
\node [] at (2.58,8.51) {\scriptsize $b_{1}$};
\draw [->-] (3.8,8.6) -- (4.1,8.1);
\node [] at (3.71,8.2) {\scriptsize $b_{2}$};
\draw [->-] (3.6,9.2) -- (4.1,9.5);
\node [] at (3.98,9.13) {\scriptsize $b_{3}$};
\node [] at (2.,8.2) {\scriptsize $j_{1}$};
\node [] at (2.8,7.6) {\scriptsize $j_{2}$};
\node [] at (3.9,7.7) {\scriptsize $j_{3}$};
\node [] at (4.6,8.3) {\scriptsize $j_{4}$};
\node [] at (4.5,9.3) {\scriptsize $j_{5}$};
\node [] at (3.9,9.9) {\scriptsize $j_{6}$};
\end{tikzpicture}
}

\newcommand{\ExpGroundStateCylinderAC}{
\Biggl|\;\begin{matrix}
\begin{tikzpicture}[scale=1]
\draw [very thick, dashed] (3.4,9.8) ..controls (3.083,9.783) and (2.767,9.767) .. (2.6,9.6)
			 ..controls (2.433,9.433) and (2.417,9.117) .. (2.4,8.8);
\draw [ultra thick] (2.4,8.8) ..controls (2.417,8.483) and (2.433,8.167) .. (2.6,8.)
			 ..controls (2.767,7.833) and (3.083,7.817) .. (3.4,7.8);
\draw [ultra thick] (3.4,7.8) ..controls (3.717,7.817) and (4.033,7.833) .. (4.2,8.)
			 ..controls (4.367,8.167) and (4.383,8.483) .. (4.4,8.8);
\draw [ultra thick] (4.4,8.8) ..controls (4.383,9.117) and (4.367,9.433) .. (4.2,9.6)
			 ..controls (4.033,9.767) and (3.717,9.783) .. (3.4,9.8);
\draw [->-] (3.4,7.2) -- (3.4,7.8);
\node [] at (3.69,7.5) {\scriptsize $a_{2}$};
\draw [->-] (5.2,8.8) -- (4.6,8.8);
\node [] at (4.9,9.02) {\scriptsize $a_{3}$};
\draw [->-] (3.4,8.4) -- (3.4,7.8);
\node [] at (3.11,8.1) {\scriptsize $b_{2}$};
\draw [->-] (3.7,8.8) -- (4.2,8.8);
\node [] at (3.95,8.58) {\scriptsize $b_{3}$};
\draw [->-] (1.6,8.8) -- (2.2,8.8);
\node [] at (1.9,9.02) {\scriptsize $a_{1}$};
\draw [->-] (3.4,10.4) -- (3.4,9.8);
\node [] at (3.69,10.1) {\scriptsize $a_{4}$};
\draw [->-] (3.1,8.8) -- (2.6,8.8);
\node [] at (2.85,8.58) {\scriptsize $b_{1}$};
\draw [->-] (3.4,9.2) -- (3.4,9.8);
\node [] at (3.11,9.5) {\scriptsize $b_{4}$};
\node [fill=white,draw,rounded corners=.06cm] at (3.4,7.8) {\scriptsize $P_M$};
\node [fill=white,draw,rounded corners=.06cm] at (4.4,8.8) {\scriptsize $P_M$};
\node [fill=white,draw,rounded corners=.06cm] at (3.4,9.8) {\scriptsize $P_M$};
\node [fill=white,draw,rounded corners=.06cm] at (2.4,8.8) {\scriptsize $P_M$};
\end{tikzpicture}\:
\end{matrix}\Biggr\rangle
}
\newcommand{\GroundStateDiskAA}{
\begin{tikzpicture}[scale=0.7]
\draw [lightgray, fill] (2.1,9.1) ..controls (2.1,8.871) and (2.157,8.643) .. (2.3,8.5)
			 ..controls (2.443,8.357) and (2.671,8.3) .. (2.9,8.3)
			 ..controls (3.129,8.3) and (3.357,8.357) .. (3.5,8.5)
			 ..controls (3.643,8.643) and (3.7,8.871) .. (3.7,9.1)
			 ..controls (3.7,9.329) and (3.643,9.557) .. (3.5,9.7)
			 ..controls (3.357,9.843) and (3.129,9.9) .. (2.9,9.9)
			 ..controls (2.671,9.9) and (2.443,9.843) .. (2.3,9.7)
			 ..controls (2.157,9.557) and (2.1,9.329) .. (2.1,9.1);
\draw [thick, dashed] (2.9,9.9) ..controls (2.667,9.867) and (2.433,9.833) .. (2.3,9.7)
			 ..controls (2.167,9.567) and (2.133,9.333) .. (2.1,9.1);
\draw [->-] (2.1,9.1) ..controls (2.133,8.867) and (2.167,8.633) .. (2.3,8.5)
			 ..controls (2.433,8.367) and (2.667,8.333) .. (2.9,8.3);
\draw [->-] (2.9,8.3) ..controls (3.133,8.333) and (3.367,8.367) .. (3.5,8.5)
			 ..controls (3.633,8.633) and (3.667,8.867) .. (3.7,9.1);
\draw [->-] (3.7,9.1) ..controls (3.667,9.333) and (3.633,9.567) .. (3.5,9.7)
			 ..controls (3.367,9.833) and (3.133,9.867) .. (2.9,9.9);
\draw [->-] (1.6,9.1) -- (2.1,9.1);
\node [] at (1.85,8.88) {\scriptsize $a_{N}$};
\draw [->-] (2.9,7.8) -- (2.9,8.3);
\node [] at (3.19,8.05) {\scriptsize $a_{1}$};
\draw [->-] (4.2,9.1) -- (3.7,9.1);
\node [] at (3.95,9.32) {\scriptsize $a_{2}$};
\draw [->-] (2.9,10.4) -- (2.9,9.9);
\node [] at (2.61,10.15) {\scriptsize $a_{3}$};
\node [] at (2.1,8.3) {\scriptsize $l_{1}$};
\node [] at (3.9,8.5) {\scriptsize $l_{2}$};
\node [] at (3.7,9.9) {\scriptsize $l_{3}$};
\end{tikzpicture}
}

\newcommand{\ExpGSbase}{
\displaystyle\sum_{a_{1},a_{2}}\mathrm{u}_{a_{1}}\mathrm{u}_{a_{2}}
\Biggl|\;\begin{matrix}
\begin{tikzpicture}[scale=0.8]
\path [lightgray, fill] (1.6,8.) -- (2.4,8.) -- (2.4,10.4) -- (1.6,10.4) -- cycle;
\draw [->-] (2.4,8.) -- (2.4,8.8);
\draw [->-] (2.4,9.6) -- (2.4,10.4);
\draw [->-, thick] (3.2,9.6) -- (2.4,9.6);
\node [] at (2.8,9.82) {\scriptsize $a_{2}$};
\draw [->-, thick] (3.2,8.8) -- (2.4,8.8);
\node [] at (2.8,8.58) {\scriptsize $a_{1}$};
\draw [ultra thick] (2.4,8.8) -- (2.4,9.6);
\node [fill=white,draw,rounded corners=.06cm] at (2.4,9.6) {\scriptsize $\rho$};
\node [fill=white,draw,rounded corners=.06cm] at (2.4,8.8) {\scriptsize $\rho$};
\node [] at (2.2,8.2) {\scriptsize $j_1$};
\node [] at (2.2,10.2) {\scriptsize $j_2$};
\end{tikzpicture}\:
\end{matrix}\Biggr\rangle
}
\newcommand{\ExpGSbaseAA}{
\frac{\sqrt{D}}{\mathrm{d}_A}\displaystyle\sum_{a'_{1},a'_{2}}\mathrm{u}_{a'_{1}}\mathrm{u}_{a'_{2}}
\Biggl|\;\begin{matrix}
\begin{tikzpicture}[scale=0.8]
\path [lightgray, fill] (1.6,8.) -- (2.4,8.) -- (2.4,10.4) -- (1.6,10.4) -- cycle;
\draw [->-] (2.4,8.) -- (2.4,8.8);
\draw [->-] (2.4,9.6) -- (2.4,10.4);
\draw [->-] (3.5,8.8) -- (3.,8.8);
\node [] at (3.25,8.58) {\scriptsize $a'_{1}$};
\draw [->-] (3.5,9.6) -- (3.,9.6);
\node [] at (3.25,9.82) {\scriptsize $a'_{2}$};
\draw [ultra thick] (2.4,8.8) -- (2.4,9.6);
\draw [ultra thick] (2.4,9.6) -- (3.,9.6);
\draw [ultra thick] (2.4,8.8) -- (3.,8.8);
\draw [ultra thick] (3.,9.6) -- (3.,8.8);
\draw [fill] (3.,9.6) circle [x radius=0.08, y radius=0.08];
\draw [fill] (3.,8.8) circle [x radius=0.08, y radius=0.08];
\node [fill=white,draw,rounded corners=.06cm] at (2.4,9.6) {\scriptsize $\rho$};
\node [fill=white,draw,rounded corners=.06cm] at (2.4,8.8) {\scriptsize $\rho$};
\node [] at (2.2,8.2) {\scriptsize $j_1$};
\node [] at (2.2,10.2) {\scriptsize $j_2$};
\end{tikzpicture}\:
\end{matrix}\Biggr\rangle
}
\newcommand{\ExpGSbaseAB}{
\frac{\sqrt{D}}{\mathrm{d}_A}\displaystyle\sum_{a'_{1},a'_{2}}\mathrm{u}_{a'_{1}}\mathrm{u}_{a'_{2}}
\Biggl|\;\begin{matrix}
\begin{tikzpicture}[scale=0.8]
\path [lightgray, fill] (1.6,8.) -- (2.4,8.) -- (2.4,10.4) -- (1.6,10.4) -- cycle;
\draw [->-] (2.4,8.) -- (2.4,9.2);
\draw [->-] (2.4,9.2) -- (2.4,10.4);
\draw [->-] (4.1,8.8) -- (3.6,8.8);
\node [] at (3.85,8.58) {\scriptsize $a'_{1}$};
\draw [->-] (4.1,9.6) -- (3.6,9.6);
\node [] at (3.85,9.82) {\scriptsize $a'_{2}$};
\draw [ultra thick] (2.4,9.2) -- (3.,9.2);
\draw [ultra thick] (3.,9.2) -- (3.6,9.6);
\draw [ultra thick] (3.,9.2) -- (3.6,8.8);
\draw [ultra thick] (3.6,9.6) -- (3.6,8.8);
\draw [fill] (3.6,9.6) circle [x radius=0.08, y radius=0.08];
\draw [fill] (3.6,8.8) circle [x radius=0.08, y radius=0.08];
\draw [fill] (3.,9.2) circle [x radius=0.08, y radius=0.08];
\node [fill=white,draw,rounded corners=.06cm] at (2.4,9.2) {\scriptsize $\rho$};
\node [] at (2.2,8.2) {\scriptsize $j_1$};
\node [] at (2.2,10.2) {\scriptsize $j_2$};
\end{tikzpicture}\:
\end{matrix}\Biggr\rangle
}
\newcommand{\ExpGSbaseAC}{
\displaystyle\sum_{a'_{1},a'_{2}}\mathrm{u}_{a'_{1}}\mathrm{u}_{a'_{2}}
\Biggl|\;\begin{matrix}
\begin{tikzpicture}[scale=0.8]
\path [lightgray, fill] (1.6,8.) -- (2.4,8.) -- (2.4,10.4) -- (1.6,10.4) -- cycle;
\draw [->-] (2.4,8.) -- (2.4,9.2);
\draw [->-] (2.4,9.2) -- (2.4,10.4);
\draw [->-] (3.6,9.6) -- (3.,9.2);
\node [] at (3.13,9.65) {\scriptsize $a'_{2}$};
\draw [->-] (3.6,8.8) -- (3.,9.2);
\node [] at (3.13,8.75) {\scriptsize $a'_{1}$};
\draw [ultra thick] (2.4,9.2) -- (3.,9.2);
\draw [fill] (3.,9.2) circle [x radius=0.08, y radius=0.08];
\node [fill=white,draw,rounded corners=.06cm] at (2.4,9.2) {\scriptsize $\rho$};
\node [] at (2.2,8.2) {\scriptsize $j_1$};
\node [] at (2.2,10.2) {\scriptsize $j_2$};
\end{tikzpicture}\:
\end{matrix}\Biggr\rangle
}
\newcommand{\ExpGSbaseAD}{
\displaystyle\sum_{a'_{1},a'_{2}}\mathrm{u}_{a'_{1}}\mathrm{u}_{a'_{2}}
\Biggl|\;\begin{matrix}
\begin{tikzpicture}[scale=0.8]
\path [lightgray, fill] (1.6,8.) -- (2.4,8.) -- (2.4,10.4) -- (1.6,10.4) -- cycle;
\draw [->-] (2.4,8.) -- (2.4,8.8);
\draw [->-] (2.4,9.6) -- (2.4,10.4);
\draw [->-, thick] (3.2,9.6) -- (2.4,9.6);
\node [] at (2.8,9.82) {\scriptsize $a'_{2}$};
\draw [->-, thick] (3.2,8.8) -- (2.4,8.8);
\node [] at (2.8,8.58) {\scriptsize $a'_{1}$};
\draw [ultra thick] (2.4,8.8) -- (2.4,9.6);
\node [fill=white,draw,rounded corners=.06cm] at (2.4,9.6) {\scriptsize $\rho$};
\node [fill=white,draw,rounded corners=.06cm] at (2.4,8.8) {\scriptsize $\rho$};
\node [] at (2.2,8.2) {\scriptsize $j_1$};
\node [] at (2.2,10.2) {\scriptsize $j_2$};
\end{tikzpicture}\:
\end{matrix}\Biggr\rangle
}
\newcommand{\ExpGSbaseAE}{
\displaystyle\sum_{a_{1},a_{2}}\mathrm{u}_{a_{1}}\mathrm{u}_{a_{2}}
\Biggl|\;\begin{matrix}
\begin{tikzpicture}[scale=0.8]
\path [lightgray, fill] (1.6,7.2) -- (2.4,7.2) -- (2.6,8.) -- (2.4,8.8) -- (2.6,9.6) -- (2.4,10.4) -- (1.6,10.4) -- cycle;
\draw [->-] (2.4,7.2) -- (2.6,8.);
\draw [->-] (2.6,9.6) -- (2.4,10.4);
\draw [->-, thick] (3.4,9.6) -- (2.6,9.6);
\node [] at (3.,9.82) {\scriptsize $a_{2}$};
\draw [->-, thick] (3.4,8.) -- (2.6,8.);
\node [] at (3.,7.78) {\scriptsize $a_{1}$};
\draw [->-] (1.8,8.8) -- (2.4,8.8);
\draw [ultra thick] (2.6,8.) -- (2.4,8.8);
\draw [ultra thick] (2.4,8.8) -- (2.6,9.6);
\node [] at (2.2,7.4) {\scriptsize $j_1$};
\node [] at (2.2,10.2) {\scriptsize $j_2$};
\node [] at (2.,8.5) {\scriptsize $k$};
\node [fill=white,draw,rounded corners=.06cm] at (2.4,8.8) {\scriptsize $\eta$};
\node [fill=white,draw,rounded corners=.06cm] at (2.6,8.) {\scriptsize $\rho_M$};
\node [fill=white,draw,rounded corners=.06cm] at (2.6,9.6) {\scriptsize $\rho_N$};
\end{tikzpicture}\:
\end{matrix}\Biggr\rangle
}
\newcommand{\ExpGSbaseAF}{
\displaystyle\sum_{a_{1},a_{2}}\mathrm{u}_{a_{1}}\mathrm{u}_{a_{2}}
\Biggl|\;\begin{matrix}
\begin{tikzpicture}[scale=1]
\path [lightgray, fill] (1.6,6.6) -- (2.4,6.6) -- (2.6,7.2) -- (2.4,8.5) -- (2.6,9.8) -- (2.4,10.4) -- (1.6,10.4) -- cycle;
\draw [ultra thick] (2.5,7.8) ..controls (2.7,7.911) and (2.9,8.022) .. (3.,8.2)
			 ..controls (3.1,8.378) and (3.1,8.622) .. (3.,8.8)
			 ..controls (2.9,8.978) and (2.7,9.089) .. (2.5,9.2);
\draw [->-] (2.4,6.6) -- (2.6,7.2);
\draw [->-] (2.6,9.8) -- (2.4,10.4);
\draw [->-, thick] (3.4,9.8) -- (2.6,9.8);
\node [] at (3.,10.02) {\scriptsize $a_{2}$};
\draw [->-, thick] (3.4,7.2) -- (2.6,7.2);
\node [] at (3.,6.98) {\scriptsize $a_{1}$};
\draw [->-] (1.8,8.5) -- (2.4,8.5);
\draw [ultra thick] (2.6,7.2) -- (2.5,7.8);
\draw [ultra thick] (2.5,7.8) -- (2.4,8.5);
\draw [ultra thick] (2.4,8.5) -- (2.5,9.2);
\draw [ultra thick] (2.5,9.2) -- (2.6,9.8);
\node [] at (2.2,6.8) {\scriptsize $j_1$};
\node [] at (2.2,10.1) {\scriptsize $j_2$};
\node [] at (2.,8.2) {\scriptsize $k$};
\node [fill=white,draw,rounded corners=.06cm] at (2.4,8.5) {\scriptsize $\eta$};
\node [fill=white,draw,rounded corners=.06cm] at (2.6,7.2) {\scriptsize $\rho_M$};
\node [fill=white,draw,rounded corners=.06cm] at (2.6,9.8) {\scriptsize $\rho_N$};
\node [fill=white,draw,rounded corners=.06cm] at (2.5,7.8) {\scriptsize $\rho_M$};
\node [fill=white,draw,rounded corners=.06cm] at (2.5,9.2) {\scriptsize $\rho_N$};
\end{tikzpicture}\:
\end{matrix}\Biggr\rangle
}
\newcommand{\ExpGSbaseAG}{
\Biggl|\;\begin{matrix}
\begin{tikzpicture}[scale=0.8]
\path [lightgray, fill] (1.6,7.8) -- (2.2,7.8) -- (2.6,7.8) -- (2.4,9.1) -- (2.6,10.4) -- (2.2,10.4) -- (1.6,10.4) -- cycle;
\draw [ultra thick] (2.5,8.4) ..controls (2.7,8.511) and (2.9,8.622) .. (3.,8.8)
			 ..controls (3.1,8.978) and (3.1,9.222) .. (3.,9.4)
			 ..controls (2.9,9.578) and (2.7,9.689) .. (2.5,9.8);
\draw [->-] (1.8,9.1) -- (2.4,9.1);
\draw [ultra thick] (2.5,8.4) -- (2.4,9.1);
\draw [ultra thick] (2.4,9.1) -- (2.5,9.8);
\draw [->-] (2.5,9.8) -- (2.6,10.4);
\draw [->-] (2.6,7.8) -- (2.5,8.4);
\node [] at (2.,8.8) {\scriptsize $k$};
\node [fill=white,draw,rounded corners=.06cm] at (2.4,9.1) {\scriptsize $\eta$};
\node [fill=white,draw,rounded corners=.06cm] at (2.5,8.4) {\scriptsize $\rho_M$};
\node [fill=white,draw,rounded corners=.06cm] at (2.5,9.8) {\scriptsize $\rho_N$};
\node [] at (2.3,8.) {\scriptsize $j_3$};
\node [] at (2.3,10.2) {\scriptsize $j_4$};
\end{tikzpicture}\:
\end{matrix}\Biggr\rangle
}
\newcommand{\ExpGSbaseAH}{
\Biggl|\;\begin{matrix}
\begin{tikzpicture}[scale=0.8]
\path [lightgray, fill] (1.6,7.8) -- (2.2,7.8) -- (2.5,7.8) -- (2.4,9.1) -- (2.5,10.4) -- (2.2,10.4) -- (1.6,10.4) -- cycle;
\draw [->-] (1.8,9.1) -- (2.4,9.1);
\draw [->-] (2.4,9.1) -- (2.5,10.4);
\draw [->-] (2.5,7.8) -- (2.4,9.1);
\node [] at (2.,8.8) {\scriptsize $k$};
\node [fill=white,draw,rounded corners=.06cm] at (2.4,9.1) {\scriptsize $\eta$};
\node [] at (2.3,8.) {\scriptsize $j_3$};
\node [] at (2.3,10.2) {\scriptsize $j_4$};
\end{tikzpicture}\:
\end{matrix}\Biggr\rangle
}
\newcommand{\ExpGSbaseAI}{
\Biggl|\;\begin{matrix}
\begin{tikzpicture}[scale=0.8]
\path [lightgray, fill] (1.6,7.8) -- (2.2,7.8) -- (2.6,7.8) -- (2.4,9.4) -- (2.6,10.4) -- (2.2,10.4) -- (1.6,10.4) -- cycle;
\draw [->-] (1.8,9.4) -- (2.4,9.4);
\draw [ultra thick] (2.5,8.5) -- (2.4,9.4);
\draw [->-] (2.4,9.4) -- (2.6,10.4);
\draw [->-] (2.6,7.8) -- (2.5,8.5);
\draw [->-] (3.2,8.5) -- (2.8,8.5);
\node [] at (2.,9.1) {\scriptsize $k$};
\node [fill=white,draw,rounded corners=.06cm] at (2.4,9.4) {\scriptsize $\eta$};
\node [fill=white,draw,rounded corners=.06cm] at (2.5,8.5) {\scriptsize $\rho_M$};
\node [] at (2.3,8.) {\scriptsize $j_3$};
\node [] at (2.3,10.2) {\scriptsize $j_4$};
\node [] at (3.,8.8) {\scriptsize $a$};
\end{tikzpicture}\:
\end{matrix}\Biggr\rangle
}
\newcommand{\ExpGSbaseAJ}{
\Biggl|\;\begin{matrix}
\begin{tikzpicture}[scale=0.8]
\path [lightgray, fill] (1.6,7.8) -- (2.2,7.8) -- (2.6,7.8) -- (2.4,8.9) -- (2.6,10.4) -- (2.2,10.4) -- (1.6,10.4) -- cycle;
\draw [->-] (1.8,8.9) -- (2.4,8.9);
\draw [ultra thick] (2.4,8.9) -- (2.5,9.7);
\draw [->-] (2.5,9.7) -- (2.6,10.4);
\draw [->-] (2.6,7.8) -- (2.4,8.9);
\draw [->-] (3.2,9.7) -- (2.8,9.7);
\node [] at (2.,8.7) {\scriptsize $k$};
\node [fill=white,draw,rounded corners=.06cm] at (2.4,8.9) {\scriptsize $\eta$};
\node [fill=white,draw,rounded corners=.06cm] at (2.5,9.7) {\scriptsize $\rho_N$};
\node [] at (2.3,8.) {\scriptsize $j_3$};
\node [] at (2.3,10.2) {\scriptsize $j_4$};
\node [] at (3.1,9.4) {\scriptsize $a$};
\end{tikzpicture}\:
\end{matrix}\Biggr\rangle
}
\newcommand{\ExpGSbaseAK}{
\Biggl|\;\begin{matrix}
\begin{tikzpicture}[scale=1]
\path [lightgray, fill] (1.6,6.8) -- (2.4,6.8) -- (2.4,8.) -- (2.4,9.2) -- (2.4,10.4) -- (1.6,10.4) -- cycle;
\draw [->-] (2.4,6.8) -- (2.4,8.);
\node [] at (2.69,7.4) {\scriptsize $M_{1}$};
\draw [->-] (2.4,8.) -- (2.4,9.2);
\node [] at (2.69,8.6) {\scriptsize $M_{2}$};
\draw [->-] (2.4,9.2) -- (2.4,10.4);
\node [] at (2.69,9.8) {\scriptsize $M_{3}$};
\draw [->-] (1.9,8.) -- (2.4,8.);
\node [] at (2.15,7.78) {\scriptsize $j_{1}$};
\draw [->-] (1.9,9.2) -- (2.4,9.2);
\node [] at (2.15,8.98) {\scriptsize $j_{2}$};
\draw [fill] (2.4,9.2) circle [x radius=0.05, y radius=0.05];
\draw [fill] (2.4,8.) circle [x radius=0.05, y radius=0.05];
\node [] at (2.7,8.) {\scriptsize $\eta_{1}$};
\node [] at (2.7,9.2) {\scriptsize $\eta_{2}$};
\end{tikzpicture}\:
\end{matrix}\Biggr\rangle
}
\newcommand{\ExpGSbaseAL}{
\displaystyle\sum_{a_{1},a_{2},a_{3}}\mathrm{u}_{a_{1}}\mathrm{u}_{a_{2}}\mathrm{u}_{a_{3}}
\Biggl|\;\begin{matrix}
\begin{tikzpicture}[scale=0.8]
\draw [->-] (2.2,5.6) -- (2.4,6.4);
\draw [->-] (1.6,7.2) -- (2.2,7.2);
\draw [ultra thick] (2.4,6.4) -- (2.2,7.2);
\draw [ultra thick] (2.2,7.2) -- (2.4,8.);
\draw [ultra thick] (2.4,8.) -- (2.2,8.8);
\draw [ultra thick] (2.2,8.8) -- (2.4,9.6);
\draw [->-] (2.4,9.6) -- (2.2,10.4);
\draw [->-] (1.6,8.8) -- (2.2,8.8);
\draw [->-] (3.4,6.4) -- (2.8,6.4);
\node [] at (3.1,6.62) {\scriptsize $a_{1}$};
\draw [->-] (3.4,8.) -- (2.8,8.);
\node [] at (3.1,8.22) {\scriptsize $a_{2}$};
\draw [->-] (3.4,9.6) -- (2.8,9.6);
\node [] at (3.1,9.82) {\scriptsize $a_{3}$};
\node [fill=white,draw,rounded corners=.06cm] at (2.2,7.2) {\scriptsize $\eta_{1}$};
\node [fill=white,draw,rounded corners=.06cm] at (2.2,8.8) {\scriptsize $\eta_{2}$};
\node [fill=white,draw,rounded corners=.06cm] at (2.4,6.4) {\scriptsize $\rho_{M_{1}}$};
\node [fill=white,draw,rounded corners=.06cm] at (2.4,8.) {\scriptsize $\rho_{M_{2}}$};
\node [fill=white,draw,rounded corners=.06cm] at (2.4,9.6) {\scriptsize $\rho_{M_{3}}$};
\node [] at (1.8,6.9) {\scriptsize $j_{1}$};
\node [] at (1.7,8.6) {\scriptsize $j_{2}$};
\end{tikzpicture}\:
\end{matrix}\Biggr\rangle
}
\newcommand{\ExpGScylinderBase}{
\displaystyle\sum_{a_{1}a_{2}b_{1}b_{2}}\mathrm{u}_{a_{1}}\mathrm{u}_{a_{2}}\mathrm{u}_{b_{1}}\mathrm{u}_{b_{2}}
\Biggl|\;\begin{matrix}
\begin{tikzpicture}[scale=1]
\draw [ultra thick] (2.4,8.8) -- (2.4,9.6);
\draw [->-] (2.4,8.) -- (2.4,8.8);
\node [] at (2.18,8.4) {\scriptsize $j$};
\draw [->-] (3.2,9.4) -- (2.4,9.6);
\node [] at (2.74,9.27) {\scriptsize $a_{2}$};
\draw [->-] (3.2,8.6) -- (2.4,8.8);
\node [] at (2.74,8.47) {\scriptsize $a_{1}$};
\draw [->-] (1.6,9.) -- (2.4,8.8);
\node [] at (2.06,9.13) {\scriptsize $b_{1}$};
\draw [->-] (1.6,9.8) -- (2.4,9.6);
\node [] at (2.06,9.93) {\scriptsize $b_{2}$};
\draw [->-] (2.4,9.6) -- (2.4,10.4);
\node [] at (2.62,10.) {\scriptsize $k$};
\node [fill=white,draw,rounded corners=.06cm] at (2.4,9.6) {\scriptsize $P_M$};
\node [fill=white,draw,rounded corners=.06cm] at (2.4,8.8) {\scriptsize $P_M$};
\end{tikzpicture}\:
\end{matrix}\Biggr\rangle
}
\newcommand{\ExpGScylinderBaseAB}{
\displaystyle\sum_{a'_{1}a'_{2}b'_{1}b'_{2}}\mathrm{u}_{a'_{1}}\mathrm{u}_{a'_{2}}\mathrm{u}_{b'_{1}}\mathrm{u}_{b'_{2}}
\Biggl|\;\begin{matrix}
\begin{tikzpicture}[scale=1]
\draw [ultra thick] (2.8,8.8) -- (2.8,9.6);
\draw [->-] (2.8,8.) -- (2.8,8.8);
\node [] at (2.58,8.4) {\scriptsize $j$};
\draw [->-] (2.8,9.6) -- (2.8,10.4);
\node [] at (3.02,10.) {\scriptsize $k$};
\draw [ultra thick] (3.4,8.6) -- (2.8,8.8);
\draw [ultra thick] (2.8,8.8) -- (2.2,9.);
\draw [ultra thick] (3.4,9.4) -- (2.8,9.6);
\draw [ultra thick] (2.8,9.6) -- (2.2,9.8);
\draw [ultra thick] (3.4,8.6) -- (3.4,9.4);
\draw [ultra thick] (2.2,9.) -- (2.2,9.8);
\draw [->-] (4.,8.4) -- (3.4,8.6);
\node [] at (3.78,8.75) {\scriptsize $a'_{1}$};
\draw [->-] (4.,9.2) -- (3.4,9.4);
\node [] at (3.78,9.55) {\scriptsize $a'_{2}$};
\draw [->-] (1.6,9.2) -- (2.2,9.);
\node [] at (1.82,8.85) {\scriptsize $b'_{1}$};
\draw [->-] (1.6,10.) -- (2.2,9.8);
\node [] at (1.82,9.65) {\scriptsize $b'_{2}$};
\draw [fill] (3.4,8.6) circle [x radius=0.08, y radius=0.08];
\draw [fill] (3.4,9.4) circle [x radius=0.08, y radius=0.08];
\draw [fill] (2.2,9.) circle [x radius=0.08, y radius=0.08];
\draw [fill] (2.2,9.8) circle [x radius=0.08, y radius=0.08];
\node [fill=white,draw,rounded corners=.06cm] at (2.8,9.6) {\scriptsize $P_M$};
\node [fill=white,draw,rounded corners=.06cm] at (2.8,8.8) {\scriptsize $P_M$};
\end{tikzpicture}\:
\end{matrix}\Biggr\rangle
}
\newcommand{\ExpGScylinderBaseAC}{
\displaystyle\sum_{a'_{1}a'_{2}b'_{1}b'_{2}}\mathrm{u}_{a'_{1}}\mathrm{u}_{a'_{2}}\mathrm{u}_{b'_{1}}\mathrm{u}_{b'_{2}}
\Biggl|\;\begin{matrix}
\begin{tikzpicture}[scale=1]
\draw [->-] (2.9,8.) -- (2.9,9.2);
\node [] at (2.68,8.6) {\scriptsize $j$};
\draw [->-] (2.9,9.2) -- (2.9,10.4);
\node [] at (3.12,9.8) {\scriptsize $k$};
\draw [ultra thick] (3.6,8.4) -- (3.6,9.2);
\draw [ultra thick] (2.2,9.2) -- (2.2,10.);
\draw [->-] (4.2,8.2) -- (3.6,8.4);
\node [] at (3.98,8.55) {\scriptsize $a'_{1}$};
\draw [->-] (4.2,9.) -- (3.6,9.2);
\node [] at (3.98,9.35) {\scriptsize $a'_{2}$};
\draw [->-] (1.6,9.4) -- (2.2,9.2);
\node [] at (1.82,9.05) {\scriptsize $b'_{1}$};
\draw [->-] (1.6,10.2) -- (2.2,10.);
\node [] at (1.82,9.85) {\scriptsize $b'_{2}$};
\draw [ultra thick] (3.6,8.4) -- (3.4,8.9);
\draw [ultra thick] (3.4,8.9) -- (3.6,9.2);
\draw [ultra thick] (2.2,9.2) -- (2.4,9.5);
\draw [ultra thick] (2.4,9.5) -- (2.2,10.);
\draw [ultra thick] (3.4,8.9) -- (2.9,9.2);
\draw [ultra thick] (2.9,9.2) -- (2.4,9.5);
\draw [fill] (3.6,8.4) circle [x radius=0.08, y radius=0.08];
\draw [fill] (3.6,9.2) circle [x radius=0.08, y radius=0.08];
\draw [fill] (2.2,9.2) circle [x radius=0.08, y radius=0.08];
\draw [fill] (2.2,10.) circle [x radius=0.08, y radius=0.08];
\draw [fill] (3.4,8.9) circle [x radius=0.08, y radius=0.08];
\draw [fill] (2.4,9.5) circle [x radius=0.08, y radius=0.08];
\node [fill=white,draw,rounded corners=.06cm] at (2.9,9.2) {\scriptsize $P_M$};
\end{tikzpicture}\:
\end{matrix}\Biggr\rangle
}
\newcommand{\ExpGScylinderBaseAD}{
\displaystyle\sum_{a'_{1}a'_{2}b'_{1}b'_{2}}\mathrm{u}_{a'_{1}}\mathrm{u}_{a'_{2}}\mathrm{u}_{b'_{1}}\mathrm{u}_{b'_{2}}
\Biggl|\;\begin{matrix}
\begin{tikzpicture}[scale=1]
\draw [->-] (2.8,8.) -- (2.8,9.2);
\node [] at (2.58,8.6) {\scriptsize $j$};
\draw [->-] (2.8,9.2) -- (2.8,10.4);
\node [] at (3.02,9.8) {\scriptsize $k$};
\draw [->-] (4.,8.2) -- (3.4,8.9);
\node [] at (3.95,8.77) {\scriptsize $a'_{1}$};
\draw [->-] (4.,9.) -- (3.4,8.9);
\node [] at (3.66,9.19) {\scriptsize $a'_{2}$};
\draw [->-] (1.6,9.4) -- (2.2,9.5);
\node [] at (1.94,9.21) {\scriptsize $b'_{1}$};
\draw [->-] (1.6,10.2) -- (2.2,9.5);
\node [] at (1.65,9.63) {\scriptsize $b'_{2}$};
\draw [ultra thick] (3.4,8.9) -- (2.8,9.2);
\draw [ultra thick] (2.8,9.2) -- (2.2,9.5);
\draw [fill] (2.2,9.5) circle [x radius=0.08, y radius=0.08];
\draw [fill] (3.4,8.9) circle [x radius=0.08, y radius=0.08];
\node [fill=white,draw,rounded corners=.06cm] at (2.8,9.2) {\scriptsize $P_M$};
\end{tikzpicture}\:
\end{matrix}\Biggr\rangle
}
\newcommand{\ExpGScylinderBaseAE}{
\displaystyle\sum_{a'_{1}a'_{2}b'_{1}b'_{2}}\mathrm{u}_{a'_{1}}\mathrm{u}_{a'_{2}}\mathrm{u}_{b'_{1}}\mathrm{u}_{b'_{2}}
\Biggl|\;\begin{matrix}
\begin{tikzpicture}[scale=1]
\draw [ultra thick] (2.4,8.8) -- (2.4,9.6);
\draw [->-] (2.4,8.) -- (2.4,8.8);
\node [] at (2.18,8.4) {\scriptsize $j$};
\draw [->-] (2.4,9.6) -- (2.4,10.4);
\node [] at (2.62,10.) {\scriptsize $k$};
\draw [->-] (3.2,8.6) -- (2.4,8.8);
\node [] at (2.74,8.45) {\scriptsize $a'_{1}$};
\draw [->-] (3.2,9.4) -- (2.4,9.6);
\node [] at (2.74,9.25) {\scriptsize $a'_{2}$};
\draw [->-] (1.6,9.) -- (2.4,8.8);
\node [] at (2.06,9.15) {\scriptsize $b'_{1}$};
\draw [->-] (1.6,9.8) -- (2.4,9.6);
\node [] at (2.06,9.95) {\scriptsize $b'_{2}$};
\node [fill=white,draw,rounded corners=.06cm] at (2.4,9.6) {\scriptsize $P_M$};
\node [fill=white,draw,rounded corners=.06cm] at (2.4,8.8) {\scriptsize $P_M$};
\end{tikzpicture}\:
\end{matrix}\Biggr\rangle
}
\newcommand{\ExpGSdiskAfterT}{
\Biggl|\;\begin{matrix}
\begin{tikzpicture}[scale=1]
\draw [lightgray, ->, fill] (2.4,9.9) ..controls (2.4,10.1) and (2.32,10.3) .. (2.2,10.4)
			 ..controls (2.08,10.5) and (1.92,10.5) .. (1.8,10.4)
			 ..controls (1.68,10.3) and (1.6,10.1) .. (1.6,9.9)
			 ..controls (1.6,9.7) and (1.68,9.5) .. (1.8,9.4)
			 ..controls (1.92,9.3) and (2.08,9.3) .. (2.2,9.4)
			 ..controls (2.32,9.5) and (2.4,9.7) .. (2.4,9.9);
\draw [->-] (2.4,9.9) ..controls (2.4,10.1) and (2.32,10.3) .. (2.2,10.4)
			 ..controls (2.08,10.5) and (1.92,10.5) .. (1.8,10.4)
			 ..controls (1.68,10.3) and (1.6,10.1) .. (1.6,9.9)
			 ..controls (1.6,9.7) and (1.68,9.5) .. (1.8,9.4)
			 ..controls (1.92,9.3) and (2.08,9.3) .. (2.2,9.4)
			 ..controls (2.32,9.5) and (2.4,9.7) .. (2.4,9.9);
\draw [dotted] (4.2,9.9) -- (4.8,9.9);
\draw [->] (3.,9.9) -- (3.,10.3);
\node [] at (3.29,10.1) {\scriptsize $a_{1}$};
\draw [->] (3.6,9.9) -- (3.6,10.3);
\node [] at (3.89,10.1) {\scriptsize $a_{2}$};
\draw [->] (4.2,9.9) -- (4.2,10.3);
\node [] at (4.49,10.1) {\scriptsize $a_{3}$};
\draw [->] (5.4,9.9) -- (5.8,9.9);
\node [] at (5.6,9.68) {\scriptsize $a_{N}$};
\draw [->] (5.4,9.9) -- (5.4,10.3);
\node [] at (5.84,10.1) {\scriptsize $a_{N-1}$};
\draw [->] (4.8,9.9) -- (4.8,10.3);
\draw [ultra thick] (2.4,9.9) -- (3.,9.9);
\draw [ultra thick] (3.,9.9) -- (3.6,9.9);
\draw [ultra thick] (3.6,9.9) -- (4.2,9.9);
\draw [ultra thick] (4.8,9.9) -- (5.4,9.9);
\draw [fill] (3.,9.9) circle [x radius=0.08, y radius=0.08];
\draw [fill] (3.6,9.9) circle [x radius=0.08, y radius=0.08];
\draw [fill] (4.2,9.9) circle [x radius=0.08, y radius=0.08];
\draw [fill] (4.8,9.9) circle [x radius=0.08, y radius=0.08];
\draw [fill] (5.4,9.9) circle [x radius=0.08, y radius=0.08];
\node [fill=white,draw,rounded corners=.06cm] at (2.4,9.9) {\scriptsize $\rho_M$};
\node [] at (1.8,9.9) {\scriptsize $j$};
\end{tikzpicture}\:
\end{matrix}\Biggr\rangle
}
\newcommand{\ExpGSdiskAfterTAA}{
\Biggl|\;\begin{matrix}
\begin{tikzpicture}[scale=1]
\draw [lightgray, ->, fill] (2.4,9.9) ..controls (2.4,10.1) and (2.32,10.3) .. (2.2,10.4)
			 ..controls (2.08,10.5) and (1.92,10.5) .. (1.8,10.4)
			 ..controls (1.68,10.3) and (1.6,10.1) .. (1.6,9.9)
			 ..controls (1.6,9.7) and (1.68,9.5) .. (1.8,9.4)
			 ..controls (1.92,9.3) and (2.08,9.3) .. (2.2,9.4)
			 ..controls (2.32,9.5) and (2.4,9.7) .. (2.4,9.9);
\draw [->-] (2.4,9.9) ..controls (2.4,10.1) and (2.32,10.3) .. (2.2,10.4)
			 ..controls (2.08,10.5) and (1.92,10.5) .. (1.8,10.4)
			 ..controls (1.68,10.3) and (1.6,10.1) .. (1.6,9.9)
			 ..controls (1.6,9.7) and (1.68,9.5) .. (1.8,9.4)
			 ..controls (1.92,9.3) and (2.08,9.3) .. (2.2,9.4)
			 ..controls (2.32,9.5) and (2.4,9.7) .. (2.4,9.9);
\draw [dotted] (4.2,9.9) -- (4.8,9.9);
\draw [->] (3.,9.9) -- (3.,10.3);
\node [] at (3.29,10.1) {\scriptsize $a_{1}$};
\draw [->] (3.6,9.9) -- (3.6,10.3);
\node [] at (3.89,10.1) {\scriptsize $a_{2}$};
\draw [->] (4.2,9.9) -- (4.2,10.3);
\node [] at (4.49,10.1) {\scriptsize $a_{3}$};
\draw [->] (5.4,9.9) -- (5.8,9.9);
\node [] at (5.6,9.68) {\scriptsize $a_{N}$};
\draw [->] (5.4,9.9) -- (5.4,10.3);
\node [] at (5.84,10.1) {\scriptsize $a_{N-1}$};
\draw [->] (4.8,9.9) -- (4.8,10.3);
\draw [dashed] (2.4,9.9) -- (3.,9.9);
\node [] at (2.7,9.68) {\scriptsize $0$};
\draw [ultra thick] (3.,9.9) -- (3.6,9.9);
\draw [ultra thick] (3.6,9.9) -- (4.2,9.9);
\draw [ultra thick] (4.8,9.9) -- (5.4,9.9);
\draw [fill] (3.,9.9) circle [x radius=0.08, y radius=0.08];
\draw [fill] (3.6,9.9) circle [x radius=0.08, y radius=0.08];
\draw [fill] (4.2,9.9) circle [x radius=0.08, y radius=0.08];
\draw [fill] (4.8,9.9) circle [x radius=0.08, y radius=0.08];
\draw [fill] (5.4,9.9) circle [x radius=0.08, y radius=0.08];
\node [] at (1.8,9.9) {\scriptsize $j$};
\end{tikzpicture}\:
\end{matrix}\Biggr\rangle
}
\newcommand{\ExpModuleCompleteness}{
\Biggl|\;\begin{matrix}
\begin{tikzpicture}[scale=1]
\draw [ultra thick] (1.8,8.8) ..controls (1.992,8.783) and (2.183,8.767) .. (2.3,8.8)
			 ..controls (2.417,8.833) and (2.458,8.917) .. (2.5,9.);
\draw [ultra thick] (2.,9.6) ..controls (2.158,9.5) and (2.317,9.4) .. (2.4,9.3)
			 ..controls (2.483,9.2) and (2.492,9.1) .. (2.5,9.);
\draw [->-] (1.6,8.) -- (1.8,8.8);
\node [] at (1.49,8.45) {\scriptsize $j$};
\draw [->-] (2.,9.6) -- (2.2,10.4);
\node [] at (1.89,10.05) {\scriptsize $j$};
\draw [->-] (3.,8.8) -- (2.5,9.);
\node [] at (2.83,9.1) {\scriptsize $a$};
\draw [->-] (1.8,8.8) -- (2.,9.6);
\node [] at (1.69,9.25) {\scriptsize $j$};
\draw [fill] (2.5,9.) circle [x radius=0.08, y radius=0.08];
\node [fill=white,draw,rounded corners=.06cm] at (1.8,8.8) {\scriptsize $\rho_M$};
\node [fill=white,draw,rounded corners=.06cm] at (2.,9.6) {\scriptsize $\rho_N$};
\end{tikzpicture}\:
\end{matrix}\Biggr\rangle
}
\newcommand{\ExpModuleCompletenessAA}{
\Biggl|\;\begin{matrix}
\begin{tikzpicture}[scale=1]
\draw [->-] (1.6,8.) -- (1.9,9.2);
\node [] at (1.54,8.65) {\scriptsize $j$};
\draw [->-] (1.9,9.2) -- (2.2,10.4);
\node [] at (1.84,9.85) {\scriptsize $j$};
\draw [->-] (3.,8.8) -- (1.9,9.2);
\node [] at (2.53,9.21) {\scriptsize $a$};
\node [fill=white,draw,rounded corners=.06cm] at (1.9,9.2) {\scriptsize $\rho_N$};
\end{tikzpicture}\:
\end{matrix}\Biggr\rangle
}
\newcommand{\ExpModuleCompletenessAB}{
\Biggl|\;\begin{matrix}
\begin{tikzpicture}[scale=1]
\draw [ultra thick] (1.8,8.8) ..controls (1.992,8.783) and (2.183,8.767) .. (2.3,8.8)
			 ..controls (2.417,8.833) and (2.458,8.917) .. (2.5,9.);
\draw [ultra thick] (2.,9.6) ..controls (2.158,9.5) and (2.317,9.4) .. (2.4,9.3)
			 ..controls (2.483,9.2) and (2.492,9.1) .. (2.5,9.);
\draw [->-] (1.6,8.) -- (1.8,8.8);
\node [] at (1.49,8.45) {\scriptsize $j$};
\draw [->-] (2.,9.6) -- (2.2,10.4);
\node [] at (1.89,10.05) {\scriptsize $j$};
\draw [->-] (3.,8.8) -- (2.5,9.);
\node [] at (2.83,9.1) {\scriptsize $a$};
\draw [->-] (1.8,8.8) -- (2.,9.6);
\node [] at (1.69,9.25) {\scriptsize $j$};
\draw [fill] (2.5,9.) circle [x radius=0.08, y radius=0.08];
\node [fill=white,draw,rounded corners=.06cm] at (2.,9.6) {\scriptsize $\rho_N$};
\node [fill=white,draw,rounded corners=.06cm] at (1.8,8.8) {\scriptsize $\mathds{1}$};
\end{tikzpicture}\:
\end{matrix}\Biggr\rangle
}
\newcommand{\ExpModuleDef}{
\Biggl|\;\begin{matrix}
\begin{tikzpicture}[scale=0.8]
\draw [->-] (1.6,8.) -- (1.6,8.8);
\node [] at (1.31,8.4) {\scriptsize $j_{1}$};
\draw [->-] (1.6,9.6) -- (1.6,10.4);
\node [] at (1.31,10.) {\scriptsize $j_{2}$};
\draw [->-] (2.4,8.4) -- (1.6,8.8);
\node [] at (2.11,8.83) {\scriptsize $a_{1}$};
\draw [->-] (2.4,9.2) -- (1.6,9.6);
\node [] at (2.11,9.63) {\scriptsize $a_{2}$};
\draw [ultra thick] (1.6,8.8) -- (1.6,9.6);
\node [fill=white,draw,rounded corners=.06cm] at (1.6,8.8) {\scriptsize $\rho$};
\node [fill=white,draw,rounded corners=.06cm] at (1.6,9.6) {\scriptsize $\rho$};
\end{tikzpicture}\:
\end{matrix}\Biggr\rangle
}
\newcommand{\ExpModuleDefAA}{
\Biggl|\;\begin{matrix}
\begin{tikzpicture}[scale=0.8]
\draw [->-] (1.6,8.) -- (1.6,9.2);
\node [] at (1.31,8.6) {\scriptsize $j_{1}$};
\draw [->-] (1.6,9.2) -- (1.6,10.4);
\node [] at (1.31,9.8) {\scriptsize $j_{2}$};
\draw [->-] (2.3,8.2) -- (2.2,8.8);
\node [] at (1.96,8.45) {\scriptsize $a_{1}$};
\draw [->-] (2.8,8.6) -- (2.2,8.8);
\node [] at (2.58,8.93) {\scriptsize $a_{2}$};
\draw [ultra thick] (2.2,8.8) -- (1.6,9.2);
\draw [fill] (2.2,8.8) circle [x radius=0.08, y radius=0.08];
\node [fill=white,draw,rounded corners=.06cm] at (1.6,9.2) {\scriptsize $\rho$};
\end{tikzpicture}\:
\end{matrix}\Biggr\rangle
}
\newcommand{\ExpModuleDefAB}{
\displaystyle\sum_{j'}\rho^{a_{1}}_{j_{1}j'}\rho^{a_{2}}_{j'j_{2}}
\Biggl|\;\begin{matrix}
\begin{tikzpicture}[scale=0.8]
\draw [->-] (1.6,8.) -- (1.6,8.8);
\node [] at (1.31,8.4) {\scriptsize $j_{1}$};
\draw [->-] (1.6,9.6) -- (1.6,10.4);
\node [] at (1.31,10.) {\scriptsize $j_{2}$};
\draw [->-] (2.4,8.4) -- (1.6,8.8);
\node [] at (2.11,8.83) {\scriptsize $a_{1}$};
\draw [->-] (2.4,9.2) -- (1.6,9.6);
\node [] at (2.11,9.63) {\scriptsize $a_{2}$};
\draw [->-] (1.6,8.8) -- (1.6,9.6);
\node [] at (1.31,9.2) {\scriptsize $j'$};
\end{tikzpicture}\:
\end{matrix}\Biggr\rangle
}
\newcommand{\ExpModuleDefAD}{
\displaystyle\sum_{j'}\rho^{a_{1}}_{j_{1}j'}\rho^{a_{2}}_{j'j_{2}}\displaystyle\sum_{a'}G^{j_{1} a_{1} {j'}^*}_{a_{2} j_{2}^* a'}\mathrm{v}_{{j'}}\mathrm{v}_{a'}
\Biggl|\;\begin{matrix}
\begin{tikzpicture}[scale=0.8]
\draw [->-] (1.6,8.2) -- (1.6,9.4);
\node [] at (1.31,8.8) {\scriptsize $j_{1}$};
\draw [->-] (2.7,8.8) -- (2.1,9.);
\node [] at (2.48,9.13) {\scriptsize $a_{2}$};
\draw [->-] (1.6,9.4) -- (1.6,10.4);
\node [] at (1.31,9.9) {\scriptsize $j_{2}$};
\draw [->-] (2.1,9.) -- (1.6,9.4);
\node [] at (2.02,9.41) {\scriptsize $a'$};
\draw [->-] (2.4,8.2) -- (2.1,9.);
\node [] at (1.98,8.5) {\scriptsize $a_{1}$};
\end{tikzpicture}\:
\end{matrix}\Biggr\rangle
}
\newcommand{\ExpModuleDefAE}{
\displaystyle\sum_{a'}\rho^{a'}_{j_{1}j_{2}}f_{a_{2} {a'}^* a_{1}}
\Biggl|\;\begin{matrix}
\begin{tikzpicture}[scale=0.8]
\draw [->-] (1.6,8.2) -- (1.6,9.4);
\node [] at (1.31,8.8) {\scriptsize $j_{1}$};
\draw [->-] (2.7,8.8) -- (2.1,9.);
\node [] at (2.48,9.13) {\scriptsize $a_{2}$};
\draw [->-] (1.6,9.4) -- (1.6,10.4);
\node [] at (1.31,9.9) {\scriptsize $j_{2}$};
\draw [->-] (2.1,9.) -- (1.6,9.4);
\node [] at (2.02,9.41) {\scriptsize $a'$};
\draw [->-] (2.4,8.2) -- (2.1,9.);
\node [] at (1.98,8.5) {\scriptsize $a_{1}$};
\end{tikzpicture}\:
\end{matrix}\Biggr\rangle
}
\newcommand{\ExpModuleOrtho}{
\Biggl|\;\begin{matrix}
\begin{tikzpicture}[scale=1]
\draw [ultra thick] (1.8,8.8) ..controls (1.991,8.738) and (2.182,8.676) .. (2.3,8.8)
			 ..controls (2.418,8.924) and (2.462,9.236) .. (2.4,9.4)
			 ..controls (2.338,9.564) and (2.169,9.582) .. (2.,9.6);
\draw [->-] (1.6,8.) -- (1.8,8.8);
\node [] at (1.49,8.45) {\scriptsize $j$};
\draw [->-] (1.8,8.8) -- (2.,9.6);
\node [] at (1.69,9.25) {\scriptsize $k$};
\draw [->-] (2.,9.6) -- (2.2,10.4);
\node [] at (1.89,10.05) {\scriptsize $j$};
\node [fill=white,draw,rounded corners=.06cm] at (1.8,8.8) {\scriptsize $\rho_M$};
\node [fill=white,draw,rounded corners=.06cm] at (2.,9.6) {\scriptsize $\rho_N$};
\end{tikzpicture}\:
\end{matrix}\Biggr\rangle
}
\newcommand{\ExpModuleOrthoAA}{
\Biggl|\;\begin{matrix}
\begin{tikzpicture}[scale=1]
\draw [->-] (2.,9.6) ..controls (2.133,9.767) and (2.267,9.933) .. (2.4,10.)
			 ..controls (2.533,10.067) and (2.667,10.033) .. (2.8,10.);
\draw [->-] (2.,8.) ..controls (1.917,8.133) and (1.833,8.267) .. (1.8,8.4)
			 ..controls (1.767,8.533) and (1.783,8.667) .. (1.8,8.8);
\draw [->-] (2.8,10.) ..controls (2.887,9.819) and (2.974,9.638) .. (3.,9.4)
			 ..controls (3.026,9.162) and (2.992,8.867) .. (2.9,8.6)
			 ..controls (2.808,8.333) and (2.66,8.095) .. (2.5,8.)
			 ..controls (2.34,7.905) and (2.17,7.952) .. (2.,8.);
\draw [ultra thick] (1.8,8.8) ..controls (1.991,8.751) and (2.182,8.702) .. (2.3,8.8)
			 ..controls (2.418,8.898) and (2.462,9.142) .. (2.4,9.3)
			 ..controls (2.338,9.458) and (2.169,9.529) .. (2.,9.6);
\draw [->-] (1.8,8.8) -- (2.,9.6);
\node [] at (1.69,9.25) {\scriptsize $k$};
\draw [dashed] (2.8,10.) -- (3.,10.4);
\draw [dashed] (1.8,7.6) -- (2.,8.);
\node [fill=white,draw,rounded corners=.06cm] at (1.8,8.8) {\scriptsize $\rho_M$};
\node [fill=white,draw,rounded corners=.06cm] at (2.,9.6) {\scriptsize $\rho_M$};
\node [] at (2.2,10.1) {\scriptsize $j$};
\node [] at (1.6,8.2) {\scriptsize $j$};
\node [] at (3.2,8.7) {\scriptsize $j$};
\node [] at (3.2,10.2) {\scriptsize $0$};
\node [] at (2.,7.6) {\scriptsize $0$};
\end{tikzpicture}\:
\end{matrix}\Biggr\rangle
}
\newcommand{\ExpModuleOrthoAB}{
\Biggl|\;\begin{matrix}
\begin{tikzpicture}[scale=1]
\draw [->-] (1.8,8.4) ..controls (1.707,8.569) and (1.613,8.738) .. (1.6,9.)
			 ..controls (1.587,9.262) and (1.653,9.618) .. (1.8,9.8)
			 ..controls (1.947,9.982) and (2.173,9.991) .. (2.4,10.);
\draw [->-] (2.4,10.) ..controls (2.523,9.914) and (2.647,9.828) .. (2.7,9.7)
			 ..controls (2.753,9.572) and (2.736,9.401) .. (2.7,9.2)
			 ..controls (2.664,8.999) and (2.61,8.769) .. (2.5,8.6)
			 ..controls (2.39,8.431) and (2.226,8.323) .. (2.1,8.3)
			 ..controls (1.974,8.277) and (1.887,8.338) .. (1.8,8.4);
\draw [dashed] (2.4,10.) -- (2.6,10.4);
\draw [dashed] (1.6,8.) -- (1.8,8.4);
\node [] at (1.9,9.5) {\scriptsize $j$};
\node [] at (2.9,9.1) {\scriptsize $j$};
\node [] at (2.8,10.1) {\scriptsize $0$};
\node [] at (2.,8.1) {\scriptsize $0$};
\end{tikzpicture}\:
\end{matrix}\Biggr\rangle
}
\newcommand{\ExpModuleOrthoAC}{
\Biggl|\;\begin{matrix}
\begin{tikzpicture}[scale=1]
\draw [->-] (2.,8.) ..controls (1.896,8.244) and (1.791,8.489) .. (1.8,8.8)
			 ..controls (1.809,9.111) and (1.931,9.489) .. (2.1,9.7)
			 ..controls (2.269,9.911) and (2.484,9.956) .. (2.7,10.);
\draw [->-] (2.7,10.) ..controls (2.775,9.992) and (2.85,9.983) .. (2.9,9.9)
			 ..controls (2.95,9.817) and (2.975,9.658) .. (3.,9.5);
\draw [->-] (2.6,8.3) ..controls (2.45,8.125) and (2.3,7.95) .. (2.2,7.9)
			 ..controls (2.1,7.85) and (2.05,7.925) .. (2.,8.);
\draw [->-] (3.,9.5) ..controls (3.004,9.38) and (3.009,9.26) .. (3.,9.1)
			 ..controls (2.991,8.94) and (2.969,8.74) .. (2.9,8.6)
			 ..controls (2.831,8.46) and (2.716,8.38) .. (2.6,8.3);
\draw [ultra thick] (2.6,8.3) ..controls (2.516,8.367) and (2.431,8.433) .. (2.4,8.6)
			 ..controls (2.369,8.767) and (2.391,9.033) .. (2.5,9.2)
			 ..controls (2.609,9.367) and (2.804,9.433) .. (3.,9.5);
\draw [dashed] (1.8,7.6) -- (2.,8.);
\draw [dashed] (2.7,10.) -- (2.9,10.4);
\node [fill=white,draw,rounded corners=.06cm] at (2.6,8.3) {\scriptsize $\rho_M$};
\node [fill=white,draw,rounded corners=.06cm] at (3.,9.5) {\scriptsize $\rho_M$};
\node [] at (3.2,8.7) {\scriptsize $j$};
\node [] at (2.6,10.4) {\scriptsize $0$};
\node [] at (1.6,7.8) {\scriptsize $0$};
\node [] at (3.1,10.) {\scriptsize $k$};
\node [] at (1.9,9.9) {\scriptsize $k$};
\node [] at (2.4,7.8) {\scriptsize $k$};
\end{tikzpicture}\:
\end{matrix}\Biggr\rangle
}
\newcommand{\ExpQonYdiagram}{
\Biggl|\;\begin{matrix}
\begin{tikzpicture}[scale=1]
\draw [->-] (2.2,9.2) -- (2.2,10.);
\draw [->-] (2.8,10.4) -- (2.2,10.);
\draw [->-] (1.6,10.4) -- (2.2,10.);
\node [] at (2.5,9.6) {\scriptsize $j_{1}$};
\node [] at (2.7,10.1) {\scriptsize $j_{2}$};
\node [] at (1.7,10.1) {\scriptsize $j_{3}$};
\end{tikzpicture}\:
\end{matrix}\Biggr\rangle
}
\newcommand{\ExpQPcreationUnderPiNewAA}{
\displaystyle\sum_{a'_2jka_4}\rmu_{a'_2}\rmu_{j}\rmu_k\rmu_{a_4}
\Biggl|\;\begin{matrix}
\begin{tikzpicture}[scale=0.9]
\path [lightgray, fill] (1.6,6.6) -- (2.1,6.6) -- (2.1,7.2) -- (2.6,7.8) -- (2.7,8.2) -- (2.1,8.8) -- (2.1,9.6) -- (2.1,10.4) -- (1.6,10.4) -- cycle;
\draw [ultra thick] (2.1,8.8) ..controls (2.35,8.8) and (2.6,8.8) .. (2.7,8.7)
			 ..controls (2.8,8.6) and (2.75,8.4) .. (2.7,8.2);
\draw [ultra thick] (2.1,7.2) ..controls (2.358,7.2) and (2.617,7.2) .. (2.7,7.3)
			 ..controls (2.783,7.4) and (2.692,7.6) .. (2.6,7.8);
\draw [lightgray, fill] (2.1,7.2) ..controls (2.358,7.2) and (2.617,7.2) .. (2.7,7.3)
			 ..controls (2.783,7.4) and (2.692,7.6) .. (2.6,7.8);
\draw [lightgray, fill] (2.1,8.8) ..controls (2.35,8.8) and (2.6,8.8) .. (2.7,8.7)
			 ..controls (2.8,8.6) and (2.75,8.4) .. (2.7,8.2);
\draw [snake it, ->] (3.2,8.1) ..controls (3.2,8.292) and (3.2,8.483) .. (3.3,8.6)
			 ..controls (3.4,8.717) and (3.6,8.758) .. (3.8,8.8);
\draw [snake it, ->] (3.8,7.4) ..controls (3.6,7.492) and (3.4,7.583) .. (3.3,7.7)
			 ..controls (3.2,7.817) and (3.2,7.958) .. (3.2,8.1);
\draw [ultra thick] (2.1,8.8) ..controls (2.35,8.8) and (2.6,8.8) .. (2.7,8.7)
			 ..controls (2.8,8.6) and (2.75,8.4) .. (2.7,8.2);
\draw [ultra thick] (2.1,7.2) ..controls (2.358,7.2) and (2.617,7.2) .. (2.7,7.3)
			 ..controls (2.783,7.4) and (2.692,7.6) .. (2.6,7.8);
\draw [ultra thick] (2.1,7.2) ..controls (2.358,7.2) and (2.617,7.2) .. (2.7,7.3)
			 ..controls (2.783,7.4) and (2.692,7.6) .. (2.6,7.8);
\draw [ultra thick] (2.1,8.8) ..controls (2.35,8.8) and (2.6,8.8) .. (2.7,8.7)
			 ..controls (2.8,8.6) and (2.75,8.4) .. (2.7,8.2);
\draw [->] (3.4,8.1) -- (3.8,8.1);
\draw [ultra thick] (2.6,7.8) -- (2.7,8.2);
\draw [ultra thick] (2.7,8.2) -- (3.2,8.1);
\draw [->-] (2.1,6.6) -- (2.1,7.2);
\draw [->-] (2.1,9.6) -- (2.1,10.4);
\draw [ultra thick] (2.1,7.2) -- (2.1,8.);
\draw [ultra thick] (2.1,8.) -- (2.1,8.8);
\draw [ultra thick] (2.1,8.8) -- (2.1,9.6);
\draw [->-] (2.1,9.6) -- (3.8,9.6);
\draw [ultra thick] (2.1,8.) -- (2.6,7.8);
\draw [fill] (2.6,7.8) circle [x radius=0.08, y radius=0.08];
\draw [fill] (2.7,8.2) circle [x radius=0.08, y radius=0.08];
\node [fill=white,draw,rounded corners=.06cm] at (3.2,8.1) {\scriptsize $P_M$};
\node [] at (3.3,7.3) {\scriptsize $j$};
\node [] at (3.3,8.9) {\scriptsize $k$};
\node [] at (3.7,7.8) {\scriptsize $a'_{2}$};
\node [fill=white,draw,rounded corners=.06cm] at (2.1,7.2) {\scriptsize $\rho_N$};
\node [fill=white,draw,rounded corners=.06cm] at (2.1,8.) {\scriptsize $\rho_N$};
\node [fill=white,draw,rounded corners=.06cm] at (2.1,8.8) {\scriptsize $\rho_N$};
\node [fill=white,draw,rounded corners=.06cm] at (2.1,9.6) {\scriptsize $\rho_N$};
\node [] at (3.1,9.9) {\scriptsize $a_{4}$};
\end{tikzpicture}\:
\end{matrix}\Biggr\rangle
}
\newcommand{\ExpQPcreationUnderPiNewAB}{
\displaystyle\sum_{a'_2jka'_4}\frac{\mathrm{d}_N}{\mathrm{d}_k}\rmu_{a'_2}\rmu_{j}\rmu_k\rmu_{a'_4}
\Biggl|\;\begin{matrix}
\begin{tikzpicture}[scale=0.9]
\path [lightgray, fill] (1.6,6.6) -- (2.1,6.6) -- (2.1,7.2) -- (2.6,7.8) -- (2.7,8.2) -- (2.1,8.8) -- (2.1,9.6) -- (2.1,10.4) -- (1.6,10.4) -- cycle;
\draw [lightgray, fill] (2.1,7.2) ..controls (2.358,7.25) and (2.617,7.3) .. (2.7,7.4)
			 ..controls (2.783,7.5) and (2.692,7.65) .. (2.6,7.8);
\path [lightgray, fill] (2.1,9.6) -- (2.1,8.8) -- (2.7,8.2) -- (3.2,8.) -- (3.7,8.) -- (3.5,8.8) -- (3.3,9.6) -- cycle;
\draw [ultra thick] (2.1,8.8) ..controls (2.35,8.9) and (2.6,9.) .. (2.7,8.9)
			 ..controls (2.8,8.8) and (2.75,8.5) .. (2.7,8.2);
\draw [ultra thick] (2.1,7.2) ..controls (2.358,7.25) and (2.617,7.3) .. (2.7,7.4)
			 ..controls (2.783,7.5) and (2.692,7.65) .. (2.6,7.8);
\draw [snake it, ->] (3.2,8.) ..controls (3.175,8.183) and (3.15,8.367) .. (3.2,8.5)
			 ..controls (3.25,8.633) and (3.375,8.717) .. (3.5,8.8);
\draw [snake it, ->] (3.8,7.3) ..controls (3.6,7.392) and (3.4,7.483) .. (3.3,7.6)
			 ..controls (3.2,7.717) and (3.2,7.858) .. (3.2,8.);
\draw [ultra thick] (2.1,8.8) ..controls (2.35,8.9) and (2.6,9.) .. (2.7,8.9)
			 ..controls (2.8,8.8) and (2.75,8.5) .. (2.7,8.2);
\draw [ultra thick] (2.1,7.2) ..controls (2.358,7.25) and (2.617,7.3) .. (2.7,7.4)
			 ..controls (2.783,7.5) and (2.692,7.65) .. (2.6,7.8);
\draw [snake it, ->] (3.5,8.8) ..controls (3.642,8.842) and (3.783,8.883) .. (3.9,8.9)
			 ..controls (4.017,8.917) and (4.108,8.908) .. (4.2,8.9);
\draw [ultra thick] (2.6,7.8) -- (2.7,8.2);
\draw [ultra thick] (2.7,8.2) -- (3.2,8.);
\draw [->-] (2.1,6.6) -- (2.1,7.2);
\draw [->-] (2.1,9.6) -- (2.1,10.4);
\draw [ultra thick] (2.1,7.2) -- (2.1,8.);
\draw [ultra thick] (2.1,8.) -- (2.1,8.8);
\draw [ultra thick] (2.1,8.8) -- (2.1,9.6);
\draw [ultra thick] (2.1,8.) -- (2.6,7.8);
\draw [ultra thick] (2.1,9.6) -- (3.3,9.6);
\draw [->-] (3.3,9.6) -- (4.2,9.6);
\draw [->-] (3.7,8.) -- (4.2,8.);
\draw [ultra thick] (3.2,8.) -- (3.7,8.);
\draw [ultra thick] (3.3,9.6) -- (3.5,8.8);
\draw [ultra thick] (3.5,8.8) -- (3.7,8.);
\draw [fill] (2.6,7.8) circle [x radius=0.08, y radius=0.08];
\draw [fill] (2.7,8.2) circle [x radius=0.08, y radius=0.08];
\draw [fill] (3.3,9.6) circle [x radius=0.08, y radius=0.08];
\draw [fill] (3.7,8.) circle [x radius=0.08, y radius=0.08];
\node [fill=white,draw,rounded corners=.06cm] at (3.2,8.) {\scriptsize $P_M$};
\node [] at (3.4,7.2) {\scriptsize $j$};
\node [] at (3.,8.6) {\scriptsize $k$};
\node [] at (4.,7.7) {\scriptsize $a'_{2}$};
\node [fill=white,draw,rounded corners=.06cm] at (2.1,7.2) {\scriptsize $\rho_N$};
\node [fill=white,draw,rounded corners=.06cm] at (2.1,8.) {\scriptsize $\rho_N$};
\node [fill=white,draw,rounded corners=.06cm] at (2.1,8.8) {\scriptsize $\rho_N$};
\node [fill=white,draw,rounded corners=.06cm] at (2.1,9.6) {\scriptsize $\rho_N$};
\node [] at (3.8,9.9) {\scriptsize $a'_{4}$};
\node [fill=white,draw,rounded corners=.06cm] at (3.5,8.8) {\scriptsize $P_N$};
\node [] at (4.1,8.7) {\scriptsize $k$};
\end{tikzpicture}\:
\end{matrix}\Biggr\rangle
}
\newcommand{\ExpQPcreationUnderPiNewAC}{
\Biggl|\;\begin{matrix}
\begin{tikzpicture}[scale=0.9]
\path [lightgray, fill] (1.6,6.7) -- (2.1,6.7) -- (2.1,7.3) -- (2.7,7.9) -- (2.8,8.4) -- (2.1,8.9) -- (2.1,9.6) -- (2.1,10.4) -- (1.6,10.4) -- cycle;
\draw [lightgray, fill] (2.1,7.3) ..controls (2.35,7.3) and (2.6,7.3) .. (2.7,7.4)
			 ..controls (2.8,7.5) and (2.75,7.7) .. (2.7,7.9);
\draw [lightgray, fill] (2.1,8.9) ..controls (2.342,8.942) and (2.583,8.983) .. (2.7,8.9)
			 ..controls (2.817,8.817) and (2.808,8.608) .. (2.8,8.4);
\draw [ultra thick] (2.1,8.9) ..controls (2.342,8.942) and (2.583,8.983) .. (2.7,8.9)
			 ..controls (2.817,8.817) and (2.808,8.608) .. (2.8,8.4);
\draw [ultra thick] (2.1,7.3) ..controls (2.35,7.3) and (2.6,7.3) .. (2.7,7.4)
			 ..controls (2.8,7.5) and (2.75,7.7) .. (2.7,7.9);
\draw [snake it, ->] (3.8,8.9) ..controls (3.933,8.942) and (4.067,8.983) .. (4.2,9.)
			 ..controls (4.333,9.017) and (4.467,9.008) .. (4.6,9.);
\draw [ultra thick] (3.1,8.4) ..controls (3.076,8.624) and (3.051,8.849) .. (3.1,9.)
			 ..controls (3.149,9.151) and (3.271,9.229) .. (3.4,9.2)
			 ..controls (3.529,9.171) and (3.664,9.036) .. (3.8,8.9);
\draw [snake it] (4.1,7.3) ..controls (3.9,7.35) and (3.7,7.4) .. (3.6,7.5)
			 ..controls (3.5,7.6) and (3.5,7.75) .. (3.5,7.9);
\draw [snake it] (3.5,7.9) ..controls (3.475,8.117) and (3.45,8.333) .. (3.5,8.5)
			 ..controls (3.55,8.667) and (3.675,8.783) .. (3.8,8.9);
\draw [ultra thick] (3.1,8.4) ..controls (3.067,8.242) and (3.033,8.083) .. (3.1,8.)
			 ..controls (3.167,7.917) and (3.333,7.908) .. (3.5,7.9);
\draw [ultra thick] (3.5,7.9) ..controls (3.658,7.867) and (3.817,7.833) .. (3.9,7.9)
			 ..controls (3.983,7.967) and (3.992,8.133) .. (4.,8.3);
\draw [ultra thick] (2.7,7.9) -- (2.8,8.4);
\draw [->-] (2.1,6.7) -- (2.1,7.3);
\draw [->-] (2.1,9.6) -- (2.1,10.4);
\draw [ultra thick] (2.1,7.3) -- (2.1,8.1);
\draw [ultra thick] (2.1,8.1) -- (2.1,8.9);
\draw [ultra thick] (2.1,8.9) -- (2.1,9.6);
\draw [ultra thick] (2.1,8.1) -- (2.7,7.9);
\draw [ultra thick] (3.8,8.9) -- (4.,8.3);
\draw [->-] (2.1,9.6) -- (4.6,9.6);
\draw [ultra thick] (2.8,8.4) -- (3.1,8.4);
\draw [->] (4.,8.3) -- (4.6,8.1);
\draw [fill] (2.7,7.9) circle [x radius=0.08, y radius=0.08];
\draw [fill] (2.8,8.4) circle [x radius=0.08, y radius=0.08];
\draw [fill] (4.,8.3) circle [x radius=0.08, y radius=0.08];
\node [fill=white,draw,rounded corners=.06cm] at (3.5,7.9) {\scriptsize $P_M$};
\node [] at (3.9,7.) {\scriptsize $j$};
\node [] at (3.7,8.3) {\scriptsize $k$};
\node [] at (4.5,7.8) {\scriptsize $a'_{2}$};
\node [fill=white,draw,rounded corners=.06cm] at (2.1,7.3) {\scriptsize $\rho_N$};
\node [fill=white,draw,rounded corners=.06cm] at (2.1,8.1) {\scriptsize $\rho_N$};
\node [fill=white,draw,rounded corners=.06cm] at (2.1,8.9) {\scriptsize $\rho_N$};
\node [fill=white,draw,rounded corners=.06cm] at (2.1,9.6) {\scriptsize $\rho_N$};
\node [] at (4.1,10.) {\scriptsize $a'_{4}$};
\node [] at (4.5,8.6) {\scriptsize $k$};
\node [fill=white,draw,rounded corners=.06cm] at (3.8,8.9) {\scriptsize $P_N$};
\end{tikzpicture}\:
\end{matrix}\Biggr\rangle
}
\newcommand{\ExpQPcreationUnderPiNewAD}{
\displaystyle\sum_{a'_2jka'_{4}}\rmu_{a'_2}\rmu_{j}\rmu_k\rmu_{a'_{4}}
\Biggl|\;\begin{matrix}
\begin{tikzpicture}[scale=0.9]
\path [lightgray, fill] (1.6,6.6) -- (2.1,6.6) -- (2.1,7.2) -- (2.6,7.8) -- (2.8,8.1) -- (2.1,8.8) -- (2.1,9.6) -- (2.1,10.4) -- (1.6,10.4) -- cycle;
\draw [lightgray, fill] (2.1,7.2) ..controls (2.358,7.2) and (2.617,7.2) .. (2.7,7.3)
			 ..controls (2.783,7.4) and (2.692,7.6) .. (2.6,7.8);
\draw [lightgray, fill] (2.1,8.8) ..controls (2.292,8.858) and (2.483,8.917) .. (2.6,8.8)
			 ..controls (2.717,8.683) and (2.758,8.392) .. (2.8,8.1);
\draw [snake it, ->] (3.3,8.) ..controls (3.292,8.233) and (3.283,8.467) .. (3.4,8.6)
			 ..controls (3.517,8.733) and (3.758,8.767) .. (4.,8.8);
\draw [snake it, ->] (4.,7.3) ..controls (3.758,7.392) and (3.517,7.483) .. (3.4,7.6)
			 ..controls (3.283,7.717) and (3.292,7.858) .. (3.3,8.);
\draw [ultra thick] (2.1,8.8) ..controls (2.292,8.858) and (2.483,8.917) .. (2.6,8.8)
			 ..controls (2.717,8.683) and (2.758,8.392) .. (2.8,8.1);
\draw [ultra thick] (2.1,7.2) ..controls (2.358,7.2) and (2.617,7.2) .. (2.7,7.3)
			 ..controls (2.783,7.4) and (2.692,7.6) .. (2.6,7.8);
\draw [->] (3.5,8.) -- (4.,8.);
\draw [ultra thick] (2.6,7.8) -- (2.8,8.1);
\draw [ultra thick] (2.8,8.1) -- (3.3,8.);
\draw [->-] (2.1,6.6) -- (2.1,7.2);
\draw [->-] (2.1,9.6) -- (2.1,10.4);
\draw [ultra thick] (2.1,7.2) -- (2.1,8.);
\draw [ultra thick] (2.1,8.) -- (2.1,8.8);
\draw [ultra thick] (2.1,8.8) -- (2.1,9.6);
\draw [->-] (2.1,9.6) -- (4.,9.6);
\draw [ultra thick] (2.1,8.) -- (2.6,7.8);
\draw [fill] (2.6,7.8) circle [x radius=0.08, y radius=0.08];
\draw [fill] (2.8,8.1) circle [x radius=0.08, y radius=0.08];
\node [fill=white,draw,rounded corners=.06cm] at (3.3,8.) {\scriptsize $P_M$};
\node [] at (3.6,7.2) {\scriptsize $j$};
\node [] at (3.6,9.) {\scriptsize $k$};
\node [] at (3.9,7.8) {\scriptsize $a'_{2}$};
\node [fill=white,draw,rounded corners=.06cm] at (2.1,7.2) {\scriptsize $\rho_N$};
\node [fill=white,draw,rounded corners=.06cm] at (2.1,8.) {\scriptsize $\rho_N$};
\node [fill=white,draw,rounded corners=.06cm] at (2.1,8.8) {\scriptsize $\rho_N$};
\node [fill=white,draw,rounded corners=.06cm] at (2.1,9.6) {\scriptsize $\rho_N$};
\node [] at (3.1,9.9) {\scriptsize $a_{4}$};
\end{tikzpicture}\:
\end{matrix}\Biggr\rangle
}
\newcommand{\ExpQPprojection}{
\Biggl|\;\begin{matrix}
\begin{tikzpicture}[scale=0.9]
\path [lightgray, fill] (1.8,7.) -- (2.2,7.) -- (2.4,7.6) -- (2.2,8.2) -- (2.4,8.7) -- (2.2,9.2) -- (2.4,9.8) -- (2.2,10.4) -- (1.6,10.4) -- cycle;
\draw [] (2.,9.2) -- (2.2,9.2);
\draw [] (2.,8.2) -- (2.2,8.2);
\draw [->-] (2.2,7.) -- (2.4,7.6);
\node [] at (2.03,7.39) {\scriptsize $j_{1}$};
\draw [->-] (2.4,7.6) -- (2.2,8.2);
\node [] at (2.03,7.81) {\scriptsize $j_{2}$};
\draw [->-] (2.2,8.2) -- (2.4,8.7);
\node [] at (2.03,8.56) {\scriptsize $j_{3}$};
\draw [->-] (2.4,8.7) -- (2.2,9.2);
\node [] at (2.03,8.84) {\scriptsize $j_{4}$};
\draw [->-] (2.2,9.2) -- (2.4,9.8);
\node [] at (2.03,9.59) {\scriptsize $j_{5}$};
\draw [->-] (2.4,9.8) -- (2.2,10.4);
\node [] at (2.03,10.01) {\scriptsize $j_{6}$};
\draw [->-] (2.4,7.6) -- (2.9,7.6);
\node [] at (2.65,7.38) {\scriptsize $a_{1}$};
\draw [->-] (2.4,9.8) -- (2.9,9.8);
\node [] at (2.65,10.02) {\scriptsize $a_{2}$};
\draw [snake it, ->] (2.4,8.7) -- (2.9,8.7);
\node [] at (2.65,8.48) {\scriptsize $k$};
\end{tikzpicture}\:
\end{matrix}\Biggr\rangle
}
\newcommand{\ExpQPprojectionAA}{
\Biggl|\;\begin{matrix}
\begin{tikzpicture}[scale=0.9]
\path [lightgray, fill] (1.8,7.) -- (2.2,7.) -- (2.4,7.6) -- (2.2,8.2) -- (2.4,8.7) -- (2.2,9.2) -- (2.4,9.8) -- (2.2,10.4) -- (1.6,10.4) -- cycle;
\draw [] (2.,9.2) -- (2.2,9.2);
\draw [] (2.,8.2) -- (2.2,8.2);
\draw [->-] (2.2,7.) -- (2.4,7.6);
\node [] at (2.03,7.39) {\scriptsize $j_{1}$};
\draw [->-] (2.4,7.6) -- (2.2,8.2);
\node [] at (2.03,7.81) {\scriptsize $j_{2}$};
\draw [->-] (2.2,8.2) -- (2.4,8.7);
\node [] at (2.03,8.56) {\scriptsize $j_{3}$};
\draw [->-] (2.4,8.7) -- (2.2,9.2);
\node [] at (2.03,8.84) {\scriptsize $j_{4}$};
\draw [->-] (2.2,9.2) -- (2.4,9.8);
\node [] at (2.03,9.59) {\scriptsize $j_{5}$};
\draw [->-] (2.4,9.8) -- (2.2,10.4);
\node [] at (2.03,10.01) {\scriptsize $j_{6}$};
\draw [->-] (2.4,7.6) -- (3.,7.6);
\node [] at (2.7,7.38) {\scriptsize $a_{1}$};
\draw [ultra thick] (3.,7.6) -- (3.,8.7);
\draw [ultra thick] (3.,8.7) -- (3.,9.8);
\draw [->-] (3.,7.6) -- (3.7,7.6);
\node [] at (3.35,7.38) {\scriptsize $a'_{1}$};
\draw [->-] (2.4,9.8) -- (3.,9.8);
\node [] at (2.7,10.02) {\scriptsize $a_{2}$};
\draw [->-] (3.,9.8) -- (3.7,9.8);
\node [] at (3.35,10.02) {\scriptsize $a'_{2}$};
\draw [snake it, ->] (2.4,8.7) -- (3.,8.7);
\draw [snake it, ->] (3.,8.7) -- (3.7,8.7);
\draw [fill] (3.,9.8) circle [x radius=0.08, y radius=0.08];
\draw [fill] (3.,7.6) circle [x radius=0.08, y radius=0.08];
\node [fill=white,draw,rounded corners=.06cm] at (3.,8.7) {\scriptsize $P_M$};
\node [] at (2.6,8.4) {\scriptsize $k$};
\node [] at (3.6,8.4) {\scriptsize $k$};
\end{tikzpicture}\:
\end{matrix}\Biggr\rangle
}
\newcommand{\ExpQuasiparticleBasisAA}{
\Biggl|\;\begin{matrix}
\begin{tikzpicture}[scale=1]
\path [lightgray, fill] (1.6,8.8) -- (2.2,8.8) -- (2.3,9.2) -- (2.2,9.6) -- (2.3,10.) -- (2.2,10.4) -- (1.6,10.4) -- cycle;
\draw [] (1.9,8.8) -- (2.2,8.8);
\draw [] (1.9,9.6) -- (2.2,9.6);
\draw [] (1.9,10.4) -- (2.2,10.4);
\draw [->-] (2.2,8.8) -- (2.3,9.2);
\draw [->-] (2.2,9.6) -- (2.3,10.);
\draw [->-] (2.3,9.2) -- (2.2,9.6);
\draw [->-] (2.3,10.) -- (2.2,10.4);
\draw [->-] (2.3,10.) -- (2.8,10.);
\node [] at (2.55,9.78) {\scriptsize $a$};
\draw [snake it, ->] (2.3,9.2) -- (2.8,9.2);
\node [] at (2.55,8.98) {\scriptsize $j$};
\end{tikzpicture}\:
\end{matrix}\Biggr\rangle
}
\newcommand{\ExpQuasiparticleBasisAB}{
\Biggl|\;\begin{matrix}
\begin{tikzpicture}[scale=1]
\path [lightgray, fill] (1.6,7.6) -- (2.2,7.6) -- (2.3,8.) -- (2.2,8.4) -- (2.3,8.8) -- (2.2,9.2) -- (2.3,9.6) -- (2.2,10.) -- (2.4,10.4) -- (1.6,10.4) -- cycle;
\draw [] (1.9,7.6) -- (2.2,7.6);
\draw [] (1.9,8.4) -- (2.2,8.4);
\draw [] (1.9,9.2) -- (2.2,9.2);
\draw [] (1.9,10.) -- (2.2,10.);
\draw [->-] (2.2,7.6) -- (2.3,8.);
\draw [->-] (2.2,8.4) -- (2.3,8.8);
\draw [->-] (2.2,9.2) -- (2.3,9.6);
\draw [->-] (2.3,8.) -- (2.2,8.4);
\draw [->-] (2.3,8.8) -- (2.2,9.2);
\draw [->-] (2.3,9.6) -- (2.2,10.);
\draw [->-] (2.3,8.) -- (2.7,8.);
\node [] at (2.5,7.78) {\scriptsize $a_{1}$};
\draw [->-] (2.3,8.8) -- (2.7,8.8);
\node [] at (2.5,8.58) {\scriptsize $a_{2}$};
\draw [->-] (2.3,9.6) -- (2.7,9.6);
\node [] at (2.5,9.38) {\scriptsize $a_{3}$};
\draw [->-, fill] (2.2,10.) -- (2.4,10.4);
\end{tikzpicture}\:
\end{matrix}\Biggr\rangle
}
\newcommand{\ExpQuasiparticleCreationAB}{
\Biggl|\;\begin{matrix}
\begin{tikzpicture}[scale=1]
\path [lightgray, fill] (1.7,7.6) -- (2.2,7.6) -- (2.3,8.) -- (2.2,8.4) -- (2.3,8.8) -- (2.2,9.2) -- (2.3,9.6) -- (2.2,10.) -- (2.3,10.4) -- (1.6,10.4) -- cycle;
\path [lightgray, fill] (2.3,9.6) -- (3.1,8.8) -- (2.8,8.8) -- (2.3,8.) -- (2.,8.9) -- cycle;
\draw [lightgray, fill] (2.3,8.) ..controls (2.508,8.133) and (2.717,8.267) .. (2.8,8.4)
			 ..controls (2.883,8.533) and (2.842,8.667) .. (2.8,8.8);
\draw [lightgray, fill] (2.3,9.6) ..controls (2.533,9.617) and (2.767,9.633) .. (2.9,9.5)
			 ..controls (3.033,9.367) and (3.067,9.083) .. (3.1,8.8);
\draw [snake it, ->] (3.5,8.8) ..controls (3.492,8.942) and (3.483,9.083) .. (3.6,9.2)
			 ..controls (3.717,9.317) and (3.958,9.408) .. (4.2,9.5);
\draw [snake it, ->] (4.2,8.1) ..controls (3.958,8.192) and (3.717,8.283) .. (3.6,8.4)
			 ..controls (3.483,8.517) and (3.492,8.658) .. (3.5,8.8);
\draw [->-] (2.3,9.6) ..controls (2.533,9.617) and (2.767,9.633) .. (2.9,9.5)
			 ..controls (3.033,9.367) and (3.067,9.083) .. (3.1,8.8);
\draw [->-] (2.3,8.) ..controls (2.508,8.133) and (2.717,8.267) .. (2.8,8.4)
			 ..controls (2.883,8.533) and (2.842,8.667) .. (2.8,8.8);
\draw [] (1.9,7.6) -- (2.2,7.6);
\draw [] (1.9,8.4) -- (2.2,8.4);
\draw [] (1.9,9.2) -- (2.2,9.2);
\draw [] (1.9,10.) -- (2.2,10.);
\draw [->-] (2.2,7.6) -- (2.3,8.);
\draw [->-] (2.2,8.4) -- (2.3,8.8);
\draw [->-] (2.2,9.2) -- (2.3,9.6);
\draw [->-] (2.2,10.) -- (2.3,10.4);
\draw [->-] (2.3,8.) -- (2.2,8.4);
\draw [->-] (2.3,8.8) -- (2.2,9.2);
\draw [->-] (2.3,9.6) -- (2.2,10.);
\draw [->] (3.7,8.8) -- (4.2,8.8);
\draw [->-] (2.3,8.8) -- (2.8,8.8);
\draw [ultra thick] (2.8,8.8) -- (3.1,8.8);
\draw [ultra thick] (3.1,8.8) -- (3.5,8.8);
\draw [fill] (2.8,8.8) circle [x radius=0.08, y radius=0.08];
\draw [fill] (3.1,8.8) circle [x radius=0.08, y radius=0.08];
\node [fill=white,draw,rounded corners=.06cm] at (3.5,8.8) {\scriptsize $P_M$};
\node [] at (3.8,8.) {\scriptsize $j$};
\node [] at (3.8,9.6) {\scriptsize $k$};
\node [] at (2.9,8.) {\scriptsize $a_{1}$};
\node [] at (2.6,9.) {\scriptsize $a_{2}$};
\node [] at (2.8,9.8) {\scriptsize $a_{3}$};
\node [] at (4.1,8.6) {\scriptsize $a'_{2}$};
\end{tikzpicture}\:
\end{matrix}\Biggr\rangle
}
\newcommand{\ExpQuasiparticleHoppingAA}{
\Biggl|\;\begin{matrix}
\begin{tikzpicture}[scale=1]
\draw [lightgray, fill] (2.2,9.2) ..controls (2.35,9.217) and (2.5,9.233) .. (2.6,9.3)
			 ..controls (2.7,9.367) and (2.75,9.483) .. (2.8,9.6);
\draw [lightgray, fill] (2.2,10.) ..controls (2.35,9.983) and (2.5,9.967) .. (2.6,9.9)
			 ..controls (2.7,9.833) and (2.75,9.717) .. (2.8,9.6);
\path [lightgray, fill] (1.6,8.8) -- (2.1,8.8) -- (2.2,9.2) -- (2.8,9.6) -- (2.2,10.) -- (2.1,10.4) -- (1.6,10.4) -- cycle;
\draw [->-] (2.2,10.) ..controls (2.35,9.983) and (2.5,9.967) .. (2.6,9.9)
			 ..controls (2.7,9.833) and (2.75,9.717) .. (2.8,9.6);
\draw [->-] (2.8,9.6) ..controls (2.85,9.492) and (2.9,9.383) .. (3.,9.3)
			 ..controls (3.1,9.217) and (3.25,9.158) .. (3.4,9.1);
\draw [snake it, ->] (2.2,9.2) ..controls (2.35,9.217) and (2.5,9.233) .. (2.6,9.3)
			 ..controls (2.7,9.367) and (2.75,9.483) .. (2.8,9.6);
\draw [snake it, ->] (2.8,9.6) ..controls (2.85,9.75) and (2.9,9.9) .. (3.,10.)
			 ..controls (3.1,10.1) and (3.25,10.15) .. (3.4,10.2);
\draw [] (1.8,8.8) -- (2.1,8.8);
\draw [] (1.8,9.6) -- (2.1,9.6);
\draw [] (1.8,10.4) -- (2.1,10.4);
\draw [->-] (2.1,8.8) -- (2.2,9.2);
\draw [->-] (2.1,9.6) -- (2.2,10.);
\draw [->-] (2.2,9.2) -- (2.1,9.6);
\draw [->-] (2.2,10.) -- (2.1,10.4);
\node [fill=white,draw,rounded corners=.06cm] at (2.8,9.6) {\scriptsize $P_M$};
\node [] at (2.5,9.) {\scriptsize $j$};
\node [] at (2.5,10.2) {\scriptsize $a$};
\node [] at (3.4,9.9) {\scriptsize $j'$};
\node [] at (3.4,9.3) {\scriptsize $a'$};
\end{tikzpicture}\:
\end{matrix}\Biggr\rangle
}
\newcommand{\ExpRhoOrthonormalAA}{
\Biggl|\;\begin{matrix}
\begin{tikzpicture}[scale=0.8]
\draw [->-] (1.6,8.) -- (2.2,10.4);
\node [] at (1.69,9.25) {\scriptsize $j$};
\end{tikzpicture}\:
\end{matrix}\Biggr\rangle
}
\newcommand{\ExpRhoRhoGraphics}{
\Biggl|\;\begin{matrix}
\begin{tikzpicture}[scale=1]
\draw [->-] (1.8,8.8) ..controls (1.991,8.751) and (2.182,8.702) .. (2.3,8.8)
			 ..controls (2.418,8.898) and (2.462,9.142) .. (2.4,9.3)
			 ..controls (2.338,9.458) and (2.169,9.529) .. (2.,9.6);
\draw [->-] (1.6,8.) -- (1.8,8.8);
\node [] at (1.91,8.35) {\scriptsize $j$};
\draw [->-] (2.,9.6) -- (2.2,10.4);
\node [] at (2.31,9.95) {\scriptsize $j$};
\draw [ultra thick] (1.8,8.8) -- (2.,9.6);
\node [fill=white,draw,rounded corners=.06cm] at (1.8,8.8) {\scriptsize $\rho$};
\node [fill=white,draw,rounded corners=.06cm] at (2.,9.6) {\scriptsize $\rho$};
\node [] at (2.6,9.) {\scriptsize $a$};
\end{tikzpicture}\:
\end{matrix}\Biggr\rangle
}
\newcommand{\ExpStrongFrobeniusAA}{
\frac{\sqrt{D}}{\mathrm{d}_A}\displaystyle\sum_{a,b}f_{a b c^*}f_{c b^* a^*}
\Biggl|\;\begin{matrix}
\begin{tikzpicture}[scale=0.7]
\draw [->-] (2.1,9.) ..controls (1.9,9.044) and (1.7,9.089) .. (1.6,9.2)
			 ..controls (1.5,9.311) and (1.5,9.489) .. (1.6,9.6)
			 ..controls (1.7,9.711) and (1.9,9.756) .. (2.1,9.8);
\draw [->-] (2.1,9.) ..controls (2.3,9.044) and (2.5,9.089) .. (2.6,9.2)
			 ..controls (2.7,9.311) and (2.7,9.489) .. (2.6,9.6)
			 ..controls (2.5,9.711) and (2.3,9.756) .. (2.1,9.8);
\draw [->-] (2.1,9.8) -- (2.1,10.4);
\node [] at (2.32,10.1) {\scriptsize $c$};
\draw [->-] (2.1,8.4) -- (2.1,9.);
\node [] at (2.32,8.7) {\scriptsize $c$};
\node [] at (1.8,9.4) {\scriptsize $a$};
\node [] at (2.9,9.4) {\scriptsize $b$};
\end{tikzpicture}\:
\end{matrix}\Biggr\rangle
}
\newcommand{\ExpStrongFrobeniusAB}{
\Biggl|\;\begin{matrix}
\begin{tikzpicture}[scale=0.7]
\draw [->-] (1.6,8.4) -- (1.6,10.4);
\node [] at (1.82,9.4) {\scriptsize $k$};
\end{tikzpicture}\:
\end{matrix}\Biggr\rangle
}
\newcommand{\ExpStrongFrobeniusCompactAA}{
\Biggl|\;\begin{matrix}
\begin{tikzpicture}[scale=0.8]
\draw [ultra thick] (2.2,9.) ..controls (2.44,8.978) and (2.68,8.956) .. (2.8,9.1)
			 ..controls (2.92,9.244) and (2.92,9.556) .. (2.8,9.7)
			 ..controls (2.68,9.844) and (2.44,9.822) .. (2.2,9.8);
\draw [ultra thick] (2.2,9.) ..controls (1.96,8.978) and (1.72,8.956) .. (1.6,9.1)
			 ..controls (1.48,9.244) and (1.48,9.556) .. (1.6,9.7)
			 ..controls (1.72,9.844) and (1.96,9.822) .. (2.2,9.8);
\draw [->-] (2.2,9.8) -- (2.2,10.4);
\node [] at (2.42,10.1) {\scriptsize $k$};
\draw [->-] (2.2,8.4) -- (2.2,9.);
\node [] at (2.42,8.7) {\scriptsize $k$};
\draw [fill] (2.2,9.) circle [x radius=0.08, y radius=0.08];
\draw [fill] (2.2,9.8) circle [x radius=0.08, y radius=0.08];
\end{tikzpicture}\:
\end{matrix}\Biggr\rangle
}
\newcommand{\TailLabel}[3]{
\begin{tikzpicture}[scale=0.8]
\path [lightgray, fill] (1.6,8.8) -- (2.2,8.8) -- (2.2,9.6) -- (2.2,10.4) -- (1.6,10.4) -- cycle;
\draw [->-] (2.2,8.8) -- (2.2,9.6);
\draw [->-] (2.2,9.6) -- (2.2,10.4);
\draw [->-] (2.2,9.6) -- (2.8,9.6);
\node [] at (2.6,9.4) {\scriptsize $#1$};
\node [] at (1.8,9.2) {\scriptsize $#2$};
\node [] at (1.8,10.) {\scriptsize $#3$};
\end{tikzpicture}
}

\newcommand{\ExpTBdryMoveOneTwo}{
\Biggl|\;\begin{matrix}
\begin{tikzpicture}[scale=1]
\path [lightgray, fill] (1.6,8.8) -- (2.2,8.8) -- (2.2,9.6) -- (2.2,10.4) -- (1.6,10.4) -- cycle;
\draw [->-] (2.2,9.6) -- (2.8,9.6);
\node [] at (2.5,9.38) {\scriptsize $a_{1}$};
\draw [->-] (2.2,8.8) -- (2.2,9.6);
\node [] at (1.98,9.2) {\scriptsize $i$};
\draw [->-] (2.2,9.6) -- (2.2,10.4);
\node [] at (1.98,10.) {\scriptsize $j$};
\end{tikzpicture}\:
\end{matrix}\Biggr\rangle
}
\newcommand{\ExpTBdryMoveOneTwoAC}{
\displaystyle\sum_{a_{2},a_{3}}\frac{\mathrm{u}_{a_{1}}\mathrm{u}_{a_{2}}\mathrm{u}_{a_{3}}}{\sqrt{\mathrm{d}_A}}\displaystyle\sum_{k}\mathrm{v}_{k}f_{a_{2}^* a_{3}^* a_{1}}G^{j^* i a_{1}^*}_{a_{2}^* a_{3}^* k}
\Biggl|\;\begin{matrix}
\begin{tikzpicture}[scale=1]
\path [lightgray, fill] (1.6,8.8) -- (2.2,8.8) -- (2.2,9.6) -- (2.2,10.4) -- (1.6,10.4) -- cycle;
\draw [->-] (2.2,8.8) -- (2.2,9.3);
\node [] at (1.98,9.05) {\scriptsize $i$};
\draw [->-] (2.2,9.3) -- (2.2,9.9);
\node [] at (1.98,9.6) {\scriptsize $k$};
\draw [->-] (2.2,9.3) -- (3.,9.3);
\node [] at (2.6,9.08) {\scriptsize $a_{2}$};
\draw [->-] (2.2,9.9) -- (2.2,10.4);
\node [] at (1.98,10.15) {\scriptsize $j$};
\draw [->-] (2.2,9.9) -- (3.,9.9);
\node [] at (2.6,10.12) {\scriptsize $a_{3}$};
\end{tikzpicture}\:
\end{matrix}\Biggr\rangle
}
\newcommand{\ExpTBdryMoveTwoOne}{
\Biggl|\;\begin{matrix}
\begin{tikzpicture}[scale=1]
\path [lightgray, fill] (1.6,8.8) -- (2.2,8.8) -- (2.2,9.6) -- (2.2,10.4) -- (1.6,10.4) -- cycle;
\draw [->-] (2.2,8.8) -- (2.2,9.3);
\node [] at (1.98,9.05) {\scriptsize $i$};
\draw [->-] (2.2,9.3) -- (2.2,9.9);
\node [] at (1.98,9.6) {\scriptsize $k$};
\draw [->-] (2.2,9.3) -- (3.,9.3);
\node [] at (2.6,9.08) {\scriptsize $a_{2}$};
\draw [->-] (2.2,9.9) -- (2.2,10.4);
\node [] at (1.98,10.15) {\scriptsize $j$};
\draw [->-] (2.2,9.9) -- (3.,9.9);
\node [] at (2.6,10.12) {\scriptsize $a_{3}$};
\end{tikzpicture}\:
\end{matrix}\Biggr\rangle
}
\newcommand{\ExpTBdryMoveTwoOneAC}{
\displaystyle\sum_{a_{1}}\frac{\mathrm{u}_{a_{1}}\mathrm{u}_{a_{2}}\mathrm{u}_{a_{3}}}{\sqrt{\mathrm{d}_A}}\mathrm{v}_{k}f_{a_{2} a_{3} a_{1}^*}G^{j a_{1} i^*}_{a_{2}^* k^* a_{3}}
\Biggl|\;\begin{matrix}
\begin{tikzpicture}[scale=1]
\path [lightgray, fill] (1.6,8.8) -- (2.2,8.8) -- (2.2,9.6) -- (2.2,10.4) -- (1.6,10.4) -- cycle;
\draw [->-] (2.2,8.8) -- (2.2,9.6);
\node [] at (1.98,9.2) {\scriptsize $i$};
\draw [->-] (2.2,9.6) -- (2.2,10.4);
\node [] at (1.98,10.) {\scriptsize $j$};
\draw [->-] (2.2,9.6) -- (3.,9.6);
\node [] at (2.6,9.38) {\scriptsize $a_{1}$};
\end{tikzpicture}\:
\end{matrix}\Biggr\rangle
}
\newcommand{\ExpTMoveOneThree}{
\Biggl|\;\begin{matrix}
\begin{tikzpicture}[scale=0.7]
\draw [->-] (1.6,10.4) -- (2.2,9.8);
\node [] at (1.71,9.91) {\scriptsize $j_{1}$};
\draw [->-] (2.2,9.) -- (2.2,9.8);
\node [] at (2.49,9.4) {\scriptsize $j_{2}$};
\draw [->-] (2.8,10.4) -- (2.2,9.8);
\node [] at (2.69,9.91) {\scriptsize $j_{3}$};
\end{tikzpicture}\:
\end{matrix}\Biggr\rangle
}
\newcommand{\ExpTMoveOneThreeAA}{
\displaystyle\sum_{j_{4},j_{5},j_{6},}\frac{\mathrm{v}_{j_{5}}\mathrm{v}_{j_{6}}\mathrm{v}_{j_{4}}}{\sqrt{D}}G^{j_{1} j_{2} j_{3}}_{j_{5} j_{6} j_{4}}
\Biggl|\;\begin{matrix}
\begin{tikzpicture}[scale=0.7]
\draw [->-] (1.6,10.4) -- (1.9,10.1);
\node [] at (1.56,10.06) {\scriptsize $j_{1}$};
\draw [->-] (2.3,9.) -- (2.3,9.4);
\node [] at (2.59,9.2) {\scriptsize $j_{2}$};
\draw [->-] (2.3,9.4) -- (1.9,10.1);
\node [] at (1.85,9.61) {\scriptsize $j_{4}$};
\draw [->-] (2.7,10.1) -- (1.9,10.1);
\node [] at (2.3,10.32) {\scriptsize $j_{6}$};
\draw [->-] (3.,10.4) -- (2.7,10.1);
\node [] at (3.04,10.06) {\scriptsize $j_{3}$};
\draw [->-] (2.7,10.1) -- (2.3,9.4);
\node [] at (2.75,9.61) {\scriptsize $j_{5}$};
\end{tikzpicture}\:
\end{matrix}\Biggr\rangle
}
\newcommand{\ExpTMoveOneTwo}{
\Biggl|\;\begin{matrix}
\begin{tikzpicture}[scale=1]
\path [lightgray, fill] (1.6,8.8) -- (2.2,8.8) -- (2.2,9.6) -- (2.2,10.4) -- (1.6,10.4) -- cycle;
\draw [->-] (2.2,8.8) -- (2.2,9.6);
\draw [->-] (2.2,9.6) -- (2.2,10.4);
\draw [->-] (2.2,9.6) -- (2.8,9.6);
\node [] at (2.5,9.38) {\scriptsize $a_{1}$};
\end{tikzpicture}\:
\end{matrix}\Biggr\rangle
}
\newcommand{\ExpTMoveOneTwoAA}{
\displaystyle\sum_{a_{2},a_{3}}\frac{f_{a_{2}^* a_{3}^* a_{1}}\mathrm{u}_{a_{2}}\mathrm{u}_{a_{3}}}{\mathrm{u}_{a_{1}}\sqrt{\mathrm{d}_A}}
\Biggl|\;\begin{matrix}
\begin{tikzpicture}[scale=1]
\path [lightgray, fill] (1.6,8.8) -- (2.2,8.8) -- (2.2,9.6) -- (2.2,10.4) -- (1.6,10.4) -- cycle;
\draw [->-] (2.2,8.8) -- (2.2,9.6);
\draw [->-] (2.2,9.6) -- (2.2,10.4);
\draw [->-] (2.2,9.6) -- (2.8,9.6);
\node [] at (2.5,9.38) {\scriptsize $a_{1}$};
\draw [->-] (2.8,9.6) -- (3.,10.2);
\node [] at (3.17,9.81) {\scriptsize $a_{3}$};
\draw [->-] (2.8,9.6) -- (3.,9.);
\node [] at (3.17,9.39) {\scriptsize $a_{2}$};
\end{tikzpicture}\:
\end{matrix}\Biggr\rangle
}
\newcommand{\TMovePentagonDiagram}{
\begin{tikzpicture}[scale=1]
\draw [] (1.8,7.8) -- (2.,8.2);
\draw [] (2.,8.2) -- (1.9,8.5);
\draw [] (2.,8.2) -- (2.3,8.1);
\draw [] (2.3,7.8) -- (2.3,8.1);
\draw [] (2.3,8.1) -- (2.6,8.2);
\draw [] (2.6,8.2) -- (2.7,8.5);
\draw [] (2.6,8.2) -- (2.8,7.8);
\draw [] (3.2,9.6) -- (3.4,10.);
\draw [] (3.4,10.) -- (3.2,10.3);
\draw [] (3.4,10.) -- (3.8,10.1);
\draw [] (3.8,10.1) -- (3.9,9.9);
\draw [] (3.8,10.1) -- (4.,10.4);
\draw [] (3.9,9.9) -- (3.8,9.6);
\draw [] (3.9,9.9) -- (4.3,9.8);
\draw [] (4.1,6.5) -- (4.4,6.8);
\draw [] (4.4,6.8) -- (4.5,7.1);
\draw [] (4.4,6.8) -- (4.8,6.5);
\draw [] (4.5,7.1) -- (4.2,7.4);
\draw [] (4.5,7.1) -- (4.9,7.1);
\draw [] (4.9,7.1) -- (5.3,6.5);
\draw [] (4.9,7.1) -- (5.3,7.4);
\draw [] (6.,9.8) -- (6.3,10.);
\draw [] (6.,10.3) -- (6.4,10.2);
\draw [] (6.3,10.) -- (6.4,10.2);
\draw [] (6.3,10.) -- (6.6,9.9);
\draw [] (6.4,10.2) -- (6.8,10.2);
\draw [] (6.6,9.9) -- (6.6,9.7);
\draw [] (6.6,9.9) -- (6.9,9.8);
\draw [] (6.9,7.7) -- (7.3,7.8);
\draw [] (7.3,7.8) -- (7.5,7.5);
\draw [] (7.3,7.8) -- (7.6,8.);
\draw [] (7.6,8.) -- (7.6,8.4);
\draw [] (7.6,8.) -- (8.,7.7);
\draw [] (7.6,8.4) -- (7.2,8.6);
\draw [] (7.6,8.4) -- (7.9,8.6);
\draw [->, thick] (2.6,8.9) -- (2.9,9.3);
\draw [->, thick] (3.,7.4) -- (3.4,7.1);
\draw [->, thick] (4.7,10.1) -- (5.3,10.1);
\draw [->, thick] (7.,9.3) -- (7.2,8.9);
\draw [->, thick] (6.1,7.1) -- (6.6,7.4);
\node [] at (1.6,7.8) {\scriptsize $1$};
\node [] at (1.7,8.5) {\scriptsize $2$};
\node [] at (2.1,7.8) {\scriptsize $3$};
\node [] at (2.6,7.8) {\scriptsize $4$};
\node [] at (2.5,8.5) {\scriptsize $5$};
\node [] at (3.,9.6) {\scriptsize $1$};
\node [] at (3.,10.2) {\scriptsize $2$};
\node [] at (3.6,9.6) {\scriptsize $3$};
\node [] at (4.2,9.7) {\scriptsize $4$};
\node [] at (3.8,10.4) {\scriptsize $5$};
\node [] at (5.9,9.7) {\scriptsize $1$};
\node [] at (5.8,10.3) {\scriptsize $2$};
\node [] at (6.4,9.7) {\scriptsize $3$};
\node [] at (6.9,9.6) {\scriptsize $4$};
\node [] at (6.7,10.4) {\scriptsize $5$};
\node [] at (3.9,6.6) {\scriptsize $1$};
\node [] at (4.,7.3) {\scriptsize $2$};
\node [] at (4.6,6.5) {\scriptsize $3$};
\node [] at (5.3,6.8) {\scriptsize $4$};
\node [] at (5.1,7.5) {\scriptsize $5$};
\node [] at (6.8,7.9) {\scriptsize $1$};
\node [] at (6.9,8.6) {\scriptsize $2$};
\node [] at (7.3,7.4) {\scriptsize $3$};
\node [] at (7.8,7.6) {\scriptsize $4$};
\node [] at (7.9,8.4) {\scriptsize $5$};
\end{tikzpicture}
}

\newcommand{\TMovePentagonDiagramAA}{
\begin{tikzpicture}[scale=0.5]
\draw [] (1.8,9.6) -- (2.,10.);
\draw [] (2.,10.) -- (1.9,10.3);
\draw [] (2.,10.) -- (2.3,9.9);
\draw [] (2.3,9.6) -- (2.3,9.9);
\draw [] (2.3,9.9) -- (2.6,10.);
\draw [] (2.6,10.) -- (2.7,10.3);
\draw [] (2.6,10.) -- (2.8,9.6);
\draw [] (4.1,9.6) -- (4.3,10.);
\draw [] (4.3,10.) -- (4.1,10.3);
\draw [] (4.3,10.) -- (4.7,10.1);
\draw [] (4.7,10.1) -- (4.8,9.9);
\draw [] (4.7,10.1) -- (4.9,10.4);
\draw [] (4.8,9.9) -- (4.7,9.6);
\draw [] (4.8,9.9) -- (5.2,9.8);
\draw [] (6.9,9.8) -- (7.2,10.);
\draw [] (6.9,10.3) -- (7.3,10.2);
\draw [] (7.2,10.) -- (7.3,10.2);
\draw [] (7.2,10.) -- (7.5,9.9);
\draw [] (7.3,10.2) -- (7.7,10.2);
\draw [] (7.5,9.9) -- (7.5,9.7);
\draw [] (7.5,9.9) -- (7.8,9.8);
\draw [] (8.7,9.5) -- (9.1,9.6);
\draw [] (9.1,9.6) -- (9.3,9.3);
\draw [] (9.1,9.6) -- (9.4,9.8);
\draw [] (9.4,9.8) -- (9.4,10.2);
\draw [] (9.4,9.8) -- (9.8,9.5);
\draw [] (9.4,10.2) -- (9.,10.4);
\draw [] (9.4,10.2) -- (9.7,10.4);
\draw [->, thick] (3.2,10.1) -- (3.8,10.1);
\draw [->, thick] (5.6,10.1) -- (6.2,10.1);
\draw [->, thick] (8.,10.1) -- (8.6,10.1);
\node [] at (1.6,9.7) {\scriptsize $1$};
\node [] at (1.6,10.3) {\scriptsize $2$};
\node [] at (2.1,9.6) {\scriptsize $3$};
\node [] at (2.6,9.6) {\scriptsize $4$};
\node [] at (2.5,10.3) {\scriptsize $5$};
\node [] at (3.9,9.6) {\scriptsize $1$};
\node [] at (3.9,10.3) {\scriptsize $2$};
\node [] at (4.5,9.6) {\scriptsize $3$};
\node [] at (5.1,9.6) {\scriptsize $4$};
\node [] at (5.,10.3) {\scriptsize $5$};
\node [] at (6.7,9.7) {\scriptsize $1$};
\node [] at (6.7,10.3) {\scriptsize $2$};
\node [] at (7.3,9.6) {\scriptsize $3$};
\node [] at (7.8,9.6) {\scriptsize $4$};
\node [] at (7.6,10.4) {\scriptsize $5$};
\node [] at (8.8,9.8) {\scriptsize $1$};
\node [] at (9.,9.3) {\scriptsize $2$};
\node [] at (9.6,9.4) {\scriptsize $3$};
\node [] at (8.9,10.2) {\scriptsize $4$};
\node [] at (9.7,10.2) {\scriptsize $5$};
\end{tikzpicture}
}

\newcommand{\TMovePentagonDiagramAB}{
\begin{tikzpicture}[scale=0.5]
\draw [] (1.8,9.6) -- (2.,10.);
\draw [] (2.,10.) -- (1.9,10.3);
\draw [] (2.,10.) -- (2.3,9.9);
\draw [] (2.3,9.6) -- (2.3,9.9);
\draw [] (2.3,9.9) -- (2.6,10.);
\draw [] (2.6,10.) -- (2.7,10.3);
\draw [] (2.6,10.) -- (2.8,9.6);
\draw [] (4.4,9.4) -- (4.7,9.7);
\draw [] (4.7,9.7) -- (4.8,10.);
\draw [] (4.7,9.7) -- (5.1,9.4);
\draw [] (4.8,10.) -- (4.5,10.3);
\draw [] (4.8,10.) -- (5.2,10.);
\draw [] (5.2,10.) -- (5.6,9.4);
\draw [] (5.2,10.) -- (5.6,10.3);
\draw [] (7.2,9.4) -- (7.6,9.5);
\draw [] (7.6,9.5) -- (7.8,9.2);
\draw [] (7.6,9.5) -- (7.9,9.7);
\draw [] (7.9,9.7) -- (7.9,10.1);
\draw [] (7.9,9.7) -- (8.3,9.4);
\draw [] (7.9,10.1) -- (7.5,10.3);
\draw [] (7.9,10.1) -- (8.2,10.3);
\draw [thick, ->] (3.2,9.9) -- (3.9,9.9);
\draw [thick, ->] (6.1,9.9) -- (6.9,9.9);
\node [] at (1.6,9.6) {\scriptsize $1$};
\node [] at (1.7,10.3) {\scriptsize $2$};
\node [] at (2.1,9.6) {\scriptsize $3$};
\node [] at (2.6,9.6) {\scriptsize $4$};
\node [] at (2.5,10.3) {\scriptsize $5$};
\node [] at (4.2,9.5) {\scriptsize $1$};
\node [] at (4.3,10.2) {\scriptsize $2$};
\node [] at (4.9,9.4) {\scriptsize $3$};
\node [] at (5.6,9.7) {\scriptsize $4$};
\node [] at (5.4,10.4) {\scriptsize $5$};
\node [] at (7.1,9.6) {\scriptsize $1$};
\node [] at (7.2,10.3) {\scriptsize $2$};
\node [] at (7.6,9.1) {\scriptsize $3$};
\node [] at (8.1,9.3) {\scriptsize $4$};
\node [] at (8.2,10.1) {\scriptsize $5$};
\end{tikzpicture}
}

\newcommand{\ExpTMovePentagonTransf}{
\Biggl|\;\begin{matrix}
\begin{tikzpicture}[scale=1]
\draw [->-] (1.6,9.) -- (2.,9.8);
\node [] at (2.06,9.27) {\scriptsize $j_{1}$};
\draw [->-] (2.,9.8) -- (1.8,10.4);
\node [] at (1.63,10.01) {\scriptsize $j_{2}$};
\draw [->-] (2.8,9.) -- (2.8,9.6);
\node [] at (3.09,9.3) {\scriptsize $j_{3}$};
\draw [->-] (3.6,9.8) -- (3.8,10.4);
\node [] at (3.97,10.01) {\scriptsize $j_{5}$};
\draw [->-] (3.6,9.8) -- (4.,9.);
\node [] at (4.06,9.53) {\scriptsize $j_{4}$};
\draw [->-] (2.,9.8) -- (2.8,9.6);
\node [] at (2.46,9.93) {\scriptsize $j_{6}$};
\draw [->-] (2.8,9.6) -- (3.6,9.8);
\node [] at (3.14,9.93) {\scriptsize $j_{7}$};
\end{tikzpicture}\:
\end{matrix}\Biggr\rangle
}
\newcommand{\ExpTMovePentagonTransfAA}{
\displaystyle\sum_{k_{3}}G^{j_{6} j_{3} j_{7}^*}_{j_{4}^* j_{5}^* k_{3}}\mathrm{v}_{j_{7}}\mathrm{v}_{k_{3}}
\Biggl|\;\begin{matrix}
\begin{tikzpicture}[scale=1]
\draw [->-] (1.6,9.) -- (2.,9.8);
\node [] at (2.06,9.27) {\scriptsize $j_{1}$};
\draw [->-] (2.,9.8) -- (1.8,10.4);
\node [] at (1.63,10.01) {\scriptsize $j_{2}$};
\draw [->-] (2.,9.8) -- (2.8,10.);
\node [] at (2.34,10.13) {\scriptsize $j_{6}$};
\draw [->-] (2.8,9.) -- (3.2,9.4);
\node [] at (3.19,9.01) {\scriptsize $j_{3}$};
\draw [->-] (3.2,9.4) -- (4.,9.);
\node [] at (3.71,9.43) {\scriptsize $j_{4}$};
\draw [->-] (2.8,10.) -- (3.8,10.4);
\node [] at (3.39,9.97) {\scriptsize $j_{5}$};
\draw [->-] (3.2,9.4) -- (2.8,10.);
\node [] at (2.77,9.54) {\scriptsize $k_{3}$};
\end{tikzpicture}\:
\end{matrix}\Biggr\rangle
}
\newcommand{\ExpTMovePentagonTransfAB}{
\displaystyle\sum_{k_{3}}G^{j_{6} j_{3} j_{7}^*}_{j_{4}^* j_{5}^* k_{3}}\mathrm{v}_{j_{7}}\mathrm{v}_{k_{3}}\displaystyle\sum_{k_{2}}G^{j_{2}^* j_{1} j_{6}^*}_{k_{3} j_{5}^* k_{2}}\mathrm{v}_{j_{6}}\mathrm{v}_{k_{2}}
\Biggl|\;\begin{matrix}
\begin{tikzpicture}[scale=1]
\draw [->-] (2.8,8.7) -- (3.2,9.4);
\node [] at (3.25,8.91) {\scriptsize $j_{3}$};
\draw [->-] (3.2,9.4) -- (4.,9.);
\node [] at (3.71,9.43) {\scriptsize $j_{4}$};
\draw [->-] (2.8,10.4) -- (1.8,10.4);
\node [] at (2.3,10.18) {\scriptsize $j_{2}$};
\draw [->-] (1.6,9.) -- (2.4,9.6);
\node [] at (2.16,9.09) {\scriptsize $j_{1}$};
\draw [->-] (3.2,9.4) -- (2.4,9.6);
\node [] at (2.74,9.27) {\scriptsize $k_{3}$};
\draw [->-] (2.8,10.4) -- (3.8,10.4);
\node [] at (3.3,10.18) {\scriptsize $j_{5}$};
\draw [->-] (2.4,9.6) -- (2.8,10.4);
\node [] at (2.86,9.87) {\scriptsize $k_{2}$};
\end{tikzpicture}\:
\end{matrix}\Biggr\rangle
}
\newcommand{\ExpTMovePentagonTransfAC}{
\displaystyle\sum_{k_{3}}G^{j_{6} j_{3} j_{7}^*}_{j_{4}^* j_{5}^* k_{3}}\mathrm{v}_{j_{7}}\mathrm{v}_{k_{3}}\displaystyle\sum_{k_{2}}G^{j_{2}^* j_{1} j_{6}^*}_{k_{3} j_{5}^* k_{2}}\mathrm{v}_{j_{6}}\mathrm{v}_{k_{2}}\displaystyle\sum_{k_{1}}G^{k_{2}^* j_{1} k_{3}}_{j_{3} j_{4}^* k_{1}}\mathrm{v}_{k_{3}}\mathrm{v}_{k_{1}}
\Biggl|\;\begin{matrix}
\begin{tikzpicture}[scale=1]
\draw [->-] (2.6,9.2) -- (3.2,9.6);
\node [] at (3.04,9.18) {\scriptsize $k_{1}$};
\draw [->-] (3.2,9.6) -- (3.2,10.2);
\node [] at (2.91,9.9) {\scriptsize $k_{2}$};
\draw [->-] (3.2,9.6) -- (4.,9.);
\node [] at (3.76,9.51) {\scriptsize $j_{4}$};
\draw [->-] (3.2,10.2) -- (2.4,10.4);
\node [] at (2.74,10.07) {\scriptsize $j_{2}$};
\draw [->-] (3.2,10.2) -- (3.8,10.4);
\node [] at (3.58,10.07) {\scriptsize $j_{5}$};
\draw [->-] (1.6,9.) -- (2.6,9.2);
\node [] at (2.05,9.33) {\scriptsize $j_{1}$};
\draw [->-] (2.8,8.6) -- (2.6,9.2);
\node [] at (2.43,8.81) {\scriptsize $j_{3}$};
\end{tikzpicture}\:
\end{matrix}\Biggr\rangle
}
\newcommand{\ExpTMovePentagonTransfAD}{
\displaystyle\sum_{k_{1}}G^{j_{2}^* j_{1} j_{6}^*}_{j_{3} j_{7}^* k_{1}}\mathrm{v}_{j_{6}}\mathrm{v}_{k_{1}}
\Biggl|\;\begin{matrix}
\begin{tikzpicture}[scale=1]
\draw [->-] (1.6,9.) -- (2.3,9.4);
\node [] at (1.82,9.42) {\scriptsize $j_{1}$};
\draw [->-] (2.3,9.4) -- (2.6,9.9);
\node [] at (2.69,9.5) {\scriptsize $k_{1}$};
\draw [->-] (2.6,9.9) -- (1.8,10.4);
\node [] at (2.06,9.93) {\scriptsize $j_{2}$};
\draw [->-] (2.6,9.9) -- (3.4,9.8);
\node [] at (3.03,10.08) {\scriptsize $j_{7}$};
\draw [->-] (2.8,9.) -- (2.3,9.4);
\node [] at (2.38,8.99) {\scriptsize $j_{3}$};
\draw [->-] (3.4,9.8) -- (3.8,10.4);
\node [] at (3.83,9.94) {\scriptsize $j_{5}$};
\draw [->-] (3.4,9.8) -- (4.,9.);
\node [] at (3.48,9.23) {\scriptsize $j_{4}$};
\end{tikzpicture}\:
\end{matrix}\Biggr\rangle
}
\newcommand{\ExpTMovePentagonTransfAE}{
\displaystyle\sum_{k_{1}}G^{j_{2}^* j_{1} j_{6}^*}_{j_{3} j_{7}^* k_{1}}\mathrm{v}_{j_{6}}\mathrm{v}_{k_{1}}\displaystyle\sum_{k_{2}}G^{j_{2}^* k_{1} j_{7}^*}_{j_{4}^* j_{5}^* k_{2}}\mathrm{v}_{j_{7}}\mathrm{v}_{k_{2}}
\Biggl|\;\begin{matrix}
\begin{tikzpicture}[scale=1]
\draw [->-] (1.6,9.) -- (2.6,9.2);
\node [] at (2.05,9.33) {\scriptsize $j_{1}$};
\draw [->-] (2.6,9.2) -- (3.2,9.5);
\node [] at (3.01,9.12) {\scriptsize $k_{1}$};
\draw [->-] (2.8,8.6) -- (2.6,9.2);
\node [] at (2.43,8.81) {\scriptsize $j_{3}$};
\draw [->-] (3.2,9.5) -- (4.,9.);
\node [] at (3.74,9.47) {\scriptsize $j_{4}$};
\draw [->-] (3.2,10.2) -- (2.4,10.4);
\node [] at (2.74,10.07) {\scriptsize $j_{2}$};
\draw [->-] (3.2,10.2) -- (3.8,10.4);
\node [] at (3.58,10.07) {\scriptsize $j_{5}$};
\draw [->-] (3.2,9.5) -- (3.2,10.2);
\node [] at (2.91,9.85) {\scriptsize $k_{2}$};
\end{tikzpicture}\:
\end{matrix}\Biggr\rangle
}
\newcommand{\ExpTMoveThreeOne}{
\Biggl|\;\begin{matrix}
\begin{tikzpicture}[scale=0.7]
\draw [->-] (1.6,10.4) -- (1.9,10.1);
\node [] at (1.56,10.06) {\scriptsize $j_{1}$};
\draw [->-] (2.3,9.) -- (2.3,9.4);
\node [] at (2.59,9.2) {\scriptsize $j_{2}$};
\draw [->-] (2.3,9.4) -- (1.9,10.1);
\node [] at (1.85,9.61) {\scriptsize $j_{4}$};
\draw [->-] (2.7,10.1) -- (1.9,10.1);
\node [] at (2.3,10.32) {\scriptsize $j_{6}$};
\draw [->-] (3.,10.4) -- (2.7,10.1);
\node [] at (3.04,10.06) {\scriptsize $j_{3}$};
\draw [->-] (2.7,10.1) -- (2.3,9.4);
\node [] at (2.75,9.61) {\scriptsize $j_{5}$};
\end{tikzpicture}\:
\end{matrix}\Biggr\rangle
}
\newcommand{\ExpTMoveThreeOneAA}{
\frac{\mathrm{v}_{j_{5}}\mathrm{v}_{j_{4}}\mathrm{v}    _{j_{6}}}{\sqrt{D}}G^{j_{1}^* j_{3}^* j_{2}^*}_{j_{5} j_{4}^* j_{6}^*}
\Biggl|\;\begin{matrix}
\begin{tikzpicture}[scale=0.7]
\draw [->-] (1.6,10.4) -- (2.3,9.8);
\node [] at (1.78,9.9) {\scriptsize $j_{1}$};
\draw [->-] (2.3,9.) -- (2.3,9.8);
\node [] at (2.59,9.4) {\scriptsize $j_{2}$};
\draw [->-] (3.,10.4) -- (2.3,9.8);
\node [] at (2.82,9.9) {\scriptsize $j_{3}$};
\end{tikzpicture}\:
\end{matrix}\Biggr\rangle
}
\newcommand{\ExpTMoveTwoOne}{
\Biggl|\;\begin{matrix}
\begin{tikzpicture}[scale=1]
\path [lightgray, fill] (1.6,8.) -- (2.2,8.) -- (2.2,9.6) -- (2.2,10.4) -- (1.6,10.4) -- cycle;
\draw [->-] (2.2,8.8) -- (2.8,8.8);
\node [] at (2.5,8.58) {\scriptsize $a_{2}$};
\draw [->-] (2.2,9.6) -- (2.8,9.6);
\node [] at (2.5,9.38) {\scriptsize $a_{1}$};
\draw [->-] (2.2,8.) -- (2.2,8.8);
\node [] at (1.91,8.4) {\scriptsize $j_1$};
\draw [->-] (2.2,8.8) -- (2.2,9.6);
\node [] at (1.91,9.2) {\scriptsize $j_2$};
\draw [->-] (2.2,9.6) -- (2.2,10.4);
\node [] at (1.91,10.) {\scriptsize $j_3$};
\end{tikzpicture}\:
\end{matrix}\Biggr\rangle
}
\newcommand{\ExpTMoveTwoOneAA}{
\displaystyle\sum_{a_{3}}\frac{f_{a_{2} a_{1} a_{3}^*}\mathrm{u}_{a_{3}}\sqrt{D}}{\mathrm{u}_{a_{2}}\mathrm{u}_{a_{1}}\sqrt{\mathrm{d}_A}}
\Biggl|\;\begin{matrix}
\begin{tikzpicture}[scale=1]
\path [lightgray, fill] (1.6,8.) -- (2.2,8.) -- (2.2,8.8) -- (2.8,9.2) -- (2.2,9.6) -- (2.2,10.4) -- (1.6,10.4) -- cycle;
\draw [->-] (2.2,8.8) -- (2.8,9.2);
\node [] at (2.64,8.78) {\scriptsize $a_{2}$};
\draw [->-] (2.2,9.6) -- (2.8,9.2);
\node [] at (2.64,9.62) {\scriptsize $a_{1}$};
\draw [->-] (2.8,9.2) -- (3.4,9.2);
\node [] at (3.1,8.98) {\scriptsize $a_{3}$};
\draw [->-] (2.2,8.) -- (2.2,8.8);
\node [] at (1.91,8.4) {\scriptsize $j_1$};
\draw [->-] (2.2,8.8) -- (2.2,9.6);
\node [] at (1.91,9.2) {\scriptsize $j_2$};
\draw [->-] (2.2,9.6) -- (2.2,10.4);
\node [] at (1.91,10.) {\scriptsize $j_3$};
\end{tikzpicture}\:
\end{matrix}\Biggr\rangle
}
\newcommand{\ExpTMoveTwoOneAlt}{
\Biggl|\;\begin{matrix}
\begin{tikzpicture}[scale=1]
\path [lightgray, fill] (1.6,8.8) -- (2.2,8.8) -- (2.2,9.6) -- (2.2,10.4) -- (1.6,10.4) -- cycle;
\draw [->-] (2.2,8.8) -- (2.2,9.6);
\draw [->-] (2.2,9.6) -- (2.2,10.4);
\draw [->-] (2.2,9.6) -- (2.8,9.6);
\node [] at (2.5,9.38) {\scriptsize $a_{1}$};
\draw [->-] (2.8,9.6) -- (3.,10.2);
\node [] at (3.17,9.81) {\scriptsize $a_{3}$};
\draw [->-] (2.8,9.6) -- (3.,9.);
\node [] at (3.17,9.39) {\scriptsize $a_{2}$};
\end{tikzpicture}\:
\end{matrix}\Biggr\rangle
}
\newcommand{\ExpTMoveTwoOneAltAA}{
\frac{f_{a_{2} a_{3} a_{1}^*}\mathrm{u}_{a_{2}}\mathrm{u}_{a_{3}}}{\mathrm{u}_{a_{1}}\sqrt{\mathrm{d}_A}}
\Biggl|\;\begin{matrix}
\begin{tikzpicture}[scale=1]
\path [lightgray, fill] (1.6,8.8) -- (2.2,8.8) -- (2.2,9.6) -- (2.2,10.4) -- (1.6,10.4) -- cycle;
\draw [->-] (2.2,8.8) -- (2.2,9.6);
\draw [->-] (2.2,9.6) -- (2.2,10.4);
\draw [->-] (2.2,9.6) -- (2.8,9.6);
\node [] at (2.5,9.38) {\scriptsize $a_{1}$};
\end{tikzpicture}\:
\end{matrix}\Biggr\rangle
}
\newcommand{\ExpTMoveTwoTwo}{
\Biggl|\;\begin{matrix}
\begin{tikzpicture}[scale=0.7]
\draw [->-] (1.6,8.8) -- (2.,9.6);
\node [] at (2.06,9.07) {\scriptsize $j_{2}$};
\draw [->-] (1.6,10.4) -- (2.,9.6);
\node [] at (1.54,9.87) {\scriptsize $j_{1}$};
\draw [->-] (3.2,8.8) -- (2.8,9.6);
\node [] at (3.26,9.33) {\scriptsize $j_{3}$};
\draw [->-] (3.2,10.4) -- (2.8,9.6);
\node [] at (2.74,10.13) {\scriptsize $j_{4}$};
\draw [->-] (2.8,9.6) -- (2.,9.6);
\node [] at (2.4,9.82) {\scriptsize $j_{5}$};
\end{tikzpicture}\:
\end{matrix}\Biggr\rangle
}
\newcommand{\ExpTMoveTwoTwoAA}{
\displaystyle\sum_{j'_{5}}G^{j_{1} j_{2} j_{5}}_{j_{3} j_{4} j'_{5}}\mathrm{v}_{j_{5}}\mathrm{v}_{j'_{5}}
\Biggl|\;\begin{matrix}
\begin{tikzpicture}[scale=0.7]
\draw [->-] (1.6,8.8) -- (2.4,9.2);
\node [] at (2.11,8.77) {\scriptsize $j_{2}$};
\draw [->-] (1.6,10.4) -- (2.4,10.);
\node [] at (1.89,9.97) {\scriptsize $j_{1}$};
\draw [->-] (2.4,9.2) -- (2.4,10.);
\node [] at (2.77,9.6) {\scriptsize $j'_{5}$};
\draw [->-] (3.2,10.4) -- (2.4,10.);
\node [] at (2.91,9.97) {\scriptsize $j_{4}$};
\draw [->-] (3.2,8.8) -- (2.4,9.2);
\node [] at (2.69,8.77) {\scriptsize $j_{3}$};
\end{tikzpicture}\:
\end{matrix}\Biggr\rangle
}
\newcommand{\TMoveTwoZero}{
\begin{tikzpicture}[scale=0.7]
\draw [->-] (2.1,9.) ..controls (1.9,9.044) and (1.7,9.089) .. (1.6,9.2)
			 ..controls (1.5,9.311) and (1.5,9.489) .. (1.6,9.6)
			 ..controls (1.7,9.711) and (1.9,9.756) .. (2.1,9.8);
\draw [->-] (2.1,9.) ..controls (2.3,9.044) and (2.5,9.089) .. (2.6,9.2)
			 ..controls (2.7,9.311) and (2.7,9.489) .. (2.6,9.6)
			 ..controls (2.5,9.711) and (2.3,9.756) .. (2.1,9.8);
\draw [->-] (2.1,8.4) -- (2.1,9.);
\node [] at (2.32,8.7) {\scriptsize $k$};
\draw [->-] (2.1,9.8) -- (2.1,10.4);
\node [] at (2.39,10.1) {\scriptsize $k'$};
\node [] at (1.9,9.4) {\scriptsize $i$};
\node [] at (2.9,9.4) {\scriptsize $j$};
\end{tikzpicture}
}

\newcommand{\TMoveTwoZeroAA}{
\begin{tikzpicture}[scale=0.7]
\draw [->-] (1.6,8.4) -- (1.6,10.4);
\node [] at (1.82,9.4) {\scriptsize $k$};
\end{tikzpicture}
}

\newcommand{\ZtwoTrivalentGraph}{
\begin{tikzpicture}[scale=0.8]
\draw [] (1.6,9.6) -- (2.4,9.6);
\draw [] (2.4,9.6) -- (3.2,9.6);
\draw [] (2.4,8.8) -- (2.4,9.6);
\draw [] (4.,8.8) -- (4.,9.6);
\draw [] (4.8,9.6) -- (5.6,9.6);
\draw [] (5.6,9.6) -- (6.4,9.6);
\draw [] (5.6,8.8) -- (5.6,9.6);
\draw [] (6.4,9.6) -- (6.4,10.4);
\draw [] (6.4,9.6) -- (7.2,9.6);
\draw [] (7.2,9.6) -- (8.,9.6);
\draw [] (7.2,8.8) -- (7.2,9.6);
\draw [] (8.,9.6) -- (8.,10.4);
\draw [] (8.,9.6) -- (8.8,9.6);
\draw [] (8.8,9.6) -- (9.6,9.6);
\draw [] (8.8,8.8) -- (8.8,9.6);
\draw [] (2.4,8.8) -- (3.2,8.8);
\draw [] (3.2,8.8) -- (4.,8.8);
\draw [] (3.2,8.) -- (3.2,8.8);
\draw [] (4.,8.8) -- (4.8,8.8);
\draw [] (4.8,8.8) -- (5.6,8.8);
\draw [] (4.8,8.) -- (4.8,8.8);
\draw [] (5.6,8.8) -- (6.4,8.8);
\draw [] (6.4,8.8) -- (7.2,8.8);
\draw [] (6.4,8.) -- (6.4,8.8);
\draw [] (7.2,8.8) -- (8.,8.8);
\draw [] (8.,8.8) -- (8.8,8.8);
\draw [] (8.,8.) -- (8.,8.8);
\draw [] (8.8,8.8) -- (9.6,8.8);
\draw [] (1.6,8.) -- (2.4,8.);
\draw [] (2.4,8.) -- (3.2,8.);
\draw [] (2.4,7.2) -- (2.4,8.);
\draw [] (3.2,8.) -- (4.,8.);
\draw [] (4.,8.) -- (4.8,8.);
\draw [] (4.,7.2) -- (4.,8.);
\draw [] (4.8,8.) -- (5.6,8.);
\draw [] (7.2,8.) -- (8.,8.);
\draw [] (8.,8.) -- (8.8,8.);
\draw [] (8.8,8.) -- (9.6,8.);
\draw [] (8.8,7.2) -- (8.8,8.);
\draw [] (2.4,7.2) -- (3.2,7.2);
\draw [] (3.2,7.2) -- (4.,7.2);
\draw [] (3.2,6.4) -- (3.2,7.2);
\draw [] (4.,7.2) -- (4.8,7.2);
\draw [] (4.8,6.4) -- (4.8,7.2);
\draw [] (6.4,6.4) -- (6.4,7.2);
\draw [] (8.,7.2) -- (8.8,7.2);
\draw [] (8.,6.4) -- (8.,7.2);
\draw [] (8.8,7.2) -- (9.6,7.2);
\draw [] (1.6,6.4) -- (2.4,6.4);
\draw [] (2.4,6.4) -- (3.2,6.4);
\draw [] (2.4,5.6) -- (2.4,6.4);
\draw [] (3.2,6.4) -- (4.,6.4);
\draw [] (4.,6.4) -- (4.8,6.4);
\draw [] (4.,5.6) -- (4.,6.4);
\draw [] (4.8,6.4) -- (5.6,6.4);
\draw [] (5.6,6.4) -- (6.4,6.4);
\draw [] (5.6,5.6) -- (5.6,6.4);
\draw [] (6.4,6.4) -- (7.2,6.4);
\draw [] (7.2,6.4) -- (8.,6.4);
\draw [] (7.2,5.6) -- (7.2,6.4);
\draw [] (8.,6.4) -- (8.8,6.4);
\draw [] (8.8,6.4) -- (9.6,6.4);
\draw [] (8.8,5.6) -- (8.8,6.4);
\draw [] (1.6,8.8) -- (2.4,8.8);
\draw [] (1.6,7.2) -- (2.4,7.2);
\draw [] (7.2,7.2) -- (7.2,8.);
\node [] at (7.42,7.6) {\scriptsize ${4}$};
\draw [] (3.2,9.6) -- (3.2,10.4);
\node [] at (3.42,10.) {\scriptsize $8$};
\draw [] (3.2,9.6) -- (4.,9.6);
\node [] at (3.6,9.38) {\scriptsize $9$};
\draw [] (4.,9.6) -- (4.8,9.6);
\node [] at (4.4,9.38) {\scriptsize $10$};
\draw [] (4.8,9.6) -- (4.8,10.4);
\node [] at (5.09,10.) {\scriptsize $11$};
\draw [] (6.4,8.) -- (7.2,8.);
\node [] at (6.8,8.22) {\scriptsize $5$};
\draw [] (5.6,8.) -- (6.4,8.);
\node [] at (6.,8.22) {\scriptsize $6$};
\draw [] (7.2,7.2) -- (8.,7.2);
\draw [] (5.6,7.2) -- (6.4,7.2);
\node [] at (6.,7.42) {\scriptsize ${2}$};
\draw [] (6.4,7.2) -- (7.2,7.2);
\node [] at (6.8,7.42) {\scriptsize ${3}$};
\draw [] (5.6,7.2) -- (5.6,8.);
\node [] at (5.38,7.6) {\scriptsize ${1}$};
\draw [] (4.8,7.2) -- (5.6,7.2);
\node [] at (5.2,6.98) {\scriptsize $7$};
\node [] at (6.4,7.6) {\scriptsize $p$};
\node [] at (4.1,10.3) {\scriptsize $p'$};
\end{tikzpicture}
}

\newcommand{\ZtwoTrivalentGraphAA}{
\begin{tikzpicture}[scale=0.8]
\draw [] (2.1,8.2) ..controls (2.228,7.908) and (2.471,7.538) .. (2.9,7.4)
			 ..controls (3.329,7.262) and (3.946,7.354) .. (4.5,7.4)
			 ..controls (5.054,7.446) and (5.546,7.446) .. (6.1,7.4)
			 ..controls (6.654,7.354) and (7.269,7.262) .. (7.7,7.4)
			 ..controls (8.131,7.538) and (8.377,7.908) .. (8.5,8.2)
			 ..controls (8.623,8.492) and (8.622,8.708) .. (8.5,9.)
			 ..controls (8.378,9.292) and (8.136,9.662) .. (7.7,9.8)
			 ..controls (7.264,9.938) and (6.634,9.846) .. (6.1,9.8)
			 ..controls (5.566,9.754) and (5.127,9.754) .. (4.6,9.8)
			 ..controls (4.073,9.846) and (3.458,9.938) .. (3.,9.8)
			 ..controls (2.542,9.662) and (2.243,9.292) .. (2.1,9.)
			 ..controls (1.957,8.708) and (1.972,8.492) .. (2.1,8.2);
\draw [] (2.9,7.4) -- (2.9,6.9);
\node [] at (2.68,7.15) {\scriptsize ${1}$};
\draw [] (6.1,7.4) -- (6.1,6.8);
\node [] at (5.88,7.1) {\scriptsize ${2}$};
\draw [] (8.5,8.2) -- (9.1,8.);
\node [] at (8.73,7.89) {\scriptsize ${3}$};
\draw [] (7.7,9.8) -- (7.9,10.4);
\node [] at (8.01,10.03) {\scriptsize ${4}$};
\draw [] (4.6,9.8) -- (4.7,10.4);
\node [] at (4.87,10.06) {\scriptsize ${5}$};
\draw [] (2.1,9.) -- (1.6,9.4);
\node [] at (1.99,9.37) {\scriptsize ${6}$};
\draw [] (2.1,8.2) -- (2.7,8.5);
\node [] at (2.51,8.12) {\scriptsize ${1'}$};
\draw [] (4.5,7.4) -- (4.5,8.);
\node [] at (4.79,7.7) {\scriptsize ${2'}$};
\draw [] (7.7,7.4) -- (7.5,8.);
\node [] at (7.87,7.79) {\scriptsize ${3'}$};
\draw [] (8.5,9.) -- (7.9,8.8);
\node [] at (8.12,9.13) {\scriptsize ${4'}$};
\draw [] (6.1,9.8) -- (6.1,9.2);
\node [] at (5.81,9.5) {\scriptsize ${5'}$};
\draw [] (3.,9.8) -- (3.3,9.2);
\node [] at (2.89,9.37) {\scriptsize ${6'}$};
\node [] at (1.8,8.6) {\scriptsize $0$};
\end{tikzpicture}
}

\newcommand{\ZtwoTrivalentGraphAB}{
\begin{tikzpicture}[scale=0.8]
\draw [] (2.1,8.2) ..controls (2.228,7.908) and (2.471,7.538) .. (2.9,7.4)
			 ..controls (3.329,7.262) and (3.946,7.354) .. (4.5,7.4)
			 ..controls (5.054,7.446) and (5.546,7.446) .. (6.1,7.4)
			 ..controls (6.654,7.354) and (7.269,7.262) .. (7.7,7.4)
			 ..controls (8.131,7.538) and (8.377,7.908) .. (8.5,8.2)
			 ..controls (8.623,8.492) and (8.622,8.708) .. (8.5,9.)
			 ..controls (8.378,9.292) and (8.136,9.662) .. (7.7,9.8)
			 ..controls (7.264,9.938) and (6.634,9.846) .. (6.1,9.8)
			 ..controls (5.566,9.754) and (5.127,9.754) .. (4.6,9.8)
			 ..controls (4.073,9.846) and (3.458,9.938) .. (3.,9.8)
			 ..controls (2.542,9.662) and (2.243,9.292) .. (2.1,9.)
			 ..controls (1.957,8.708) and (1.972,8.492) .. (2.1,8.2);
\draw [] (2.9,7.4) -- (2.9,6.9);
\node [] at (2.68,7.15) {\scriptsize ${1}$};
\draw [] (6.1,7.4) -- (6.1,6.8);
\node [] at (5.88,7.1) {\scriptsize ${2}$};
\draw [] (8.5,8.2) -- (9.1,8.);
\node [] at (8.73,7.89) {\scriptsize ${3}$};
\draw [] (7.7,9.8) -- (7.9,10.4);
\node [] at (8.01,10.03) {\scriptsize ${4}$};
\draw [] (4.6,9.8) -- (4.7,10.4);
\node [] at (4.87,10.06) {\scriptsize ${5}$};
\draw [] (2.1,9.) -- (1.6,9.4);
\node [] at (1.99,9.37) {\scriptsize ${6}$};
\node [] at (1.9,8.) {\scriptsize $0$};
\end{tikzpicture}
}

\begin{document}

\title{Boundary Hamiltonian theory for gapped topological phases on an open surface} 

\date{\today}

\author{Yuting Hu}
\email{yuting.phys@gmail.com}
\affiliation{Department of Physics and Center for Field Theory and Particle Physics, Fudan University, Shanghai 200433, China}

\author{Zhu-Xi Luo}
\affiliation{Department of Physics and Astronomy, University of Utah, Salt Lake City, Utah, 84112, U.S.A.}
\author{Ren Pankovich}
\affiliation{Department of Physics and Astronomy, University of Utah, Salt Lake City, Utah, 84112, U.S.A.}

\author{Yidun Wan}
\email{ydwan@fudan.edu.cn}
\affiliation{Department of Physics and Center for Field Theory and Particle Physics, Fudan University, Shanghai 200433, China}
\affiliation{Collaborative Innovation Center of Advanced Microstructures, Nanjing 210093, China}

\author{Yong-Shi Wu}
\email{yswu@fudan.edu.cn}
\affiliation{State Key Laboratory of Surface Physics, Fudan University, Shanghai 200433, China}
\affiliation{Department of Physics and Center for Field Theory and Particle Physics, Fudan University, Shanghai 200433, China}
\affiliation{Collaborative Innovation Center of Advanced Microstructures, Nanjing 210093, China}
\affiliation{Department of Physics and Astronomy, University of Utah, Salt Lake City, Utah, 84112, U.S.A.}

\begin{abstract}
In this paper we propose a Hamiltonian approach to gapped topological phases on an open surface with boundary. Our setting is an extension of the Levin-Wen model to a 2d graph on the open surface, whose boundary is part of the graph. We systematically construct a series of boundary Hamiltonians such that each of them, when combined with the usual Levin-Wen bulk Hamiltonian, gives rise to a gapped energy spectrum which is topologically protected; and the corresponding wave functions are robust under changes of the underlying graph that maintain the spatial topology of the system. We derive explicit ground-state wavefunctions of the system and show that the boundary types are classified by Morita-equivalent Frobenius algebras. We also construct boundary quasiparticle creation, measuring and hopping operators. These operators allow us to characterize the boundary quasiparticles by bimodules of Frobenius algebras. Our approach also offers a concrete set of tools for computations. We illustrate our approach by a few examples.

\end{abstract} 


\maketitle

\section{Introduction}

Two important characteristic properties of a 2d matter phase with an intrinsic topological order\cite{Wen1990a}, crucial for topological quantum computation, are a finite set of topologically protected, degenerate ground states\cite{TaoWu1984,Niu1985,Wen1990c} and the corresponding anyon excitations obeying braid statistics\cite{Wu1984}. While the former furnishes robust quantum memories\cite{Dennis2002}, the latter delineates a logical Hilbert space that supports topological quantum computation via the unitary braiding of the anyons\cite{Kitaev2003a,Freedman2003,Stern2006,Nayak2008}. It is also hopeful to realize or simulate Abelian\cite{Luo2016,Luo2016a} and non-Abelian anyons\cite{Lesanovsky2012,Li2017}. On closed spatial 2-surfaces, its genus number and the fusion rules between anyon excitations determine the ground state degeneracy (GSD)\cite{Wen1990,Wen1990c,Wen1991,Wen1990a, Nayak2008}. And in particular, on a torus, the GSD equals the number of anyon species. Nevertheless, realizing closed-surface material with topological order is difficult in experiments; it is much more natural to make finite open systems. Yet, it is necessary that any boundary massless modes that often appear can be gapped to have a well defined GSD. The gapping conditions of Abelian topological orders have recently been understood in terms of the concept of Lagrangian subsets\cite{Kitaev2012,Levin2013,Kong2013,Barkeshli,Barkeshli2013c,Lan2014b,Wang2012,HungWan2013a}, and subsequently the GSD of these Abelian phases on open surfaces with multiple boundaries were computed\cite{Wang2012,Iadecola2014}, based on the idea of anyon transport across boundaries. Experiments detecting and applying the topological degeneracy with gapped boundaries were proposed in \cite{Barkeshli2014,Barkeshli2014a}. The gapping conditions of non-Abelian topological orders have recently been tackled by the mechanism of anyon condensation\cite{HungWan2014} and equivalently by solving certain algebraic equations\cite{Lan2014}. Gapped boundaries of topological orders can also be classified by Frobenius algebras\cite{HungWan2015a}, using the mechanism of anyon condensation. 

Nevertheless, unfortunately, the rich studies and classifications of gapped boundaries of topological orders are not practical enough because they are based on abstract mathematical theories rather than explicit Hamiltonian models consisting of both bulk and boundary terms. This causes consequent issues. For example, Ref.\cite{HungWan2014} offers a closed-form formula of computing the GSD of a topological order on a $n$-hole surface, in terms of the condensed anyons at the holes and their fusion rules. This formula, though mathematically complete and beautiful, cannot tell us how to realize the boundary conditions on a given Hamiltonian model of the topological order. Although it is known that a gapped boundary of a topological order corresponds to certain anyon condensation, it is not clear how bulk anyons interact with the boundary excitations. Hence, the proposal of using the generalized Laughlin-Tao-Wu charge-pumping argument\cite{Laughlin1983,TaoWu1984} to braid anyons and realize topological quantum computation in Ref.\cite{HungWan2014} would be impossible unless there is a concrete Hamiltonian model in which pumping a boundary excitation from one boundary to another can be studied in terms of physical operators.

More broadly, a dynamical theory in which one is not able to specify boundary conditions is not a complete dynamical theory. Therefore, Hamiltonian models of topological orders such as the Levin-Wen model\cite{Levin2004}, the Kitaev Model\cite{Kitaev2003a}, and the twisted quantum double model\cite{Hu2012a} as a generalization of the Kitaev model are not complete dynamical models on open surfaces because they do not include boundary terms. 
\begin{figure}
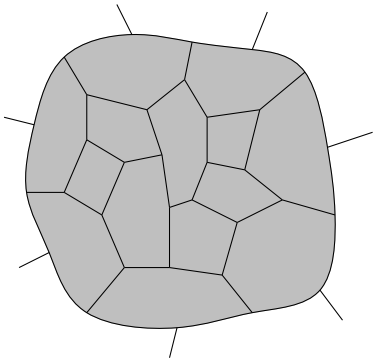

\centering
\FigDiskGraph
\caption{Trivalent graph with boundary}\label{fig:TriGraphBdry}
\end{figure}

In this paper, we develop a new approach -- in the framework of discrete models on a 2d graph with boundary -- to gapped topological phases on open surfaces. Our approach is based on three physical ansatze as follows.
\begin{enumerate}
\item \textit{Locality}: The boundary Hamiltonian is a local one.
\item \textit{Asympototics}: The boundary (interacting) theory is the asymptotics of the bulk (interacting) theory; hence, the boundary degrees of freedom would come from the bulk degrees of freedom.
\item \textit{Topological invariance}: The ground-state Hilbert space is invariant under topology-preserving mutations of the graph. This leads to Frobenius algebras charactering boundary interactions, as  emergent structures of the bulk degrees of freedom.  
\end{enumerate}
The Locality ansatz is rather natural. The Asymptotics ansatz and the emergence of Frobenius algebras may deserve more justification, as we now elaborate.

It is well-known that for a continuum Chern-Simons gauge theory on a bounded spatial region in a plane, besides the usual Chern-Simons bulk term, the action contains an additional term on the boundary \cite{elitzur1989}, to ensure the gauge invariance of the total action. The physical interpretation for adding a boundary action term is holography, i.e. a holographic correspondence between the bulk and the boundary, respected in a topological phase in two  dimensions. Motivated by the desired holography in the discrete framework, the central idea of ours is to construct explicitly and systematically an appropriate boundary Hamiltonian, to be added to the Levin-Wen bulk Hamiltonian that was originally designed for a closed surface. Thus, our new approach features the new Hamiltonian below, defined on a trivalent graph as in Fig. \ref{fig:TriGraphBdry} : 
\begin{equation}
H=H_{\mathrm{LW}}+H_{\mathrm{bdry}},
\end{equation}
where $H_{\mathrm{LW}}$ is the usual Levin-Wen model Hamiltonian in the bulk of the graph, while $H_{\mathrm{bdry}}$ is our boundary Hamiltonian defined along the boundary of the graph. As in the original Levin-Wen model, the bulk degrees of freedom, i.e., the string types labeled on the bulk edges, are objects of a unitary fusion category (UFC), such as the representations of a finite group or a quantum group. The boundary degrees of freedom in our new approach however are elements of a Frobenius algebra that is a composite object in the bulk UFC. This consideration of Frobenius algebra is again motivated by the demand of exact solubility and by holography. First, when we restrict the boundary degrees of freedom to live in a subset of those in bulk described by the input UFC, and require the commutativity between the boundary operators and the bulk operators, we find that the boundary degrees of freedom would have to form a Frobenius algebra---an algebra with certain associativity. Second, restricted to bulk ground states only, the boundary theory of a topological order may be thought of as a $(1+1)$-d topological quantum field theory (TQFT) of the bulk $(2+1)$-d TQFT. It has been shown that $(1+1)$-d TQFTs are in one-to-one correspondence with commutative Frobenius algebras\cite{Kock2004}. Third, in the case with finite groups, the Levin-Wen model is dual to the Kitaev quantum double model. Shor \textit{et al} have shown that the boundary degrees of freedom in the Kitaev model defined by a finite group $G$ live in a subgroup of $G$\cite{Beigi2011}, which in the dual Levin-Wen model corresponds to a Frobenius algebra in the UFC of the representations of $G$. 

The main result of this paper is an explicit and systematic construction of the boundary Hamiltonian $H_{bdry}$, using the data of a Frobenius algebra in the UFC that defines the bulk Levin-Wen model. It turns out that the new Hamiltonian model for  two dimensional systems with boundary maintains exact solubility and yields a gapped spectrum that is protected by the spatial topology, as in the original Levin-Wen model. In this way, our model offers a way of computing the ground state degeneracy of a topological phase on an open surface in terms of the input data, which is more fundamental than the output data, i.e. anyon species, which are consequences of the dynamics of the system. We will use several examples to illustrate how our approach works and to check the agreement of the concrete results with the existing approaches (if available).      
 
Compared to the existing approaches to gapped topological phases on open surfaces, our boundary Hamiltonian approach has the following advantages. First, the usual classification of topological boundaries by Lagrangian subsets and/or by anyon condensation, as mentioned above, is based on the data of the anyon species of excitations in topological orders, which is the output of the dynamics of the system. In contrast, our classification is characterized by the input data of the dynamical model. This is more in line with the spirit of the usual Hamiltonian dynamics. Secondly, for a given Levin-Wen bulk Hamiltonian, the boundary terms of our model may not be unique, and the gapped energy spectrum of the whole bounded system depends on the choice of the boundary term accordingly. Hence, it is obvious that the boundary Hamiltonians can be used to characterize and classify the boundary conditions that give rise to gapped topological boundaries. Moreover, by solving the total Hamiltonian (bulk plus boundary terms) we can obtain the explicit wave functions of the ground and excited states, all in the form of tensor network states. This will provide us a very detailed, dynamic understanding of the stationary topological states of the whole bounded system, especially of what happens on and near the boundary. For example, our model enables us to study the boundary excitations explicitly. Also anyon condensation may be understood at more microscopic scales. These studies will be reported later separately. 

Two of us also report in a companion paper\cite{Bullivant2017} a similar approach of  constructing the boundary Hamiltonian of the twisted quantum double model. It is worth noting that there have been a few studies of the boundary Hamiltonians in the Kitaev model\cite{Beigi2011,Cong2016a,Cong2017a}, as well as in the Levin-Wen model\cite{Kitaev2012} in the language of module categories. While our approach is systematic and easier to access by the condensed matter community, we shall discuss in Section \ref{sec:KK} the relation between our approach and the one taken by Kitaev and Kong\cite{Kitaev2012}. 

This paper is a much expanded version, with many details and more results, of a paper of three of us\cite{Hu2017}, in which the main ideas and some results were briefly reported.      




\section{Review of the Levin-Wen model}\label{S:LWModels}

Let us briefly review the Levin-Wen model. The model is a lattice Hamiltonian model that is defined by a set of input data---a unitary fusion category (UFC) $\mathcal{C}$---that specifies the Hilbert space of the model. In this work, we will use the tensor description of $\mathcal{C}$ in terms of $6j$-symbols.

\begin{figure}[ht!]
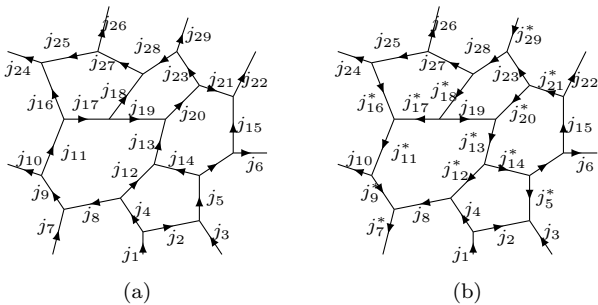

  \centering
  \subfigure[]{\FigTrivalentGraphAA}
  \label{fig:TrivalentGraphA}\quad
  \subfigure[]{\FigTrivalentGraphAB}
  \label{fig:TrivalentGraphB}
  \caption{A configuration of string types on a directed trivalent graph. The configuration (b) and hence the associated Hilbert space is regarded the same as (a), with some of the directions of some edges reversed and the corresponding labels $j$ conjugated $j^*$.}
  \label{fig:TrivalentGraph}
 \end{figure}

The model is defined on a trivalent graph embedded in a closed, oriented surface. The Hilbert space is spanned by the degrees of freedom on edges (See Fig. \ref{fig:TrivalentGraph}), which are the objects (called string types) in $\mathcal{C}$ and are labeled by $j$ that runs over a finite set of integers $L=\{j=0,1,...,N\}$. The Hilbert space is spanned by all configurations of the labels on edges. Each label $j$ has a ``conjugate'' $j^*$, which is also an integer and satisfies $j^{**}=j$. If we reverse the direction of one edge labeled by $j$ and replace the label by $j^*$, we require the state remains the same. See Fig. \ref{fig:TrivalentGraph}. There is a unique ``trivial'' label $j=0$ with $0^*=0$.
 
As objects in a UFC $\mathcal{C}$, the string types are subject to fusion rules. A \textit{fusion rule} on $L$ is a function $N:L \times L \times L\rightarrow \ds{N}$ such that for $a,b,c,d\in L$,
\begin{align}
&N_{0a}^b=N_{a0}^b=\delta_{ab},\label{eq:N:a0b}\\
&N_{ab}^0=\delta_{ab^*},\label{eq:N:ab0}\\
&\sum_{x\in L}N_{ab}^xN_{xc}^d=\sum_{x\in L}N_{ax}^dN_{cd}^x.\label{eq:NN=NN}
\end{align}
A fusion rule is multiplicity-free if $N_{ab}^c\in \{0,1\}$ for all $a,b,c\in L$.  We restrict to the multiplicity-free case throughout this paper unless otherwise stated. We define $\delta_{abc}:=N_{ab}^{c^*}$, with the symmetric properties:  $\delta_{abc}=\delta_{bca}$ and $\delta_{abc}=\delta_{c^*b^*a^*}$. A triple $(a,b,c)$ is admissible if $\delta_{abc}=1$.

Given a fusion rule on $L$, a \textit{quantum dimension} is a map $\rmd:L\rightarrow \ds{R}$ such that $\rmd_{a^*}=\rmd_{a}$ and
\begin{equation}
\sum_{c\in L}\rmd_c\delta_{abc^*}=\rmd_a\rmd_b.
\label{eq:dimcond}
\end{equation}
In particular, $\rmd_0=1$. Let $\alpha_j=\mathrm{sgn}(\rmd_j)$, which take values of $\pm 1$ for each label $j$ and satisfy
\begin{equation}
\alpha_i\alpha_j\alpha_k=1, \quad \text{if }\delta_{ijk}=1.
\end{equation}

Given fusion rules and quantum dimensions on $L$, we may define $6j$-symbols, often denoted as $G$. A \textit{tetrahedral symmetric unitary} $6j$-symbol is a map $G:L^6\rightarrow \ds{C}$ satisfying these conditions:
\begin{align}
\label{eq:6jcond}
\begin{array}{ll}
&G^{ijm}_{kln}=G^{mij}_{nk^{*}l^{*}}
=G^{klm^{*}}_{ijn^{*}}=\alpha_m\alpha_n\,\overline{G^{j^*i^*m^*}_{l^*k^*n}},\\
&\sum_{n}{\rmd_{n}}G^{mlq}_{kp^{*}n}G^{jip}_{mns^{*}}G^{js^{*}n}_{lkr^{*}}
=G^{jip}_{q^{*}kr^{*}}G^{riq^{*}}_{mls^{*}},\\
&\sum_{n}{\rmd_{n}}G^{mlq}_{kp^{*}n}G^{l^{*}m^{*}i^{*}}_{pk^{*}n}
=\frac{\delta_{iq}}{\rmd_{i}}\delta_{mlq}\delta_{k^{*}ip},\\
\end{array}
\end{align}
where the second equation above is the \textit{pentagon identity}.

The input data of the Levin-Wen model is such a set: $\{\rmd_j,\delta_{ijk},G_{klm}^{ijm}\}$ that can be derived from the representation theory of a group or a quantum group, and more generally, such a set of data is from a UFC. For instance, we may take the labels $j$ to be the irreducible representations of a finite group $H$. The trivial label $0$ is the trivial representation. The fusion rules indicate whether the tensor product $j_1\otimes{j_2}\otimes{j_3}$ contains the trivial representation or not. Each number $\alpha_j$ is the Frobenius-Schur indicator telling if the representation $j$ is real, complex, or pseudoreal. The relation $\rmd_j=\alpha_j\mathrm{dim}(j)$ holds, where $\mathrm{dim}(j)$ is the dimension of the corresponding representation space. The $6j$-symbols $G_{kln}^{ijm}$ are identified with the (symmetrized) Racah $6j$-symbols of the group $H$. In this example, the Levin-Wen model can be mapped to the Kitaev quantum double model. 

One important property of the $6j$-symbols is that
\begin{equation}
\label{eq:G=Gdelta}
G^{ijm}_{kln}=G^{ijm}_{kln}\delta_{ijm}\delta_{klm^*}\delta_{lin}\delta_{nk^*j^*}.
\end{equation}
To prove this, one can rewrite the orthogonality condition as
\begin{equation}
\sum_{n}\left({\rmv_{n}\rmv_{q}}G^{mlq}_{kp^{*}n}\right)\overline{\left({\rmv_{n}\rmv_{i}}G^{mlq}_{kp^{*}n}\right)}
=\delta_{iq}\delta_{mli}\delta_{k^{*}ip},
\end{equation}
where $\bar{c}$ stands for the complex conjugate of a complex number $c$. When $q=i$, the above equality implies that $G^{mli}_{kp^*n}$ must vanish unless $\delta_{mlq}\delta_{k^{*}ip}=1$. By the tetrahedral symmetry, one arrives at Eq. \eqref{eq:G=Gdelta}, where $\rmv_j=\sqrt{\rmd_j}$ is a choice of a square root of the quantum dimension. The number $\rmv_j$ is either real or pure imaginary, depending on the $\alpha_j=\mathrm{sgn}(\rmd_j)$, and is determined up to a sign that can be fixed as follows. From the conditions \eqref{eq:6jcond}, we have
$(G^{ijk}_{0kj}\rmv_j\rmv_k)^2=\delta_{ijk}$, and it is possible to fix the sign of $\rmv_j$ such that $G^{ijk}_{0kj}\rmv_j\rmv_k=\delta_{ijk}$. We define
\begin{equation}
\label{eq:v::Definition}
\rmv_j:=\frac{1}{G^{j^*j0}_{0\,0\,j}}.
\end{equation}
In particular, $\rmv_0=1$ because $\rmd_0=1$ (from Eq. \eqref{eq:dimcond}), and thus $G^{000}_{000}=1$ from Eq. \eqref{eq:6jcond}. Indeed, we can verify $\rmv_j^2=\rmd_j$ directly from the orthogonality condition in Eq. \eqref{eq:6jcond} together with $\rmd_0=1$. The definition in Eq. \eqref{eq:v::Definition} also implies that
\begin{equation}
\label{eq:Gvv=delta::vConvention}
G^{ijk}_{0kj}\rmv_j\rmv_k=\delta_{ijk},
\end{equation}
which is due to the pentagon identity $\rmd_0G^{ijk}_{0kj}G^{j^*j0}_{0\,0\,j}G^{k^*k0}_{0\,0\,k}=G^{ijk}_{0kj}G^{j^*i^*k^*}_{k^*0j}$ and the orthogonality $\rmd_jG^{ijk}_{0kj}G^{j^*i^*k^*}_{k^*0j}=\frac{1}{\rmd_k}\delta_{ijk}$.

There are two types of local operators, $Q_v$ defined at vertices $v$ and $B_p^s$ (indexed by the string types $s=0,1,...,N$) on plaquettes $p$. On a trivalent graph, a ${Q}_v$ acts on the labels $j_1,j_2$, and $j_3$ on three edges incident at the vertex $v$, such that
\begin{equation}
  {Q}_v \ExpQonYdiagram
  =\delta_{j_1j_2j_3}
  \ExpQonYdiagram,
\end{equation}
where the tensor $\delta_{j_1j_2j_3}$ determines whether the triple $\{j_1,j_2,j_3\}$ is admissible or not at $v$. Since $\delta_{j_1j_2j_3}$ is
invariant under permutations of its three indices, the ordering in the
triple $\{j_1,j_2,j_3\}$ is irrelevant.

An operator $B_p^s$ acts on the boundary edges of the plaquette $p$, and has the following matrix elements on a triangular plaquette.
\begin{align}
\label{eq:Bps::InLW}
&\bbra{\BpsActionAA} B_p^s \bket{\BpsActionAB}\nonumber\\
=&
\rmv_{j_1}\rmv_{j_2}\rmv_{j_3}\rmv_{j'_1}\rmv_{j'_2}\rmv_{j'_3}
G^{j_5j^*_1j_3}_{sj'_3j^{\prime*}_1}G^{j_4j^*_2j_1}_{sj'_1j^{\prime*}_2}G^{j_6j^*_3j_2}_{sj'_2j^{\prime*}_3}.
\end{align}
The action of $B_p^s$ on a quadrangle, a pentagon,
or a hexagon, etc, is similar. Note that the matrix of $B_p^s$ is non-diagonal only on the labels of the boundary edges (such as $j_1$, $j_2$, and $j_3$ in the above graph). The operators $B_p^s$ have the properties
\begin{align}
  &B_p^{s\dagger}=B_p^{s^*},
  \label{eq:BpsDagger}\\
  &B_p^rB_{p}^s=\sum_{t}\delta_{rst^*}B_p^t,
  \label{eq:BpsAlgebra}
\end{align}
which can be verified by using conditions \eqref{eq:6jcond}.

The operators defined above comprise the Hamiltonian of the model: 
\begin{equation}
  \label{eq:HamiltonianLW}
  {H}=-\sum_{v}{Q}_v-\sum_{p}B_p,
  \quad
  B_p=\frac{1}{D}\sum_{s}\rmd_sB_p^{s}
\end{equation}
where the sum run over vertices $v$ and plaquettes $p$ of the trivalent graph, and $D=\sum_j{\rmd}_j^2$ is the total quantum dimension.

It turns out that all ${Q}_v$ and $B_p$ involved are mutually-commuting projectors. Namely, (1) $[Q_v,Q_{v^{\prime}}]=0=[B_p,B_{p^{\prime}}]$, $[Q_v,B_p]=0$; (2) ${Q}_v{Q}_{v}={Q}_v$ and $B_pB_{p}=B_p$. Thus the Hamiltonian is exactly soluble. The elementary energy eigenstates are given by common eigenvectors of all these projectors. The ground states satisfies the constraints ${Q}_v=B_p=1$ for all $v$ and $p$, while the excited states violate these constraints for certain plaquettes and/or vertices.

In cases where the input data $\{\rmd,\delta,G\}$ arises from the representations of groups or quantum groups, we have $\delta_{rst^*}=\delta_{srt^*}$. Then the operators $B_p^s$ also meet the following commutation relation,
\begin{equation}
  [B_{p_1}^r,B_{p_2}^s]=0,
\end{equation}
which can be verified by the conditions \eqref{eq:6jcond}, with $p_1$ and $p_2$
two neighboring plaquettes, and by Eq. \eqref{eq:BpsAlgebra}, together with $\delta_{rst^*}=\delta_{srt^*}$ when $p_1=p_2$.

\subsection{topological feature}\label{sec:topofeature}

We briefly review the bulk topological features of the ground state. Any two given trivalent graphs
$\Gamma^{(1)}$ and $\Gamma^{(2)}$ can be mutated into each other by a composition of three elementary moves, call Pachner moves. There unitary linear maps \cite{Hu2012} associated to each move:
\begin{align}
\label{T1T2T3}
&T_{2\rightarrow 2}\ExpTMoveTwoTwo
=\ExpTMoveTwoTwoAA\nonumber\\
&T_{1\rightarrow 3}\ExpTMoveOneThree
=\ExpTMoveOneThreeAA\nonumber\\
&T_{3\rightarrow 1}\ExpTMoveThreeOne
=\ExpTMoveThreeOneAA
\end{align}
This provides a linear transformation $\mathcal{H}^{(1)}\rightarrow\mathcal{H}^{(2)}$ between the two correspondence Hilbert spaces. Instead of $T_{3\rightarrow 1}$, one may use another move: squeeze a ``bubble'',
\begin{equation}\label{eq:TMoveAlternativeSqueezeBubble}
T'_{2\rightarrow 0}\bket{\,\TMoveTwoZero}=\delta_{kk'}\frac{\rmv_i\rmv_j}{\rmv_k\sqrt{D}} \bket{\,\TMoveTwoZeroAA},
\end{equation}
which is a composition of $T_{3\rightarrow 1}$ and $T_{2\rightarrow 2}$.

The bulk topological feature of the ground states is described as follows. The ground-state Hilbert space is invariant under arbitrary composition of $T_{2\rightarrow 2},T_{1\rightarrow 3}$, and $T_{3\rightarrow 1}$. Moreover, given two trivalent graphs there are multiple ways to compose $T_{2\rightarrow 2},T_{1\rightarrow 3},T_{3\rightarrow 1}$, but different ways results in a unique transformation on the ground-state Hilbert space.

\begin{figure}[!ht]
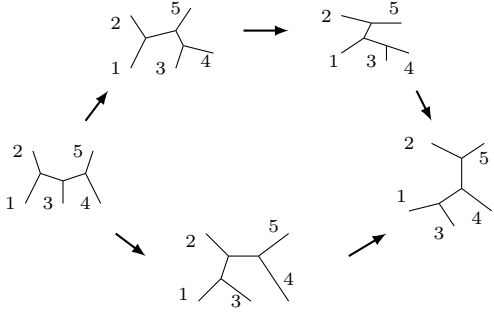

        {\centering
                \TMovePentagonDiagram}
        \caption{Two ways of composing $T_{2\rightarrow 2}$ moves to transform from the left most graph to the right most graph.}
        \label{fig:TMovePentagonDiagram}
\end{figure}

The topological feature is similar in the Hilbert space $\mathcal{H}^{Q=1}$ of simultaneous eigenvectors of $Q=1$ at all vertices. 
All states in $\mathcal{H}^{Q=1}$ are invariant under any transformation composed of $T_{2\rightarrow 2}$. For example, consider composition of $T_{2\rightarrow 2}$ illustrated in Fig. \ref{fig:TMovePentagonDiagram} involving 5 external edges. There are two ways to compose  $T_{2\rightarrow 2}$ moves to transform from the left most graph to the right most graph. Explicitly, they are
\begin{widetext}
\begin{align}\label{eq:TMovePentagonOne}
&T\left(\bmm\TMovePentagonDiagramAB\emm\right):\nonumber\\
&\ExpTMovePentagonTransf\mapsto\ExpTMovePentagonTransfAA\nonumber\\
\mapsto&\ExpTMovePentagonTransfAD
\mapsto\ExpTMovePentagonTransfAE
\end{align}
and
\begin{align}\label{eq:TMovePentagonTwo}
&T\left(\bmm\TMovePentagonDiagramAA\emm\right):\nonumber\\
&\ExpTMovePentagonTransf\mapsto\ExpTMovePentagonTransfAA\nonumber\\
&\mapsto\ExpTMovePentagonTransfAB\nonumber\\
&\mapsto\ExpTMovePentagonTransfAC
\end{align}
\end{widetext}

By the pentagon identity \eqref{eq:6jcond}, Eq. \eqref{eq:TMovePentagonOne} is identified with Eq. \eqref{eq:TMovePentagonTwo}. We can simply write both transformations as $T$. In fact, any transformation composed by $T_{2\rightarrow 2}$ moves only depends on the initial and final graphs (with the same topology) and can be written as $T$, without specifying the choice of the sequence of $T_{2\rightarrow 2}$ moves.

\begin{figure}[!ht]
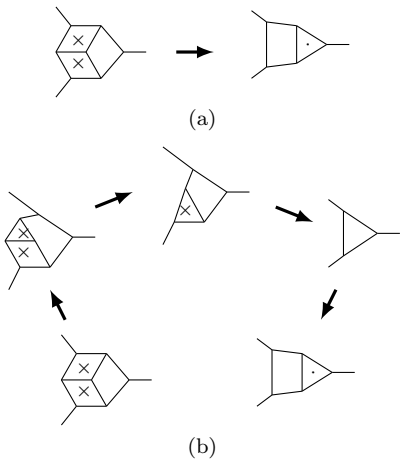

        \centering
        \subfigure[]{\FigTmoveInOutAA}
        \subfigure[]{\FigTmoveInOutAB}
        \caption{(a) Initial and final graphs with marked plaquettes. (b) A possible sequence of Pachner moves taking the initial graph to the final.}
        \label{fig:TmoveInOut}
\end{figure}

Transformations involving $T_{2\rightarrow 2}$, $T_{1\rightarrow 3}$ and $T_{3\rightarrow 1}$ moves from one given graph to another are generally not unique; however, there are subsets of all possible sequences of $T_{1\rightarrow 3}$ and $T_{3\rightarrow 1}$ moves, each of which leads to a unique transformation. Such a subset is obtained by specifying (using a \textquoteleft$\times$') in the initial graph the plaquettes to be killed by $T_{3\rightarrow 1}$ moves, and in the final graph the plaquettes (using a \textquoteleft$\cdot $') to be created by $T_{1\rightarrow 3}$ moves. See Fig. \ref{fig:TmoveInOut} for an example. In Fig. \ref{fig:TmoveInOut}(a), the initial graph has two plaquettes marked by $\times$, while the final graph has one plaquette marked by $\cdot$. In this case, all possible sequences of $\Tmv{2}{2},\Tmv{1}{3}$, and $\Tmv{3}{1}$ moves result in the same transformation between the Hilbert spaces associated with the initial and final graphs. one such sequence is shown in Fig. \ref{fig:TmoveInOut}(b). 

In the rest of the paper, we will follow the above convention. That is, we suppress the choice of sequence of Pachner moves behind any transformation, and denote the transformation by $\mathcal{T}$ hereafter. In this convention, the $B_p$ operator is compactly expressed by
\begin{equation}\label{eq:BpGraphicalAA}
\Tact\ExpBpGraphicalAA.
\end{equation}

\section{Boundary Hamiltonian}
In this section, we construct the boundary Hamiltonian explicitly. To understand our systematic construction, we will first introduce certain necessary mathematical structures.

\subsection{Frobenius algebra as input data}

The topological feature in the bulk is a consequence of the conditions \eqref{eq:6jcond} on the $6j$-symbols. We expect the gapped boundaries to have similar topological feature; hence, it is natural to look for ingredients in the input data that may play a role similar to that of the $6j$-symbols in the bulk.

We shall need a mathematical structure---Frobenius algebra objects in a UFC to construct the boundary terms to extend the LW Hamiltonian. Let $G$ be a symmetric $6j$-symbol over the label set $L$.  A \textit{Frobenius algebra} is a subset $L_A \subset L$ equipped with a multiplication $f_{ijm}$ satisfying
\begin{align}
\label{eq:fcond}
&\text{(association) }  \sum_{c}f_{abc^*}f_{cde^*}G^{abc^*}_{de^*g}\rmv_c\rmv_g,
=f_{age^*}f_{bdg^*},\nonumber\\
&\text{(non-degeneracy)}\quad   f_{bb^*0}\neq 0, \quad\forall b\in L_A,
\end{align}

We can normalize (as described in appendix \ref{app:FrobeniusAlgebra}) the non-degeneracy condition to
\begin{equation}\label{nondegeneracy}
f_{bb^*0}=1,\quad \forall b\in L_A.
\end{equation}
Due to the symmetry conditions \eqref{eq:6jcond} of symmetric $6j$-symbols\cite{Hu2012}, the multiplication meets the following defining properties.
\begin{align}\label{eq:fsymmetry}
&\text{(unit)}\quad f_{bb^*0}=f_{b0b^*}=f_{0bb^*}=1,\nonumber\\
&\text{(cyclic)}\quad f_{abc}=f_{cab},\nonumber\\
&\text{(strong)}\quad \sum_{ab}f_{abc}f_{c^*b^*a^*}\rmv_a\rmv_b=\rmd_A\rmv_c,
\end{align}
where $\rmd_A=\sum_{a\in L_A}\rmd_a$ is the quantum dimension of $A$.

If we set $g=0,d=b^*,e=a$ in Eq. \eqref{eq:fcond}, we get
\begin{equation}\label{eq:FrobeniusDeltaBRaw}
\sum_{c}f_{abc^*}f_{cb^*a^*}\rmv_c=\rmv_a\rmv_b
\end{equation}
Define
\begin{equation}\label{eq:DefineDelta}
\Delta_{abc}=f_{abc}f_{c^*b^*a^*}
\end{equation}
Then the above equation is expressed by
\begin{equation}\label{eq:Deltacond}
\sum_{c}\Delta_{abc}\rmv_c=\rmv_a\rmv_b,
\end{equation}
which, when compared  with Eq. \eqref{eq:dimcond}, implies that symbols $\Delta_{abc}$ play a role similar to that of the fusion coefficients $\delta_{ijk}$.
The definition above can be illustrated graphically by the Pachner moves. The association condition \eqref{eq:fcond} graphically reads
\begin{align}\label{eq:AssociationGraphical}
&\Tact\ExpFrobeniusAlgebraAA\nonumber\\
=&\ExpFrobeniusAlgebraAB\nonumber\\
=&\ExpFrobeniusAlgebraAC,
\end{align}
where in the initial and final states, each vertex is associated with a multiplication $f$.

The strong condition \eqref{eq:fsymmetry}
may also be understood graphically as
\begin{equation}\label{eq:StrongFrobeniusAA}
\Tact\ExpStrongFrobeniusAA
=\ExpStrongFrobeniusAB,
\end{equation}
or, equivalently, as
\begin{align}\label{eq:EquivalentStrongFrobeniusAA}
&\Tact\ExpEquivalentStrongFrobeniusAA
\nonumber\\
=& \ExpEquivalentStrongFrobeniusAB\nonumber\\
=& \ExpEquivalentStrongFrobeniusAC.
\end{align}

To save writing, we can suppress the coefficients $f$ and related summation. For example, let us express the association and strong conditions compactly:
\begin{equation}\label{eq:FrobeniusAlgebraCompact}
\Tact\ExpFrobeniusAlgebraCompactAA\;=\;
\ExpFrobeniusAlgebraCompactAB
\end{equation}
\begin{equation}\label{eq:StrongConditionCompact}
\Tact{\frac{\sqrt{D}}{\rmd_A}\ExpStrongFrobeniusCompactAA}
\;=\;
\ExpStrongFrobeniusAB
\end{equation}
\begin{align}\label{eq:EquivalentStrongFrobeniusCompactAA}
\Tact{\frac{\sqrt{D}}{\rmd_A}\ExpEquivalentStrongFrobeniusCompactAA}
\;=\;
\ExpEquivalentStrongFrobeniusCompactAB
\end{align}
The rule is to put a thick dot at any vertex associated with an $f$ and draw an unlabeled thick line indicating a summation. We shall call this rule the thick-line convention and follow it hereafter. It is sometimes natural and handy to refer to the multiplication $f$ of a Frobenius algebra $A$ as the Frobenius algebra without causing any confusion. 

To summarize, a Frobenius algebra $f$ determines a state (on a trivalent graph) with an $f$ at each vertex, and a factor $\sqrt{D}/{\rmd_A}$ at an internal plaquette. Such a state is invariant under Pachner moves $T_{2\rightarrow 2}$, and $T_{3\rightarrow 1}$ but not invariant under $T_{1\rightarrow 3}$ that creates plaquettes. In other words, it is invariant under moves from bigger/dense graphs to smaller/sparse graphs.

For computational convenience, we set $f_{ijk}=0$ for any $i,j,k$ in $L\backslash L_A$, so that $f$ is defined for all labels.

\subsection{Boundary Hamiltonian}
A section of a generic boundary of our model is depicted in Fig. \ref{fig:Boundary}. The boundary is a domain wall separating the bulk (in gray in the figure) and the vacuum. The bulk edges are labeled by $j_1,j_2,\dots$, which take value in $L$, the set of objects of the input UFC. The boundary degrees of freedom, also taking value in $L$, inhabit the tails (dangling edges) $a_1,a_2\dots$. In ground states, the boundary degrees of freedom are restricted to a Frobenius algebra $L_A\subseteq L$, as implemented by projection operators comprising the boundary Hamiltonian to be explained shortly. The Hilbert space of the model thus consists of all possible configurations of the bulk and boundary degrees of freedom.
\begin{figure}[h]
        \centering
        $\FigBoundary$
        \caption{Boundary is a wall carrying tails. $j$'s are bulk labels and $a$ are tail labels.}
        \label{fig:Boundary}
\end{figure}

The boundary Hamiltonian comprises two sets of operators as follows.
\begin{equation}\label{eq:bdryHamiltonian}
H_{\text{bdry}}=-\sum_n\Qb_n-\sum_p \Bb_{p}.
\end{equation}
Here, $\Qb_n$ is a boundary edge operator acting on open edge $n$, which projects the boundary degrees of freedom to $L_A\subseteq L$: 
\begin{equation}\label{eq:bdryQn}
\Qb_n\bket{\TailLabel{a_n}{j_1}{j_2}}=\delta_{a_n\in L_A}\bket{\TailLabel{a_n}{j_1}{j_2}}.
\end{equation}
And $\Bb_{p}$ is an operator comprised of operators $\Bb_p^t$:
\begin{equation}
\Bb_{p}=\dfrac{1}{\rmd_A}\sum_{t\in L_A}\rmv_t \Bb^t_{p},
\quad
\rmd_A=\sum_{t\in L_A} \rmd_t,
\end{equation}
where $\Bb_p^t$ acts on a boundary open plaquette hold between two nearest neighboring open edges:
\begin{widetext}
\begin{equation}
\Bb^t_{p}: \ExpBoundaryBbpAA
\mapsto
\sum_{a'_{1},a'_{2},j'_{2},j'_{3}}
f_{t^* {a'_{2}}^* a_{2}} f_{a_{1} t {a'_{1}}^*}\mathrm{u}_{a_{1}}
        \mathrm{u}_{a_{2}}\mathrm{u}_{a'_{1}}
        \mathrm{u}_{a'_{2}}
G^{j_{4}^* j_{3} a_{2}^*}_{t^* {a'_{2}}^* j'_{3}}
G^{j_{5} j_{2} j_{3}^*}_{t^* {j'_{3}}^* j'_{2}}
G^{t^* {j'_{2}}^* j_{2}}_{j_{1} a_{1}^* a'_{1}}
\mathrm{v}_{j_{2}}\mathrm{v}_{j_{3}}\mathrm{v}_{j'_{2}}\mathrm{v}_{j'_{3}}
\bket{\BoundaryBbpAF}.
\end{equation}
\end{widetext}

If one of the neighboring open edges $a_n$ or $a_{n+1}$ $\notin L_A$, then
\begin{equation}\label{BbpisZero}
\Bb_p^t=0.
\end{equation}

\subsection{Graphical presentation}

We reviewed unitary transformations associated with Pachner moves on 2D graphs, in terms of $6j$-symbol. These transformations quantitatively describe the topological feature in the bulk. Likewise, we can use Frobenius algebras to associate unitary transformations to Pachner moves on 1D boundary part of graphs.

Similar to transformations \eqref{T1T2T3}, we can use the Frobenius algebra $A$ to define unitary transformations associated with 1D Pachner moves on the boundaries of a graph: (with $u_a=\sqrt{v_a}$)         
\begin{align}\label{eq:TMoveBoundary}
&T_{1\rightarrow 2} \ExpTBdryMoveOneTwo
\nonumber\\
=&\ExpTBdryMoveOneTwoAC\nonumber\\
&T_{2\rightarrow 1} \ExpTBdryMoveTwoOne\nonumber\\
=&\ExpTBdryMoveTwoOneAC.
\end{align} 
where $\rmu_a=\sqrt{\rmv_a}$ (sign of square root may be arbitrarily chosen but if fixed once then for all).

Alternative to $\Tmv{1}{2}$ and $\Tmv{2}{1}$, the boundary Pachner moves can be defined as
\begin{align}\label{eq:TMoveBoundaryAlt}
&T'_{1\rightarrow 2} \ExpTMoveOneTwo
=\ExpTMoveOneTwoAA ,\nonumber\\
&T'_{2\rightarrow 1}\ExpTMoveTwoOne
=\ExpTMoveTwoOneAA,
\end{align}
or
\begin{equation}\label{eq:TMoveTwoOneAlternative}
T''_{2\rightarrow 1} \ExpTMoveTwoOneAlt
=\ExpTMoveTwoOneAltAA.
\end{equation}
The  $\Tmv{1}{2}$ and $\Tmv{2}{1}$ can be derived by composing the bulk Pachner moves and these alternative $T'$ moves.

The action of $\Bb_p$ can be expressed in terms of $T$ moves: $\Bb_p= T_{2\rightarrow 2}\circ\dots\circ T_{2\rightarrow 2}\circ T'_{2\rightarrow 1}\circ T_{1\rightarrow 2}$. 
To see this, we expanded in terms of $f$ and $G$:
\begin{widetext}
\begin{align}
&\Bb_p \ExpBoundaryBbpAA
= \Tact{\frac{\sqrt{D}\rmu_{a'_{1}}\rmu_{a'_{2}}}{\rmd_A\rmu_{a_1}\rmu_{a_2}}\ExpBoundaryBbpAG}
\nonumber\\
=&\Tact\ExpBoundaryBbpAH\nonumber\\
=&\Tact\ExpBoundaryBbpAI\nonumber\\
=&\Tact\ExpBoundaryBbpAJ\nonumber\\
=&\ExpBoundaryBbpAM.\label{eq:ExpBoundaryBbpAM}
\end{align}
\end{widetext}
Recall the thick-line convention that a thick dot stands for an $f$, an unlabeled thick line stands for a summation, and a $\times$ marks the plaquette to be killed by Pachner moves.
More compactly, we can write $\Bb_p$ as
\begin{equation}\label{eq:BbpIsTmoves}
\Bb_p \ExpBoundaryBbpAA
= \Tact{T'_{1\rightarrow 2}T'_{2\rightarrow 1}\ExpBoundaryBbpAA},
\end{equation}
where by Eq. \eqref{eq:TMoveBoundaryAlt} the move sequence $T'_{1\rightarrow 2}T'_{2\rightarrow 1}$ would generate an extra bulk plaquette, which would then be killed by the transformation $\mathcal{T}$ due to Eq. \eqref{eq:EquivalentStrongFrobeniusCompactAA}, such that the final graph would be the one in the last row of Eq. \eqref{eq:ExpBoundaryBbpAM}.
This sequence of moves and transformations would generate all the coefficients and summations in the last row of Eq. \eqref{eq:ExpBoundaryBbpAM}.
On a simple open plaquette,
\begin{equation}\label{eq:SimpleBbp}
\Bb_p\ExpBoundaryBbpSimple
=\Tmv{1}{2}\Tmv{2}{1}\ExpBoundaryBbpSimple
\end{equation}

On a generic open plaquette, the formula above should be sandwiched between a sequence of moves $\Tmv{2}{2}...$ that turns the generic open plaquette to a simple plaquette and another sequence of moves $\Tmv{2}{2}...$ that turns the simple plaquette back to the shape of the original generic plaquette. That is, we have $\Bb_p=\Tmv{2}{2}...\Tmv{1}{2}\Tmv{2}{1}\Tmv{2}{2}...$ on a generic open plaquette.

\subsection{Property of Boundary terms}

\begin{figure}[!ht]
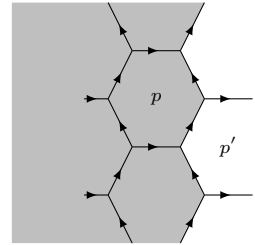

        \centering
        \FigBBcommute
        \caption{A bulk plaquette $p$ and a boundary plaquette $p'$ neighboring to each other.}
        \label{fig:BBcommute}
\end{figure}

We state that the \emph{Boundary terms $\Qb_n$ and $\Bb_p$ commute with bulk terms $Q_v$ and $B_p$.} The proof is straightforward though somewhat tedious. Here we offer only a sketch of the proof below.

\noindent Sketchy proof: We first consider the nontrivial case of $B_p$ and $\Bb_{p'}$ for neighboring plaquettes $p$ and $p'$. See Fig. \ref{fig:BBcommute}. Since both $B_p$ and $\Bb_{p'}$ can be expressed as a composition of $T$ moves, the composition of $B_p$ and $\Bb_{p'}$ does not depend on the order of the $T$ moves involved:
\begin{align}\label{eq:BBTmove}
&B_p\Bb_{p'}\nonumber\\
=\:&\Tact\FigBBcommuteAA\nonumber\\
=\:&\Bb_{p'}B_p.
\end{align}

Alternatively, We can compare $\Bb_{p'}$ with $B_p$ and use the known properties of $B_p$ to argue for the their commutativity.
Observe that the matrix element of $\Bb_p^t$ can be divided into two parts: $f$ and $\mathrm{u}$ factors acting only on open edge labels $a$'s, while $G$ and $\rmv$ factors on labels $j$'s and $a$'s. The part involving $G$ and $v$ factors is exactly the same as those of $B_p^t$ acting on $j$ labels. Loosely speaking, $\Bb_p^t$ is a combination of half of $B_p^t$ operator acting on half plaquette involving bulk labels, and the action on the tails determined by $f$.
Since $B_p^s$ is commuting with $B_{p'}^s$ for any neighboring plaquette $p,p'$, $\Bb_p^t$ should commute with all $B_p^s$.

Two boundary plaquette operators $\Bb_p$ and $\Bb_{p'}$ on two different boundary plaquettes $p$ and $p'$ also commute, i.e., $\Bb_p\Bb_{p'}=\Bb_{p'}\Bb_p$. This commutativity obviously holds if $p$ and $p'$ are far away, since $\Bb_p$ is defined locally.
If $p$ and $p'$ are neighboring to each other, we have
\begin{align}\label{eq:BBCommuteAB}
&\Bb_p\Bb_{p'}\nonumber\\
=\:&\Tact\FigBBcommuteAB\nonumber\\
=\:&\Tact\FigBBcommuteAC\nonumber\\
=\:&\Bb_{p'}\Bb_p
\end{align}
where the second equality follows from the association condition.

We also state that \emph{boundary plaquette operators $\Bb_{p}$ are mutual commuting projections: }
\begin{equation}\label{eq:BpBpisBp}
\Bb_{p}\Bb_{p}=\Bb_{p,}
\end{equation}
which is verified by directly computation:
\begin{align}
&\Bb_p\Bb_p\ExpBpBpeqBp
=\Tact{\frac{D}{\rmd_A^2}\displaystyle\sum_{a'_{1},a'_{2}}\frac{\rmu_{a'_1}\rmu_{a'_2}}{\rmu_{a_1}\rmu_{a_2}}
\ExpBpBpeqBpAA}\nonumber\\
=&\Tact{\frac{D}{\rmd_A^2}\displaystyle\sum_{a'_1,a'_2}\frac{\rmu_{a'_1}\rmu_{a'_2}}{\rmu_{a_1}\rmu_{a_2}}
\ExpBpBpeqBpAB}\nonumber\\
=&\Tact{\frac{\sqrt{D}}{\rmd_A}\displaystyle\sum_{a'_1,a'_2}\frac{\rmu_{a'_1}\rmu_{a'_2}}{\rmu_{a_1}\rmu_{a_2}}
\ExpBpBpeqBpAC}\nonumber\\
=&\Bb_p\ExpBpBpeqBp
\end{align}
where in the second equality uses the association condition, and third the strong condition.

\section{Ground state on a Disk}

We consider a disk without quasiparticles in the bulk. Effectively, on such a disk, we can apply the $\mathcal{T}$ transformation to shrink the bulk graph to a single plaquette, bounded by a circle with outward open edges, as in the equation below. We label the wall edges by $l$'s taking value in $L$ and the open edges by $a$'s taking value in $L_A\subseteq L$, with $A$ a Frobenius algebra. Denote by $N$ the length of the boundary.

Restricting to the Hilbert space $H^{Q=1}$, the bulk Hamiltonian reads $H_{\text{bulk}}=-\frac{1}{D}\sum_{s\in L}\rmd_s B^s_p$ with
\begin{equation}
\begin{aligned}
&\bbra{\BoundaryDiskEff{l'_1}{l'_2}{l'_N}}B^s_p\bket{\BoundaryDiskEff{l_1}{l_2}{l_N}}
\\
=&\rmv_{l_1}\rmv_{l'_1}\rmv_{l_2}\rmv_{l'_2}\dots\rmv_{l_N}\rmv_{l'_N}
G^{a^*_2l^*_2l_1}_{sl'_1l^{\prime*}_{2}}
G^{a^*_3l^*_3l_2}_{sl'_2l^{\prime*}_{3}}
\dots
G^{a^*_1l^*_1l_N}_{sl'_Nl^{\prime*}_{1}}.
\end{aligned}
\end{equation}

The boundary Hamiltonian takes the form $H_{\text{bdry}}=-\sum_n\Bb_{(n,n+1)}$, where $(n,n+1)$ labels the boundary plaquette $p$ sandwiched by the links $(a_n, a_{n+1})$, and $\Bb_{(n,n+1)}=\frac{1}{\rmd_A}\sum_t \Bb^t_{(n,n+1)}$, 

\begin{equation}
\label{eq:BbpEff}
\begin{aligned}
&\bbra{\BpBasisEff{a'_n}{a'_{n+1}}{l_{n-1}}{l'_n}{l_{n+1}}}\Bb^t_{(n,n+1)}\bket{\BpBasisEff{a_n}{a_{n+1}}{l_{n-1}}{l_n}{l_{n+1}}}
\\
=&
\rmu_{a_n}\rmu_{a_{n+1}}\rmu_{a'_n}\rmu_{a'_{n+1}}
\rmv_{l_n}\rmv_{l'_n}\nonumber\\
&\qquad \times f_{a_{n+1}t^{a'_{n+1}}}f_{ta_n{a'_{n}}}
G^{l^*_{n+1}l_na^*_{n+1}}_{ta^{\prime*}_{n+1}l'_n}
G^{l^*_nl_{n-1}a^*_n}_{a^{\prime*}_nt^*l'_n}.
\end{aligned}
\end{equation}

A topologically ordered system on the disk has exactly one ground state, which is the simultaneous $+1$ eigenvector of $B_p$ and $\prod_{n=1}^N\Bb_{(n,n+1)}$. To find the unique ground state, we need to first understand the notion of local ground states on the boundary, which boils down to solving the eigen-problem of $\prod_n\Bb_{(n,n+1)}=1$. It turns out that the local eigenvectors are characterized by $A$-modules over the Frobenius algebra $A$, which is defined as follows.

A (right) module over Frobenius algebra $A$ (or, a $A$-module) is a subset $L_M\subseteq L$  of labels equipped with an action tensor $\rho^a_{j_1j_2}$, with $a\in L_A$ and $j_1,j_2\in L_M$. Note that $L_M$ is not anything \textit{ad hoc} but to be obtained by solving the tensor equations of $\rho^a_{j_1j_2}$. The tensor $\rho^a_{j_1j_2}$ vanishes outside of these subsets and satisfies the following condition.
\begin{equation}\label{eq:ModuleDefProperty}
\sum_{j'}
\rho^{a_{1}}_{j_{1}j'}
\rho^{a_{2}}_{j'j_{2}}
G^{j_{1} a_{1} {j'}^*}_{a_{2}j_{2}^*a'}
\mathrm{v}_ {{j'}}\mathrm{v}_{a'}
=
\rho^{a'}_{j_{1}j_{2}}f_{a_{2} {a'}^* a_{1}},
\end{equation}
which can be understood in terms of Pachner moves:
\begin{align}\label{eq:ModuleDefTensor}
&\Tmv{2}{2}\ExpModuleDefAB\nonumber\\
=&\ExpModuleDefAD\nonumber\\
=&\ExpModuleDefAE.
\end{align}

Let us again take the thick-line convention. We also suppress the indices of the action tensor $\rho$ and put it in a box. In this boxed notation, condition \eqref{eq:ModuleDefProperty} takes the following compact form.
\begin{equation}\label{eq:ModuleDef}
\Tmv{2}{2}\ExpModuleDef=
\ExpModuleDefAA.
\end{equation}
Here a boxed $\rho$ at a vertex means that the tensor $\rho$ is associated with the vertex (e.g., $\rho^c_{j_1j_2}$ on the RHS, with a thick-line summation). Let $Mod_{A}$ collects all the modules over $A$.

The unit condition on the Frobenius algebra $A$ implies the unit condition on $A$-modules:
\begin{equation}\label{eq:unitRho}
\rho^0_{jj}=1.
\end{equation}
If we set $a'=0$, $a_1=a^*_2=a$, $j_2=j_1=j$, and $j'=k$ in Eq. \eqref{eq:ModuleDefProperty}, we get
\begin{equation}\label{eq:RhoRho}
\sum_{k}\rho^a_{jk}\rho^{a^*}_{kj}\rmv_k=\rmv_a\rmv_j,
\end{equation}
which is presented graphically using $T$ move \eqref{eq:TMoveAlternativeSqueezeBubble},
\begin{equation}\label{eq:RhoRhoGraphics}
\Tact\ExpRhoRhoGraphics
=
\frac{\rmd_a}{\sqrt{D}}
\ExpRhoOrthonormalAA.
\end{equation}
Each Frobenius algebra $A$ has a trivial module $M_0=A$. The action tensor of $M_0$ is thus $[\rho_{M_0}]^a_{jk}=f_{ak^*j}$. We shall denote the minimal set of inequivalent $A$-modules by $\{(M,\rho_M)\}$.

Using $\rho$, the local eigenvectors of $\Bb_p=1$, i.e., the basis of local ground states on the boundary, are expressed as
\begin{equation}\label{eq:GSbasis}
\ExpGSbase,
\end{equation}
as can be verified directly
\begin{align}\label{eq:GSbaseProof}
&\Bb_p\ExpGSbase\nonumber\\
=&\Tact\ExpGSbaseAA\nonumber\\
=&\Tact\ExpGSbaseAB\nonumber\\
=&\Tact\ExpGSbaseAC\nonumber\\
=&\ExpGSbaseAD.
\end{align}

The ground state on a disk is non-degenerate. Using the local basis found above, the unique ground state on the disk is expressed as
\begin{equation}
\label{eq:GroundStateDisk}
\ket{\Phi}=\sum_M\frac{\rmd_M}{\rmd_A\sqrt{D}} \ket{\Phi_M},
\end{equation}
where $d_M=\sum_{j\in L_M} d_j$ and $\Phi_M$ is the wavefunction corresponding to the local eigenvector characterized by the $A$-module $(M,\rho_M)$:
\begin{align}
\label{eq:GroundStateWF}
\Phi_M\bpm\GroundStateDiskAA\epm
=
\rmu_{a_1}\rmu_{a_2}\dots
[\rho_{M}]^{a_1}_{l_1l_2}[\rho_{M}]^{a_2}_{l_2l_3}\dots
\end{align}
Here $\{\left (M,\rho_M\right )\}$ are all (inequivalent) irreducible modules over the algebra $A$.

The unique ground state on a disk can also be expressed in terms of $f_{ijk}$ and $6j$-symbols:
\begin{align}\label{eq:GSonDiskfG}
&\ket{\Phi}=\Bb_p\ket{\Phi_{M_0}}
\nonumber\\
&=\sum_{a_1a_2\dots l_1l_2\dots}
\rmu_{a_1}\rmu_{a_2}\dots\rmv_{l_1}\rmv_{l_2}\dots\rmv_{l'_1}\rmv_{l'_2}\dots
\nonumber\\
\times&G^{a_1l_2^*l_1}_{sl'_1l_2^{\prime\star}}
G^{a_2l_3^*l_2}_{sl'_2l_3^{\prime\star}}
\dots
f_{a_1l_2^*l_1}f_{a_2l_3^*l_2}\dots
\bket{\GroundStateDiskAA}.
\end{align}

We now prove that $\ket{\Phi}$ in Eq. \eqref{eq:GroundStateDisk} is a ground state on  the disk. It suffices to show that $B_p\ket{\Phi}=\ket{\Phi}$.
We apply unitary Pachner moves on $\ket{\Phi}$ and get
\begin{align}\label{eq:BoundaryMuttate}
&\sum_{M}\sum_{j}\frac{\rmd_M}{\rmd_A\sqrt{D}}\sum_{a_1a_2\dots}\rmu_{a_1}\rmu_{a_2}\dots\times
\nonumber\\
&\quad\ExpGSdiskAfterT
\end{align}
which can be evaluated using eq. \eqref{eq:RhoCompleteness}
\begin{align}\label{eq:CircleZero}
&\frac{1}{\sqrt{D}}\sum_{j}\rmd_j\sum_{a_1a_2\dots}\rmu_{a_1}\rmu_{a_2}\dots\times
\nonumber\\
&\quad\ExpGSdiskAfterTAA
\end{align}
which is a $B_p=1$ eigenvector. Hence $\ket{\Phi}$ is a ground state on disk.

Certain useful proofs can be found in Appendix \ref{app:GSonDisk}.
We often abuse the notation by referring to $M$ as an $A$-module.

\section{Topological feature of ground states}

In this section, we study the topological feature of the ground state. Namely, we show that the ground states of our Hamiltonian is invariant under Pachner moves.

The bulk topological feature in the case with boundaries is the same as that in the case without boundaries.
We then need only to show the boundary topological feature via the 1+1D Pachner moves on boundaries, which are defined in Eqs. \eqref{eq:TMoveBoundary} through \eqref{eq:TMoveTwoOneAlternative}. 

With boundary, the topological feature can be described as follows. The ground state is invariant under any transformation composed by $T_{2\rightarrow 2},T_{1\rightarrow 3},T_{3\rightarrow 1}$ in the bulk and $\Tmv{1}{2},\Tmv{2}{1}$ on the boundary. Moreover, such transformation is unique: different ways to composing $T$'s results in the same transformation.

To show the uniqueness of the transformation, we consider boundary Pachner moves $\Tmv{1}{2},\Tmv{2}{1}$. Take example of a transformation from $N_1$ tails to $N_2$ tails. The composition of $\Tmv{1}{2},\Tmv{2}{1}$ amounts a graph structure with $N_1$ input edges and $N_2$ output edges, where each trivalent vertex is attached with a multiplication $f$.
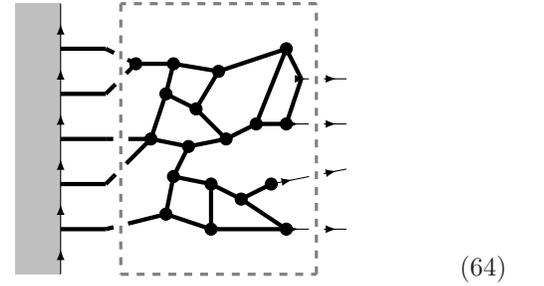
\begin{equation}\label{diag:BoundaryTransform}
\BoundaryTransform
\end{equation}
From the Frobenius condition, the transformation presented by the graph in the dashed box is unique.

\section{Ground states on a cylinder}
\label{sec:EffectiveCylinder}

A topologically ordered system on a cylinder has two boundaries. We can specify the two boundary Hilbert spaces and define the two boundary Hamiltonians by two Frobenius algebras over $L_A$ and $L_B$, respectively. The corresponding multiplications are denoted by $f^{k}_{ij}$ for $i,j,k\in L_A$ and $g^{c}_{ab}$ for $a,b,c\in L_B$.

\begin{figure}[h]
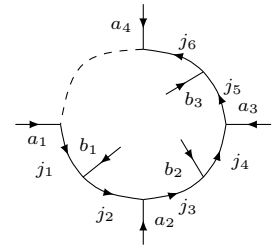

        \centering
        \GroundStateCylinderAA
        \caption{Effective Hilbert space of on the cylinder.}
        \label{fig:EffectiveTheoryCylinder}
\end{figure}

If we consider the states without any bulk quasiparticles, we can completely shrink the bulk graph by Pachner moves, such that the cylinder graph becomes a ring with open edges on both sides of the ring, as in Fig. \ref{fig:EffectiveTheoryCylinder}. Consider the Hilbert subspace spanned by all the labels in the graph. The total Hamiltonian contains two boundary Hamiltonians defined by the two Frobenius algebras. 

The ground states are characterized by the $A$-$B$-bimodules, as will be defined shortly. Each bimodule $P_M$ gives rise to a ground-state wavefunction: 
\begin{equation}\label{eq:cylinderWavefunction}
\begin{aligned}
&\Phi^{\text{cyl}}_M\bpm\GroundStateCylinderAA\epm
\\
=&\left (\prod_{n}\rmu_{a_n}\rmu_{b_n}\right )
\sum_{M} \frac{\rmd_M}{\rmd_A}
\prod_n[P_{M}]^{a_nb_n}_{j_{n-1}j_nk_n}
\end{aligned}
\end{equation}

An $A$-$B$-bimodule is a subset $L_M$ equipped with an action tensor $P^{ab}_{ijk}$, satisfying
\begin{equation}\label{eq:BimoduleDef}
\Tact\ExpBimoduleDef
=
\ExpBimoduleDefAA
\end{equation}

Here the tensor $P_M$ is expressed by a box,
whose meaning is as follows.
\begin{equation}\label{eq:BimoduleBoxConvention}
\ExpBimoduleDefAB
\equiv
\sum_ {j} P^{ab} _ {ijk}\ExpBimoduleDefAC.
\end{equation}
The RHS is independent of $j$ which is summed; hence, $j$ does not appear on LHS. Note that in this work modules and bimodules are multiplicity free (see Appendix \ref{app:multiplicity} for details).

The $A$-$B$-bimodules are subject to the orthonormality and completeness conditions, respectively as follows.  
\begin{equation}\label{eq:BimoduleOrtho}
\frac{D}{\rmd_A^2}\Tact\ExpBimoduleOrtho
=
\delta_{M,N}\frac{\rmd_j}{\rmd_M}\ExpBimoduleOrthoAA.
\end{equation}
\begin{equation}\label{eq:BimoduleCompleteness}
\sum_{M}\frac{\rmd_M}{\rmd_j}[P_M]^{ab}_{jij}
=
\delta_{a,0}\delta_{b,0}\delta_{i,j}\rmd_A^2.
\end{equation}

Both conditions can be proved in a fashion similar to that in the case of $A$-modules.
Now the ground state $\ket{\Phi^{\text{cyl}}_M}$ characterized by the bimodule $P_M$ can be expressed graphically as
\begin{align}\label{eq:GScylinderGrapihcs}
\ket{\Phi^{\text{cyl}}_M}
=&\sum_{a_{1}\dots b_{1}\dots}\mathrm {u}_{a_{1}}\mathrm{u}_{a_{2}}\dots
\mathrm{u}_{b_{1}}\mathrm{u}_{b_{2}}\dots\times
\nonumber\\
&\ExpGroundStateCylinderAC
\end{align}
Let us prove this. First, similar to the disk case, we study the local basis of the ground states on a cylinder.
By local we mean a piece of the ring comprising two neighboring tensors as follows.\begin{align}\label{eq:GScylinderBase}
&\Bb_p\Bb_{p'}\ExpGScylinderBase\nonumber\\
=
&\frac{D}{\rmd_A^2}\Tact\ExpGScylinderBaseAB
\nonumber\\
=&\frac{D}{\rmd_A^2}\Tact\ExpGScylinderBaseAC\nonumber\\
=&\Tact\ExpGScylinderBaseAD\nonumber\\
=&\ExpGScylinderBaseAE.
\end{align}
Hence, $\ket{\Phi^{\text{cyl}}_M}$ is a ground state for the $A$-$B$-bimodule $M$.

Each Frobenius algebra $A$ also has a trivial bimodule $M_0=A$. The action tensor of $M_0$ is thus $[P_{M_0}]^{ab}_{ijk}=f_{aj^*i}f_{bjk^*}$.

\section{Boundary Excitations}

Boundary elementary excitations are $\Bb_p=0$ eigenstates for certain  boundary plaquettes $p$ (we only consider $\Bb_P$ since $\Qb_n=0$ implies $\Bb_p=0$). Boundary elementary excitations support topological quasiparticles. In this section we characterize the excitations and topological quasiparticles, by studying the algebra of local operators $\Bb^t_{p}$. We show that topological quasiparticles are classified by the bimodules over $A$.

The main result is that topological quasiparticles are classified by $A$-$B$-bimodules, which are solutions to Eq. \eqref{eq:BimoduleDef}.
Particularly, if $A=B$, the topological quasiparticles and cylinder ground states are classified by the $A$-$A$-bimodules, which are also solutions to Eq. \eqref{eq:BimoduleDef} for $A=B$. The GSD on the cylinder is identical to the number of quasiparticle species on the boundaries. 

There are three kinds of important operators to characterize quasiparticles. One is a set of orthonormal projection operators as measuring operators to identify quasiparticles. Another is the set of creation operators to create quasiparticle pairs (quasiparticles can not be singly created). The third is a set of hopping operators that can hop a quasiparticle along a boundary. We will construct these three kinds of operators in the following three subsections. Then we discuss the topological feature of quasiparticles in terms of hopping operators. We also discuss fluxons as a special subset of quasiparticles.

\subsection{Measure Quasiparticles}

In this subsection we construct a set of orthonormal projection operators as measuring operators to identify quasiparticles. 

Given a bimodule $M$, define the corresponding measuring operator $\Pi_M$ by
\begin{align}\label{eq:ProjectionM}
&\Pi_{M}\ExpQPprojection\nonumber\\
=
&
\frac{\rmd_M}{\rmd_k}\sum_{a'_{1}a'_{2}}
\frac{\rmu_{a'_{1}}\rmu_{a'_{2}}}{\rmu_{a_{1}}\rmu_{a_{2}}}
\Tact\ExpQPprojectionAA.
\end{align}

Using the orthonormal condition \eqref{eq:BimoduleOrtho} and completeness condition \eqref{eq:BimoduleCompleteness}, we verify that the set $\{\Pi_{Mj}\}$ is orthonormal
\begin{equation}\label{eq:ProjectionsOrthogonal}
\Pi_{M}\Pi_{N}=\delta_{M,N}\Pi_{M},
\end{equation}
\begin{equation}\label{eq:ProjectionComplete}
\sum_{M}\Pi_{M}=\mathds{1}.
\end{equation}
A boundary elementary excitation is a $+1$ eigenstate of certain $\prod_M$.
In particular, however, when $M$ is the trivial module $M_0=A$,
\begin{equation}\label{eq:ParticularProjection}
\Bb_p\Bb_{p'}=\Pi_{M_0};
\end{equation}
hence, the eigenstate of $\prod_{M_0}=1$ is in fact a state without any quasiparticles in the boundary region it acts on. This verifies that $\Pi_{M}$ commutes with $\Bb_p$, and hence indeed identifies the good quantum numbers of elementary excitations.

\subsection{Creation Operators}

The elementary boundary excitations are characterized by topological quasiparticles. On an $A$-boundary component, quasiparticle species are identified with the $A$-$A$-bimodules $M$. Consequently, a boundary elementary excitation with quasiparticles carrying an $A$-$A$-bimodule $M$ would be a $+1$ eigenstate of the measuring operator $\prod_M$. We construct a creation operator $W_{M}$ to create a pair of quasiparticles carrying the bimodule $M$. Below shows how such an operator acts on a boundary section. 
\begin{align}\label{eq:creation}
&W_{Mj}
\ExpQuasiparticleBasisAB
\nonumber\\
=
&\sum_{a'_1ja'_3ka'_5}
\frac{\rmu_{j}\rmu_{a'_2}\rmu_{k}}{\rmu_{a_1}\rmu_{a_2}\rmu_{a_3}}
\Tact{\frac{D}{\rmd_A^2}\ExpQuasiparticleCreationAB}.
\end{align}
for $M\in Mod_{A|A}$. In this example, the operator $W_M$ creates an $M$-type and $M^*$-type quasiparticles on both neighboring open edges $j$ and $k$ of the middle open edge, which becomes $a_3'$. We use wavy lines to indicate the quasiparticles. If $j=k=0$, the quasiparticles become fluxons (to be defined in Section \ref{subsec:fluxon}) residing in the corresponding plaquettes. By acting creation operators on ground states, we get an elementary boundary excitation basis $W_{M}\ket{\Phi}$. 

We now verify that $W_{M}\ket{\Phi}$ is an eigenvector of $\Pi_{M}=1$ in the following. It suffices to verify that
\begin{align}\label{eq:WonPhi}
&\Pi_N\Tact\ExpQPcreationUnderPiNewAA
\nonumber\\
=&\frac{D}{\rmd_A^2}\Tact\ExpQPcreationUnderPiNewAB
\nonumber\\
=&\frac{D}{\rmd_A^2}\sum_{a'_ 2jka'_ 4}\frac{\mathrm{d}_N}{\mathrm{d}_k}\mathrm{u}_{a'_2}\mathrm{u}_ {j}\mathrm{u}_k\mathrm{u}_{a'_ 4}
\Tact\ExpQPcreationUnderPiNewAC
\nonumber\\
=&\delta_{M,N}\Tact\ExpQPcreationUnderPiNewAD
\end{align}
where the last equality is due to the completeness condition \eqref{eq:BimoduleCompleteness}.
Hence $W_{M}\ket{\Phi}$ is a $\Pi_M=1$ eigenvector.

\subsection{Hopping operators}

Quasiparticles can move along the boundary under the hopping operator $H_M$ defined by

\begin{equation}\label{eq:braiding}
H_M\ExpQuasiparticleBasisAA=
\frac{\sqrt{D}}{\rmd_A}
\sum_{j'a'}\frac{\rmu_{a'}\rmu_{j'}}{\rmu_{a}\rmu_{j}}\Tact{\ExpQuasiparticleHoppingAA}.
\end{equation}
that hops an $M$-type quasiparticle initially at the bottom open edge upward across the edge.

The topological feature of elementary excitations can be described using hopping of quasiparticles.

\subsection{Fluxons}
\label{subsec:fluxon}
We consider a subclass of quasiparticles called fluxons. Thus, we can restrict to the Hilbert subspace of $\prod_v\Qb_v=1$. We find that the local operators $\Bb^t_{p}$  form an algebra
\begin{equation}
\label{eq:BbarAlgebra}
\Bb^r_{p}\Bb^s_{p}=\sum_t \dfrac{\rmv_r\rmv_s}{\rmd_A\rmv_t}f_{r^*s^*t}f_{t^*sr} \Bb^t_{p}.
\end{equation}

The quasiparticles occupied at plaquette $p$ are then identified by the orthonormal projection operators
\begin{align}
&\nbar^x_{p}=\sum_t \overline{Y^x_t}\, \Bb^t_{p},
\\
&\nbar^x_{p}\nbar^y_{p}=\delta_{x,y}\nbar^x_{p},
\end{align}
where $Y^x_t$ satisfies the following conditions, as can be derived from Eq. \eqref{eq:BbarAlgebra}:
\begin{equation}\label{eq:YYDef}
\dfrac{1}{\rmd_A}\sum_{rs}\dfrac{\rmv_r\rmv_s}{\rmv_t}f_{r^*s^*t}f_{t^*sr} Y^x_r Y^x_s=Y^x_t.
\end{equation}
Particularly, $\nbar^{x=0}_{p}=\Bb_{p}$ with $ Y^0_t=1 $ for all $ t\in L_A $.

For an excitation $\psi$ with $\nbar^x_{p}=1$ we say $\psi$ supports an $x$-type fluxon at position $p$. Fluxons are a subclass of the full set of topological quasiparticles identified by the bimodules. Indeed, let $Y^x_t=[P_M]^{tt^*}_{0t0}$, then Eq. \eqref{eq:YYDef} is identified with \eqref{eq:BimoduleOrtho}. Hence, fluxon is a special type of quasiparticle identified by those modules $(M,P_M)$ in which $M$ contains $0$.

\section{Examples}

\subsection{Charge boundary}
For any input fusion category, there is always a trivial Frobenius algebra $A_0=0$, such that $\Bb_p$ is trivial and hence the boundary Hamiltonian reduces to
\begin{equation}\label{eq:chargeHamiltonian}
H=-\sum_{v}\Qb_v,
\end{equation}
\begin{equation}\label{eq:chargeBdryQn}
\Qb_n\bket{\TailLabel{a_n}{j_1}{j_2}}=\delta_{a_n,0}\bket{\TailLabel{a_n}{j_1}{j_2}}.
\end{equation}

The $A_0$-modules (and $A_0$-$A_0$-bimodules) are the entire label set $ L$, with $[\rho_j]^0_{jj}=1$ ($[\rho_j]^{00}_{jjj}=1$), $j\in L$.
Boundary quasiparticles are then characterized by labels $j\in L$.

\subsection{LW $\Z_2$ model}
The input fusion category is $\Z_2$; hence the label set is $L=\{0,1\}$, with $\rmd_0=\rmd_1=1$, and $0^*=0,1^*=1$. Fusion rules are the $\Z_2$ group multiplication rule $\delta_{011}=1$, and the $6j$-symbols are
\begin{equation}
G^{{i}{j}{m}}_{{k}{l}{n}}=
\delta_{{i}{j}{m}}
\delta_{{k}{l}\dual{{m}}}
\delta_{{j}{k}\dual{{n}}}
\delta_{{i}{n}{l}}.
\end{equation}

There are two Frobenius algebras, one is the trivial one $A_0=0$, which defines a charge boundary condition. Quasiparticles on the charge boundary are identified with $1$ and $e$, with $e$ a $\mathds{Z}_2$ charge.

The nontrivial Frobenius algebra is $A_1=0\oplus 1$, with $L_A=\{0,1\}=L$. This is a flux boundary. The boundary quasiparticles are identified with $1, m$ with $m$ a $\mathds{Z}_2$-flux.

Cylindrical model has $\mathrm{GSD}=2$ with the charge-charge or flux-flux boundary conditions, and $\mathrm{GSD}=1$ with the charge-flux boundary condition.

\begin{figure}[h]
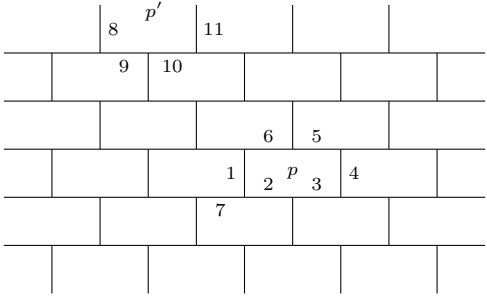

        \centering
        \ZtwoTrivalentGraph
        \caption{A trivalent graph on disk.}
        \label{fig:DiskGrid}
\end{figure}

Consider the model with the flux boundary condition (with the algebra $A_1$) on a disk illustrated in Fig. \ref{fig:DiskGrid}. The total Hamiltonian is
\begin{equation}
H=H_{\text{bulk}}+\epsilon H_{\text{bdry}},
\end{equation}
where $\epsilon$ is a positive constant, and
\begin{equation}
H_{\text{bulk}}=-\sum_v A_v-\sum_p B_p,~H_{\text{bdry}}=-\sum_{p'} \Bb_{p'}.
\end{equation}
Seen in Fig. \ref{fig:DiskGrid}, examples of these operators are
\begin{equation}
A_v=\sigma^z_1\sigma^z_2\sigma^z_7,
B_p=\sigma^x_1\sigma^x_2\sigma^x_3\sigma^x_4\sigma^x_5\sigma^x_6\sigma^x_7,
\end{equation}
and
\begin{equation}
\Bb_{p'}=\sigma^x_8\sigma^x_9\sigma^x_{10}\sigma^x_{11}.
\end{equation}

If we consider states without quasiparticles in the bulk, we can simplify the problem with the effective theory on the disk as defined on a chain, see Fig. \ref{fig:DiskZ2}. The Hilbert space is spanned by $N+1$ spins: on $N$ external edges and one internal edge denoted by $0$ (the spins on all other internal edges are determined by the fusion rules). We require a global constraint
\begin{equation}
\prod_{n=1}^N \sigma^z_n=1.
\end{equation}

\begin{figure}[h]
        \centering
        \ZtwoTrivalentGraphAB
        \caption{$\Z_2$ effective boundary theory.}
        \label{fig:DiskZ2}
\end{figure}

The bulk Hamiltonian is reduced to
\begin{equation}
H_{\text{bulk}}=-\sigma^x_{0}.
\end{equation}

The boundary Hamiltonian is
\begin{equation}
H_{\text{bdry}}=-\sum_{n=1}^{N-1}\sigma^x_n\sigma^x_{n+1}-\sigma^x_{0}\sigma^x_{N}\sigma^x_1.
\end{equation}
The extra $\sigma^x_0$ in last term is due to the nontrivial action of $\Bb_{p'}$ on the spin at edge $0$.

Similarly, we have an effective theory for the cylinder as defined on a chain with the flux-flux boundary condition (Fig. \ref{fig:CylinderZ2}). Suppose we have $N$ external edges on both boundaries of the cylinder. Then the Hilbert space is spanned by the $2N+1$ spins, with the global constraint
\begin{equation}
\prod_{n}^N\sigma^z_{n}\prod_{n'}^N\sigma^z_{n'}=1.
\end{equation}
Here we denote external edges on one boundary by $n=1,\dots N$ and on the other boundary by $n'=1',\dots N'$.

The plaquettes in the bulk region are completely removed by the Pachner moves. Hence we have only two boundary Hamiltonians:
\begin{equation}
\begin{aligned}
H_{\text{bdry}}
=&-\sum_{n=1}^{N-1} \sigma^x_n\sigma^x_{n+1}-\sigma^x_0\sigma^x_N\sigma^x_1
\\
&\qquad
-\sum_{n'=1}^{N-1} \sigma^x_{n'}\sigma^x_{(n+1)'}-\sigma^x_0\sigma^x_{N'}\sigma^x_{1'}.
\end{aligned}
\end{equation}

\begin{figure}[h]
        \centering
        \ZtwoTrivalentGraphAA
        \caption{$\Z_2$ effective boundary theory.}
        \label{fig:CylinderZ2}
\end{figure}

\subsection{LW Fibonacci model}

The input fusion category is the Fibonacci category with string types $L=\{0,2\}$, 
also denoted by $\{\mathbf{1},\tau\}$. Let $\phi=\frac{1+\sqrt{5}}{2}$ be the golden ratio. The quantum dimensions of $0$ and $2$ are $\rmd_0=1$ and $\rmd_2=\phi$. The fusion rules are 
\begin{equation}
\label{Fibbranchingrule}
\delta_{000}=\delta_{022}=\delta_{222}=1,\delta_{002}=0,
\end{equation}
and the nonzero independent $6j$-symbols $G$ are given by
\begin{align}
\label{Fib6js}
G^{000}_{000}=1,
G^{022}_{022}=G^{022}_{222}=1/\phi,
\nonumber\\
G^{000}_{222}=1/\sqrt{\phi},
G^{222}_{222}=-1/{{\phi}^2}.
\end{align}

The Fibonacci category has two Frobenius algebras: the trivial one $A_0=0$  and the nontrivial one $A_1=0\oplus 2$. The $A_0$-modules are $N_0=0$ and $N_1=2$, with action tensor $[\rho_j]^0_{jj}=1$. This defines a charge boundary condition.

For $A_1=0\oplus 2$, set $L_{A_1}=\{0,2\}$. The only nontrivial multiplication reads
\begin{equation}
f_{222}=\phi^{-3/4}.
\end{equation}
$A_1$has two modules: (1) $M_0=0\oplus 2$, i.e., $A_1$ itself, with action morphism being the multiplication $\rho^a_{jk}=f_{ak^*j}$; (2). $M_1=2$, with action morphism given by
\begin{equation}
[\rho_1]^2_{22}=-\phi^{-1/4}.
\end{equation}

The two algebras $A_0$ and $A_1$ are Morita equivalent, hence giving rise to the same boundary condition.

\subsection{LW Ising model}
The input fusion category is the Ising category, with $L=\{0,1,2\}$, also denoted by $\{\mathbf{1},\sigma,\psi\}$. The quantum dimensions  are $\rmd_0=1$, $\rmd_1=\sqrt{2}$, and $\rmd_2=1$. The fusion rules are 
\begin{equation}
\label{eq:Isingbranchingrule}
\delta_{000} = 1,
\delta_{011} = 1, 
\delta_{022} = 1,
\delta_{112} = 1, 
\end{equation}
and the nonzero $6j$-symbols $G$ are 
\begin{align}
\label{eq:IsingG}
\begin{aligned}
&G_{000}^{000}=1,G_{111}^{000}=\frac{1}{\sqrt[4]{2}},
G_{222}^{000}=1,G_{011}^{011}=\frac{1}{\sqrt{2}},\\
&G_{122}^{011}=\frac{1}{\sqrt[4]{2}},G_{211}^{011}=\frac{1}{\sqrt{2}},G_{022}
^{022}=1,G_{112}^{112}=-\frac{1}{\sqrt{2}}.
\end{aligned}
\end{align}
There are two Frobenius algebras: the trivial one $A_0=0$, giving rise to the charge boundary condition, and $A_1=0\oplus 2$.

The Frobenius algebra $A_0=0$ has three modules $N_0=0$, $N_1=1$, and $N_2=2$, which are labels in $L$.

The Frobenius algebra $A_1=0\oplus 2$ has three modules: (1) $M_0=0\oplus 2$, with $[\rho_0]^2_{20}=[\rho_0]^2_{02}=1$.
(2). $M_1=1$, with $[\rho_1]^2_{11}=1$.
(3). $M_2=1$, with $[\rho_2]^2_{11}=-1$.

\section{Equivalent boundary conditions}

The boundary conditions are classified by $A$-$A$-bimodules, in the sense that boundary elementary excitations with good quantum numbers are identified  with equivalent bimodules.

In this section, however, we discuss a situation where two different Frobenius algebras in a unitary fusion category give rise to equivalent boundary conditions. Two Frobenius algebras $A$ and $B$ are Morita equivalent if category $Mod_A$ of $A$-modules is equivalent to $Mod_B$.\cite{Ostrik2001} By the previous analysis, the local ground state basis is characterized by modules. Hence, Morita equivalent Frobenius algebras define equivalent boundary conditions.

For any $A$-module $M$, $k\otimes M$ is also a right module. But $k\otimes M$ is reducible; hence, we can decompose $k\otimes M$ into a direct sum of irreducible modules. To do so, we need to study the equivalence between $k\otimes M$ and some other irreducible module $N$. Define a morphism $\eta:k\otimes M\rightarrow N$ as a tensor $\eta$ satisfying
\begin{equation}\label{eq:EtaDefAA}
\Tact\ExpGSbaseAI=\ExpGSbaseAJ
\end{equation}
We denote the number of independent solutions of $\eta$ by $N_{jM}^N$, called the fusion rule in $Mod_A$.

In the following, we give a practical check of the Morita equivalence between $Mod_A$ and $Mod_B$. Frobenius algebra $A$ is equivalent to $B$ if
the following two conditions hold.
(1) All irreducible $A$-modules $M_0,M_1,\dots,M_n$ are mapped to irreducible $B$-modules $M'_0,M'_1,\dots,M'_n$.
(2) The fusion rule is preserved by the mapping.

For example, in Fibonacci case, the two Frobenius algebras are Morita equivalent. One can easily verify that the fusion rules
\begin{equation}\label{eq:equivalentModules}
2\otimes M_0=M_0\oplus M_1,2\oplus M_1=M_0
\end{equation}
are equivalent to
\begin{equation}\label{eq:equivalentModulesAA}
2\otimes N_0=N_1,2\otimes N_1=N_0\oplus N_1,
\end{equation}
by mapping $M_0\rightarrow N_1$ and $M_1\rightarrow N_0$.
Hence, the two Frobenius algebras are Morita equivalent and give rise to the same boundary conditions.

In the Ising case, the two Frobenius algebras are also Morita equivalent by mapping $M_1\rightarrow N_2,M_2\rightarrow N_1$ and $M_3\rightarrow N_3$. One verifies the fusion rules
\begin{align}\label{eq:MoritaEquivalenceVerifyIsing}
&1\otimes M_1=M_2\oplus M_3,1\otimes M_2=M_1,1\otimes M_3=M_1,
\nonumber\\
&2\otimes M_1 = M_1, 2\otimes M_2=M_3,2\otimes M_3=M_2,
\end{align}
are preserved under the mapping.
Hence, the two Frobenius algebras are Morita equivalent and give rise to the same boundary conditions.

\section{Relation to the Kitaev-Kong formulation}\label{sec:KK}

We used a Frobenius algebra to define the Boundary theory in this paper. This formulation is closely related to Kitaev and Kong's work\cite{Kitaev2012} that formulates boundary theories using module categories over $\mathcal{C}$. In this section we will discuss the relation between our approach and the Kitaev-Kong (KK) formulation.

In our approach, we take boundary degrees of freedom from the labels of the input UFC---the same degrees of freedom as in the bulk, and we start with local boundary Hamiltonians. To write down a ``good'' boundary Hamiltonian we examine the (unitary representation of) 1+1D boundary Pachner moves. The desired form of the Hamiltonian will be one such that the ground-state Hilbert space is invariant under all bulk and boundary Pachner moves. This invariance leads to a Frobenius algebra structure appearing in the boundary Hamiltonian operators.

With Hilbert space spanned by labels of the input UFC, all operators are explicitly expressed using these labels. Our approach is convenient for computational purposes yet rigorous in characterizing the topological properties. The reader can compute the ground states and excitations by solving the Hamiltonian eigen-problems without knowledge of categories.

Given the bulk Levin-Wen model with input fusion category $\mathcal{C}$, in the KK formulation, the input data to specify the boundary degrees of freedom and boundary operators is a module category $\mathcal{M}$ over $\mathcal{C}$. The topological feature of the boundary ground states comes from the compatibility conditions between the bulk degrees of freedom in $\mathcal{C}$ on the left side and boundary degrees of freedom in $\mathcal{M}$. Here and after we assume the bulk is on the left of a boundary. This is always possible if one tracks along the boundary clockwise.

In the following we will build up the correspondence between the Hilbert space structures in KK formulation and our formulation, by studying the eigen-problem of $\Bb_p=1$, where $p$ label the boundary plaquettes.

Given a boundary $\Gamma$, the local basis of boundary ground states (i.e., the $\prod_p\Bb_p=1$ eigenstates) has been discussed in previous sections and has the form
\begin{equation}\label{eq:GSbasisAA}
\ExpGSbase
\end{equation}
for some  $M\in Mod_A$.
This basis is defined for simple boundary plaquettes but can be generalized to cover the cases with generic boundary plaquettes, where bulk edges must also be taken into account. We write the generic form of potential basis vectors as
\begin{equation}\label{eq:GSbaseAE}
\ExpGSbaseAE,
\end{equation}
where we assume two potentially different modules $M$ and $N$ and a tensor $\eta$ connected to bulk edge $k$, to be determined by the condition $\Bb_p=1$.

Acting $\Bb_p$ on such states yields
\begin{equation}\label{eq:GSbaseAF}
\frac{\sqrt{D}}{\rmd_A}\Tact\ExpGSbaseAF.
\end{equation}

The above is a $\Bb_p=1$ eigenvector if and only if
\begin{equation}\label{eq:EtaDef}
\frac{\sqrt{D}}{\rmd_A}\Tact\ExpGSbaseAG
=
\ExpGSbaseAH.
\end{equation}
This condition is equivalent to the defining property \eqref{eq:EtaDefAA} of the morphisms in $Hom(k\otimes M,N)$ of the category $Mod_A$.

\begin{figure}[!ht]
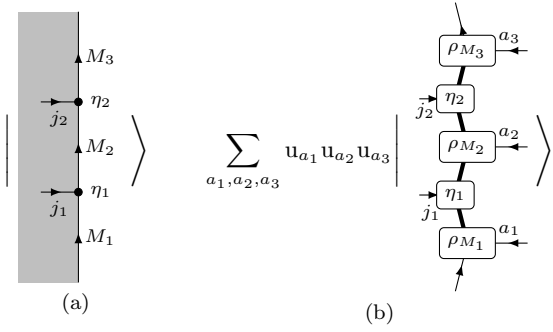

        \centering
        \subfigure[]{$\ExpGSbaseAK$}
        \qquad
        \subfigure[]{$\ExpGSbaseAL$}
        \caption{(a)The Hilbert space structure in the KK formulation. The boundary degrees of freedom live in a module category. (b) The Hilbert space structure in our formulation. With $Mod_A$ treated as the module category, the local basis of boundary ground states recovers the KK Hilbert space structure.}
        \label{fig:KKcorrespondenc}
\end{figure}

Therefore, on a generic boundary graph, the local basis of boundary ground states is characterized by modules $M$'s in $Mod_A$ and morphisms $\eta$ in $Mod_{A}$. It is known that $Mod_A$ is equivalent to a module category. We can use such module category data $M$ and $\eta$ as input degrees of freedom to describe the ground states. See Fig. \ref{fig:KKcorrespondenc}. Hence, we build up the mapping between the Hilbert space in our formulation and that in the KK formulation on the level of ground states. 

This mapping is two-way, which follows from a mathematical theorem: the category of right modules over an algebra $A$ in $\mathcal{C}$ is equivalent to the right module category over (unitary fusion) $\mathcal{C}$\cite{Ostrik2001}. The mapping is many to one. Namely, two Frobenius algebras $A$ and $B$ are Morita equivalent if $Mod_A$ is equivalent to $Mod_{B}$, and they specify the same boundary condition.

In the KK formulation, the boundary excitations are constructed using module functors $Fun(\mathcal{M},\mathcal{M})$ of the input module category $\mathcal{M}$. In our formulation, however, the elementary excitations are identified with the bimodules $Mod_{A|A}$. In this paper, we are not going to directly prove the equivalence of elementary excitations in the two formulations. Nevertheless, it is known that\cite{Ostrik2001} if $\mathcal{M}$ is taken to be $Mod_A$ then the category $Fun(\mathcal{M},\mathcal{M})$ is equivalent to $Mod_{A|A}$. Therefore, we expect our formulation also agrees with the KK formulation on boundary elementary excitations.

By above analysis, we show that our approach with an Frobenius algebra $A$ is equivalent to the KK formulation with input module category $\mathcal{M}=Mod_A$.

\appendix

\section{Some proofs and details}

\subsection{Frobenius algebra}\label{app:FrobeniusAlgebra}

\noindent {\bf Property:} The unit and cyclic conditions are consequences of the association condition, via appropriate choice of normalization.

\noindent {Proof}: We present a choice of normalization of $f$. 
The $f_{abc}$ is determined up to a continuous transformation. If $f_{abc}$ is a solution, then for any nonzero complex function $\xi_a$
\begin{equation}
f'_{abc}=f_{abc}\frac{\xi_{c^*}}{\xi_a\xi_b}
\end{equation}
is also a solution. 

Since $f_{bb^*0}\neq 0$ for all $b\in L_A$. For each dual pair $a=b^*$, set $\xi_a=\xi_b=\sqrt{f_{ab0}}$ (the order of $a,b$ and the sign of the square root are randomly chosen but fixed once for all). We have
        \begin{equation}
        f_{jj^*0}=1.
        \end{equation}
        Particularly, we have
        \begin{equation}
        f_{000}=1.
        \end{equation}
        
        Now we prove the property with this normalization choice.
        
        Let $a=0$ in eq. \eqref{eq:fcond}. Using $G^{0bc^*}_{de^*g}{\rmv_c\rmv_g}=\delta_{b,c}\delta_{e,g}$, the equation implies
        \begin{equation}
        f_{0bb^*}=f_{0gg^*}
        \end{equation}
        since $f_{000}=1$ we get
        \begin{equation}
        f_{0bb^*}=1.
        \end{equation}
        Similarly, setting $c=0$ in Eq. \eqref{eq:AssociationGraphical}  results in
        \begin{equation}
        f_{a0a^*}=1.
        \end{equation}

        By setting $e=0$ in Eq. \eqref{eq:AssociationGraphical}, together with that $G^{abc^*}_{d0g}{\rmv_c\rmv_g}=\delta_{a,g^*}\delta_{d,c^*}$, we obtain the cyclic symmetry condition
        \begin{equation}
        f_{abc^*}=f_{bc^*a}.
        \end{equation}

\subsection{Ground states on a disk}\label{app:GSonDisk}

\textit{Orthonormality and Completeness}
\begin{equation}\label{eq:orthonormal}
\frac{1}{\rmd_A}\sum_{a}[\rho_M]^a_{jk}[\rho_M]^{a^*}_{kj}\frac{\rmv_a\rmv_k}{\rmv_j}
=
\frac{\rmd_k}{\rmd_M}
\end{equation}
\begin{equation}\label{eq:RhoCompleteness}
\sum_M\frac{\rmd_M}{\rmd_j}[\rho_M]^a_{jj}=\delta_{a,0}\rmd_A
\end{equation}
Orthonormality condition \eqref{eq:orthonormal} is expressed graphically as
\begin{equation}\label{eq:orthonormalGraphics}
\frac{\sqrt{D}}{\rmd_A}\Tact\ExpModuleOrtho
=\delta_{M,N}\frac{\rmd_k}{\rmd_M}\ExpRhoOrthonormalAA
\end{equation}

Sketch proof: Given a minimal set of modules $\{\rho_M\}$, we have
\begin{align}\label{eq:ProofOrthonormal}
\Tact\ExpModuleOrtho
=
\delta_{M,N}\beta \ExpRhoOrthonormalAA
\end{align}
for $\beta\in\mathds{C}$ independent of $j$. To compute $\beta$, we evaluate
\begin{align}\label{eq:ProofOrthonormalAA}
&\sum_{j}\Tact\ExpModuleOrthoAA
=\sum_{j}\Tact\ExpModuleOrthoAC\nonumber\\
&\quad=\beta\sum_{j\in L_M}\Tact\ExpModuleOrthoAB
\end{align}
Using eq. \eqref{eq:RhoRhoGraphics}, the last equality is evaluated explicitly as
\begin{equation}\label{eq:ProofOrthonormalAB}
\rmd_k \rmd_A/\sqrt{D}=\beta \rmd_M
\end{equation}
Hence we proved eq. \eqref{eq:orthonormal}.

We now prove completeness condition. using eq. \eqref{eq:orthonormal} and \eqref{eq:ModuleDef}, we have
\begin{equation}\label{eq:ProofRhoCompleteness}
\sum_{M}\frac{\rmd_M}{\rmd_j}\frac{\sqrt{D}}{\rmd_A}\Tact\ExpModuleCompleteness
=
\ExpModuleCompletenessAA
\end{equation} 
Introduce tensor $\mathds{1}$ with $\mathds{1}^a_{jj}=\sqrt{D}\delta_{a,0}$, which satisfies
\begin{equation}\label{eq:unitOne}
\Tact\ExpModuleCompletenessAB
=
\ExpModuleCompletenessAA.
\end{equation}

Compare Eq. \eqref{eq:unitOne} and Eq. \eqref{eq:ProofRhoCompleteness}, we arrive at Eq. \eqref{eq:RhoCompleteness}.

\subsection{Bimodules with multiplicity}\label{app:multiplicity}
In general, the action tensor of modules $\rho$ and bimodules $P$ carries extra indices, say, $\alpha$ and $\beta$. The action tensor of a bimodule $P_M$ is now expressed by
\begin{equation}\label{eq:BimoduleBoxConventionExtraIndices}
\ExpBimoduleDefAD
\equiv
\sum_ {j} P^{ab,\alpha\beta} _ {ijk}\ExpBimoduleDefAC.
\end{equation}
In defining property \eqref{eq:BimoduleDef}, for a thick line on the LHS we should also sum over appropriate $\alpha$ indices of the two action tensors. Similarly, in general we need to put extra indices to a module tensor action and follow the same convention. Nevertheless, the discussion and derivation throughout the paper remains true when we add the extra indices to tensor actions and add the corresponding summation rule to the thick line convention. Therefore, in the rest of paper, we suppress the $\alpha$ indices for simplicity.

 \acknowledgements{We thank Zhenghan Wang, Liang Kong, Davide Gaiotto for helpful discussions. YDW is also supported by the Shanghai Pujiang program.}

\vskip 3cm
\bibliographystyle{apsrev}
\bibliography{StringNet}

\end{document}